\documentclass[rmp,amsmath,amssymb,graphicx,twocolumn,superscriptaddress,nolongbibliography]{revtex4-1}

\usepackage{bm}

\usepackage{soul}
\usepackage{url}
\usepackage{color}
\usepackage{graphicx}
\usepackage{bm}
\usepackage{amsmath}
\usepackage{amssymb}
\usepackage{gensymb}
\usepackage{tikz}
\newcommand*\circled[1]{\tikz[baseline=(char.base)]{
            \node[shape=circle,draw,inner sep=2pt] (char) {#1};}}

\newcommand\notn{/\mkern-11mu {{\bm{n}}}}

\begin{document}

\title{Helfrich-Hurault elastic instabilities driven by geometrical frustration}

\author{Christophe Blanc}
\affiliation{UMR CNRS 5221, Laboratoire Charles Coulomb, Universit\'{e} Montpellier, place Eug\`{e}ne Bataillon, 34095 Montpellier, Cedex 5, France}
\author{Guillaume Durey}
\affiliation{School of Engineering, Brown University, Providence RI 02912, USA}
\affiliation{UMR CNRS 7083 Gulliver, \'{E}cole Sup\'{e}rieure de Physique et de Chimie Industrielles de la Ville de Paris, PSL Research University, 75005 Paris, France}
\author{Randall D. Kamien}
\affiliation{Department of Physics and Astronomy, University of Pennsylvania, Philadelphia PA 19104, USA}
\author{Teresa Lopez-Leon}
\affiliation{UMR CNRS 7083 Gulliver, \'{E}cole Sup\'{e}rieure de Physique et de Chimie Industrielles de la Ville de Paris, PSL Research University, 75005 Paris, France}
\author{Maxim O. Lavrentovich}
	\email{mlavrent@utk.edu}
\affiliation{Department of Physics \& Astronomy, University of Tennessee, Knoxville TN 37996-1200, USA}
\author{Lisa Tran}
	\email{l.tran@uu.nl}
\affiliation{Department of Physics, Soft Condensed Matter \& Biophysics, Debye Institute for Nanomaterials Science, Utrecht University, Utrecht 3584 CC, The Netherlands}

\date{\today{}}

\begin{abstract}
The Helfrich-Hurault (HH) elastic instability is a well-known mechanism behind patterns that form as a result of strain upon liquid crystal systems with periodic ground states. In the HH model, layered structures undulate and buckle in response to local, geometric incompatibilities, in order to maintain the preferred layer spacing. Classic HH systems include cholesteric liquid crystals under electromagnetic field distortions and smectic liquid crystals under mechanical strains, where both materials are confined between rigid substrates. However, richer phenomena are observed when undulation instabilities occur in the presence of deformable interfaces and variable boundary conditions. Understanding how the HH instability is affected by deformable surfaces is imperative for applying the instability to a broader range of materials. In this review, we re-examine the HH instability and give special focus to how the boundary conditions influence the mechanical response of lamellar systems to geometrical frustration. We use lamellar liquid crystals confined within a spherical shell geometry as our model system. Made possible by the relatively recent advances in microfluidics within the past 15 years, liquid crystal shells are composed entirely of fluid interfaces and have boundary conditions that can be dynamically controlled at will. We examine past and recent work that exemplifies how topological constraints, molecular anchoring conditions, and boundary curvature can trigger the HH instability in liquid crystals with periodic ground states. We then end by identifying similar phenomena across a wide variety of materials, both biological and synthetic. With this review, we aim to highlight that the HH instability is a generic and often overlooked response of periodic materials to geometrical frustration. 

\vspace{5cm}
\end{abstract}

\maketitle

\tableofcontents{}

\section{Introduction}

\begin{figure}[!ht]
\includegraphics[width=.9\columnwidth]{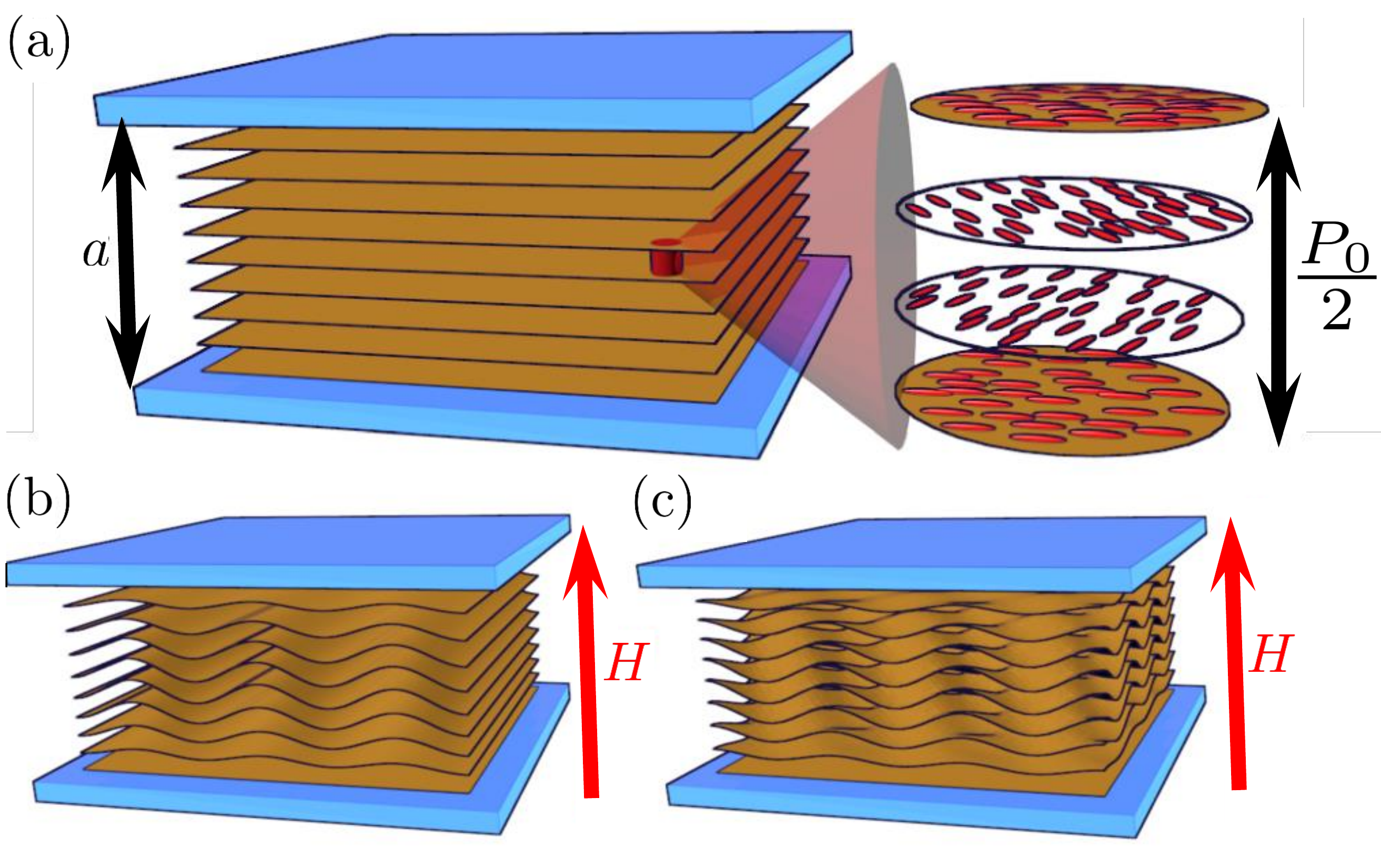}
\caption{ The classic Helfrich-Hurault instability in a short-pitched cholesteric. The mesophase is confined between solid substrates with planar anchoring and can be described as a lamellar system of period $P_0/2$ (a). Undulation in the periodic layers of the cholesteric along a single (b) or, more realistically, along two perpendicular directions (c) develops under an applied magnetic field ($H$) of sufficient magnitude. }
\label{fig:cholHH}
\end{figure}

Subjected to shear, solids strain but fluids flow: what else can happen?  Between solid and liquid lie the liquid crystalline phases of matter: like a crystal they transmit torque and shear stresses but only in {\sl some} directions and geometries.  For instance, the long-range orientational order of a nematic liquid crystal, a phase where the rod-like constituents tend to point in the same direction (the director), implies that if a rod is rotated away from its preferred direction in one region, its surroundings will rotate with it.  Nematics do not have translational order, so they do not support shear stresses. However, smectic and cholesteric liquid crystals do.  Smectics break translational symmetry by having the rod-like molecules sort into layers, resulting in a density modulation. Cholesterics form ``pseudolayers", maintaining a constant density throughout the material, but break translational symmetry due to a helical twisting of the director. The thickness of a cholesteric pseudolayer is defined by a rotation of the molecules by $\pi$, as represented in Fig.~\ref{fig:cholHH}a. Both smectics and cholesterics have one-dimensional periodicity in three-dimensional samples, like a messy stack of cards.  When extensional shear is applied to a structure with a preferred layer spacing, the layers can undulate in order to maintain their preferred distance.  This sort of response was studied by W. Helfrich and J.P. Hurault in the early 1970s within the context of electromagnetic instabilities, depicted in Fig.~\ref{fig:cholHH}b and 1c for a cholesteric \cite{helfrich_electrohydrodynamic_1971,hurault1}.  Today, we refer to all of these undulating instabilities in layered systems as the ``Helfrich-Hurault'' (HH) instability.  

The undulatory deformations of the HH instability are similar in spirit to the martensitic patterns seen in crystals, where changes in a crystal structure require accompanying volumetric changes \cite{ball1992proposed}. Indeed, smectic liquid crystals have even been described as ``the weirdest martensite.'' \cite{PhysRevLett.116.147802}.  Within a smectic, layers can break and rejoin, creating topological defects, localized regions of disorder, such as dislocations and disclinations.  In general, topological defects result from system \textit{frustration} that can arise from either local geometrical effects, \textit{i.e.}, geometrical frustration, or global geometrical effects, \textit{i.e.}, topological frustration. 

For example, Frank-Kasper phases, which catalogue the numerous possible arrangements of atoms in complex alloys, are a renowned, historical illustration of structure from frustration \cite{frank-kasper}. The most locally compact packing of four rigid, identical spheres is tetrahedral, in which each corner of the tetrahedron represents the center of each sphere. However, imperfections, \textit{i.e.} defects, arise when the tetrahedron is the unit structure for tiling space. Tetrahedrons cannot fill space completely without distortion --- their symmetry conflicts with a translation symmetric tessellation since the dihedral angle of a tetrahedron is not commensurable with 2$\pi$ \cite{sadoc2006geometrical, klem-curvedcryst}. Defects are necessarily present in the system because the packing is ``limited'' by the shape of the packing unit. Although tetrahedrons are unable to fill space, this ``limitation'' actually has higher local densities and greater vibrational entropy compared to face-centered-cubic or hexagonal close-packings. This enables a wide array of possible configurations that gives the packings of Frank-Kasper phases freedom to deform in order to accommodate neighboring atoms \cite{klem-curvedcryst}. Frank-Kasper phases demonstrate that not only are defects often necessary to stabilize systems, but that they can also be constructed from a purely geometrical argument. The regular network of disclinations in Frank-Kasper phases requires only the tiling of polytetrahedra to be constructed.

It is not a coincidence that Sir F. Charles Frank is the same ``Frank'' of both Frank-Kasper phases and the Frank free energy density of a liquid crystal  --- underlying both formulations is the importance of geometry in the description of material properties. Classic examples of geometrical frustration in liquid crystals are  blue phases --- states that emerge when it is favorable to introduce defects to minimize the chiral elastic energy of the bulk. As it is for the network of defects in Frank-Kasper phases, the defect networks in blue phases also emerge from geometrical frustration \cite{sethna-bluephase}. 
Similar to how the imperfect packing of pentagons on a plane can be made perfect when the plane is \textit{curved} into a spherical topology, the disclination line networks of blue phases are removed from blue phases in the curved space of $\mathbb{S}^3$. The defects in blue phases can be thought of as the consequence of ``folding out" the 3-sphere onto Euclidean space. The conflict between local and global order, as demonstrated by Frank-Kasper and blue phases, is a signature of geometrical frustration.   For a thorough review of blue phases, we recommend \cite{rmp-bluephases}.

However, in these and countless other systems, geometrical frustration is often accompanied by topological frustration, depending on the global structure of the phase.  Using the Gauss-Bonnet theorem, for instance, it is possible to locally measure the Gaussian curvature of a patch of surface just by studying the curvature of closed loops.  If you can measure the curvature everywhere, it is then possible to deduce the global topology of the surface only if it has no boundaries or if somehow the boundary conditions are precisely defined.  In some cases, the boundary can be interpreted as yet another defect at infinity. However, setting aside considerations of the boundaries for now, the important issue here is that geometrical frustration causes problems in {\sl your} neighborhood: even if the Earth were a hemisphere that ended with a precipice at the equator, we would still not be able to draw perfect polygons on it.  Either the angles would not be quite right, the edge lengths would be unequal, or you could not get it to lie directly against the Earth.  {\sl This} is geometrical frustration: a fundamental incompatibility between one set of shapes (the polygons) and the others (the Earth).   Topological frustration needs to be solved {\sl somewhere}; geometrical frustration has to be solved {\sl everywhere}.

Liquid crystals are the ideal systems to differentiate geometrical frustration from topological frustration. Most liquid crystal systems have open boundaries and the notion of global topology is moot -- defects can end on the interfaces between phases or at the sample wall, and they can transform from bulk defects to boundary defects.  The frustration can come about because the geometric parameters do not match (too many sardines in the can), the shapes do not match (square peg, round hole), or, as in the blue phase, there is a local geometry (double-twist) that cannot be extended into the whole volume.  The softness of liquid crystals, the ability to control and monitor their boundary conditions, and the relatively straightforward method of real-space detection of defects allows us to explore and differentiate local from global frustration.  

Here, we focus specifically on lamellar liquid crystal phases, where geometrical frustration is often relieved through the HH instability. To apply the HH instability beyond the classical systems, we also give additional scrutiny to boundary conditions. Lamellar liquid crystal phases that exhibit this instability are pervasive in nature, seen within a wide array of biological materials ranging from plant cell walls to arthropod cuticles \cite{bouligand,plantcellwall,nacre,rey-lcmodelsbio,beetle,mitov-softmatter}. In the first studies by Helfrich and Hurault in the 1970s, the HH instability was examined in lamellar liquid crystals confined between two solid substrates, with undulation in the layers of liquid crystal induced by electromagnetic fields \cite{helfrich_electrohydrodynamic_1971,hurault1}. Yet, many lamellar liquid crystals, including those found in living matter, often have deformable boundaries at fluid interfaces, where periodic layer undulations can occur in the absence of external driving. Elucidating the coupling between such boundaries and bulk deformations is necessary to apply the HH instability to a broader class of materials that undulate to relieve geometrical frustration, including in the morphogenesis of biological liquid crystals. 

To isolate the effects of a fluid boundary on liquid crystals within the laboratory, a synthetic system must have both deformable interfaces and tunable thicknesses to control the balance between bulk and surface forces. An experimental system ideal for this purpose is a liquid crystal shell, made possible in 2005 by the seminal work of Utada \textit{et al.} on microfluidics \cite{Utada_DoubleEmulsions}. As first demonstrated by Fernandez-Nieves \textit{et al.}, using a liquid crystal as the middle phase in the production of water-in-liquid-crystal-in-water double emulsions, a thin film of liquid crystal can be made free-standing and stable \cite{fernandez-nieves-shell}. With simple adjustments of flow rates and/or the addition of solutes in the surrounding aqueous phases, both the thickness of the liquid crystal shell and the molecular anchoring at the shell interfaces can be dynamically varied at will. Liquid crystal shells are then model systems for probing the role of curved and possibly deformable boundaries both in triggering the HH instability and in stabilizing the resultant defect structures. 

The purpose of this review is two-fold. First, the HH instability is detailed as a mechanism of pattern formation that results from frustration in lamellar liquid crystals, taking special care to distinguish geometrical versus topological effects. Second, the HH instability is not only historically reviewed, but recent work on cholesteric and smectic liquid crystal shells is presented, to illustrate the mechanisms through which deformable boundaries can influence and trigger bulk layer undulations. 

In the following section, we briefly review the elasticity of liquid crystals. In Section III, we dive into the history of the HH instability and detail the classic HH systems, where lamellar liquid crystals are confined between solid substrates. In Section IV, we consider liquid crystals with free, deformable interfaces and describe our model system: the liquid crystal shell. We then characterize the HH instability in cholesteric shells in Section V, where undulations can arise due to topological frustration and surface anchoring. We then move to smectic shells in Section VI, where the HH instability is triggered by geometrical frustration due to boundary curvature. We end by identifying the HH instability across a wide range of elastic materials, both synthetic and biological.

\section{The \textbf{\textit{Dramatis Person\ae}}}
Before plunging in, we pause briefly to outline liquid crystal elasticity.  There are any number of excellent and thorough textbooks \cite{dgennes-prost,cha95,kleman-odl}, that cover this but here we offer the reader a highly abridged review.  The simplest of the liquid crystalline phases is the nematic.  In this phase, a preferred, ``long''-axis of the molecules aligns along a local direction, represented by a unit vector $\bm{n}$.  At first glance this would appear to be equivalent to a magnet where $\bm{n}$ would take the place of the local spin, $\bm{m}$, but the nematic phase has an additional symmetry: $\bm{n}$ and $-\bm{n}$ represent the same structure -- the nematic is a {\sl line field} not a vector field.  According to Frank \cite{Frank}, distortions away from the uniform nematic phase are measured through four geometric quantities that are invariant under the nematic symmetry:
$\vec{S}=\bm{n}(\nabla\!\cdot\!\bm{n})$, $T=\bm{n}\!\cdot\!(\nabla\!\times\!\bm{n})$, $\vec{B}=(\bm{n}\!\cdot\!\nabla)\bm{n}$, and $G=\nabla\!\cdot\!(\vec{B}-\vec{S})$ -- splay, twist, bend, and saddle-splay, respectively.  The Frank free energy density is a rotationally-invariant expression in terms of these two vectors ($\vec{S}$,$\vec{B}$), pseudoscalar ($T$), and scalar ($G$):
\begin{equation}\label{Fmaster}
f=\tfrac{1}{2}K_1 \vec{S}^2 + \tfrac{1}{2}{K_2}(T+q_0)^2 +\tfrac{1}{2}{K_3}\vec{B}^2 +K_{24}G.
\end{equation}
We note that $\vec S \cdot \vec B=0$ so there are no cross terms.  This free energy exhausts all the rotationally-invariant groupings of terms up to quadratic order in single gradients of $\bm{n}$.  The four elastic constants inherit their names from the expressions they multiply so, for instance, $K_2$ is the ``twist'' elastic constant, and stability implies that $K_1$, $K_2$, and $K_3$ are positive. Finally, because $T$ is a pseudoscalar, $q_0$ must be as well, and the existence of a pseudoscalar quantity would imply that the material is chiral.  To rationalize the names of these distortions, one can evaluate the splay for the two-dimensional texture
$\bm{n} = \hat\rho$ ($\vec{S}=\hat\rho/\rho$) and evaluate the bend for the two-dimensional texture $\bm{n}=\hat\theta$ ($\vec{B}=-\hat\rho/\rho$), where $\rho$ and $\theta$ are the standard polar co\"ordinates.  Both twist and saddle-splay measure three-dimensional textures: if $\bm{n}=[\cos(qz),\sin(qz),0]$ then $T=-q$, while if $\bm{n}=[x,-y,1]/\sqrt{1+x^2+y^2}$, then $G=2/(1+x^2+y^2)$.  The result for $G$ can be understood by viewing $\bm{n}$ as the unit normal to the saddle surfaces of the surface family $z=\frac{1}{2}(y^2-x^2)$ and then $G$ is the negative of the Gaussian curvature at each point \cite{RMPGeometry}.

In the absence of boundaries, the saddle-splay does not contribute to the energy (via Stokes' theorem).  When $q_0=0$, a ground state is $\bm{n}=[0,0,1]$, while if $q_0\ne 0$ then it is straightforward to check that $\bm{n}_c=[\cos(q_0z),\sin(q_0z),0]$ is a ground state.  We call this helically twisting ground-state the {\sl cholesteric} or the chiral nematic, and in this case, it has a pitch axis along $\hat z$.   By rotational invariance, these ground states can be rotated in space, leading to a whole manifold of degenerate ground states. The cholesteric ground state can then be viewed in terms of pseudolayers of constant orientation.  Moreover, since a global rotation of $\bm{n}$ around the $z$ axis by an angle $\phi$ cannot change the energy, we know that at long length scales, this global symmetry is promoted to a Goldstone mode so that small ground state fluctuations can be viewed as deformations of the pseudolayers.  The HH effect distorts these pseudolayers when the preferred spacing $\pi/q_0$ differs from a spacing imposed by fields or boundary conditions.

We will also, in the following, discuss smectic phases.  In the smectic phase, translational symmetry is broken and the molecules arrange themselves into actual layers, creating a one-dimensional density wave with the ground state being a set of uniformly-spaced, flat layers.  These layers generate a field of unit layer normals $\bm{N}$.  Since the normals are only defined up to sign, the symmetry of $\bm{N}$ is precisely that of the nematic director discussed above.  We can thus create an energy in complete analogy with the Frank free energy, where we substitute $\bm{n}$ with $\bm{N}$ in the discussion of the last paragraphs.  However, it is important to note that twist necessarily vanishes if $\bm{N}$ is normal to a surface from the Frobenius integrability condition \cite{cartan2012differential}.  Yet, there must also be an energy penalty for deformations away from the preferred spacing.  To measure this, we introduce a phonon field, $u(x,y,z)$ that measures the deviation from the ground state.\footnote{Unlike a crystal, however, $u(x,y,z)$ only has one component reflecting the one-dimensional density wave.} Because the sign of $u$ has the same $u\rightarrow -u$ ambiguity as the vector $\bm{n}$, we will measure deviations of $u$ along $\bm{N}$ and define the strain as $e=\bm{N}\!\cdot\!\nabla u$, invariant under $(\bm{N},u)\rightarrow -(\bm{N},u)$.  The free energy density is
\begin{equation}
f= \tfrac{1}{2}{B} e^2 + \tfrac{1}{2}{\overline{K}_1} \vec{S}_{\hbox{sm}}^2 + \tfrac{1}{2}{\overline{K}_3} \vec{B}_{\hbox{sm}}^2 + \overline{K}_{24} G_{\hbox{sm}},
\end{equation}
where the subscript $\hbox{sm}$ refers to the quantities with $\bm{n}$ replaced with $\bm{N}$ and 
where $B$ is known as the bulk modulus (and should not be confused with the bend vector $\vec B$!).  Since we can parameterize the smectic layers as level sets of $\phi=z-u(x,y,z)$, $\bm{N}=\nabla\phi/\vert\nabla\phi\vert$ can be calculated from $u$.  For instance, $\bm{N}\approx (-\partial_x u, -\partial_y u,1)$ to lowest order in gradients of $u$.  Because the phase is comprised of nematogenic molecules, we must also include the Frank free energy for the nematic director and there is a coupling between $\bm{N}$ and $\bm{n}$.  In the smectic-A phase, $\bm{n}$ prefers to align with $\bm{N}$, while in the smectic-C phase the layer normal and director prefer a fixed, nonzero angle between them.  This leads to yet another director-like field which is the component of $\bm{n}$ perpendicular to $\bm{N}$: the $\mathbf{c}$-director.

The deformations of the smectic-A layers and of the cholesteric pseudolayers are controlled by the same free energy density  \cite{oswald2005smectic,kleman_parodi_1975}:
\begin{equation}
f_e=\frac{B}{2}\left(1-\frac{1}{|\nabla \phi|}\right)^2+\frac{K}{2} (\mathbf{\nabla}\cdot  {\bm{N}})^2
\label{EqFS}
\end{equation}
where the first term accounts for relative dilation of the layers and the second term is the curvature energy of the layers. In the long distance limit when the layers are nearly-planar, this free energy reduces to 
 \begin{equation}
f_e={B\over 2}\left({\partial u\over\partial z}\right)^2 + {K\over 2} \left({\partial^2 u\over\partial x^2}+{\partial^2 u\over \partial y^2}\right)^2,
\label{ener-displ}
\end{equation}
where  the average layer normal (or the pitch axis) is along $\hat z$. The $\overline{K}_3$ and $\overline{K}_{24}$ contributions are higher degree in a gradient expansion and are, in this simplest case, neglected. 
In the case of the cholesteric, we would replace $u$ with the deviation of the angle of the director field in the plane perpendicular to the pitch axis.  This basic free energy is the starting point for this review. Note that this elastic free energy density applies to any system with one-dimensional, periodic ground states.   Without loss of generality, the periodicity is along the $\hat z$-direction, and we can write the density (or pseudo-density) as 
\begin{equation}\label{eq:u} \rho({\bf x}) = \rho_0 + \rho_1\cos\big[q\left(z-u({\bf x})\right)\big],
\end{equation}
where $q$ is the ground state wavevector magnitude.  The first term in (\ref{ener-displ}) measures the energy penalty for changing the periodicity while the second term measures the energy cost of bending the ``layers.''  

We now dive into the history of the HH instability.

\section{The classic Helfrich-Hurault instability \label{History}}

As has been made evident from the success of liquid crystals in the display industry, liquid crystal technology relies upon the material's interaction with external fields. Recall that the simplest liquid crystalline phase, the nematic, is characterized by long-range order of the orientation of anisotropic molecules with one ``long'' axis and two, equivalent ``short'' axes.\footnote{These are the so called, ``calamitic'' nematics.  Discotic phases are also nematic though they have one short axis and two, equivalent long axes.}  These axes are geometric, dielectric, {\sl and} optical, leading to birefringent optics. The dielectric anisotropy of liquid crystals enables their manipulation with electromagnetic fields, and their birefringence renders optically detectable responses. Systematic investigations of liquid crystals under these external fields became of special interest in the 1960s, the decade when liquid crystal displays were first conceptually conceived, and regular textures were soon experimentally observed and identified. Some patterns were related to flows or to other dynamical aspects --- such as  electrohydrodynamic convection in nematics \cite{helfrich_conductioninduced_1969} --- but  others, found especially in layered or quasi-layered systems, remained static and exhibited well-defined wavelengths that resulted from direct competition between liquid crystal elasticity and its anisotropic, electromagnetic properties.

The possibility of such an instability was predicted by Helfrich in the case of cholesteric liquid crystals, where the molecules have a tendency to twist in a helical fashion, with the pitch defined as the distance required for a $2\pi$ rotation of the molecule along the pitch axis (Fig. \ref{fig:cholHH}) \cite{helfrich_deformation_1970}. Note again that cholesterics have a periodic ground state, with no density modulation, but rather, a modulation in orientation and consequently in the dielectric tensor. Because of this, the periodicity in the system is often referred to as ``pseudolayers''. 

Experimental data for the instability in cholesterics emerged almost simultaneously in the early 1970s \cite{helfrich_deformation_1970} and were followed by two successive theoretical papers, first by Helfrich in 1971 and later refined by J.P. Hurault in 1973 \cite{helfrich_electrohydrodynamic_1971,hurault1}. Initially associated with the cholesteric phase, as depicted in Fig.~\ref{fig:cholHH}, the HH buckling instability was rapidly identified as a generic mechanism to relieve stresses and strains due to external stimuli in one-dimensional and two-dimensional, periodic systems.

\subsection{Cholesteric layer distortions from electric and magnetic fields}
\begin{figure}[!ht]
\includegraphics[width=.95\columnwidth]{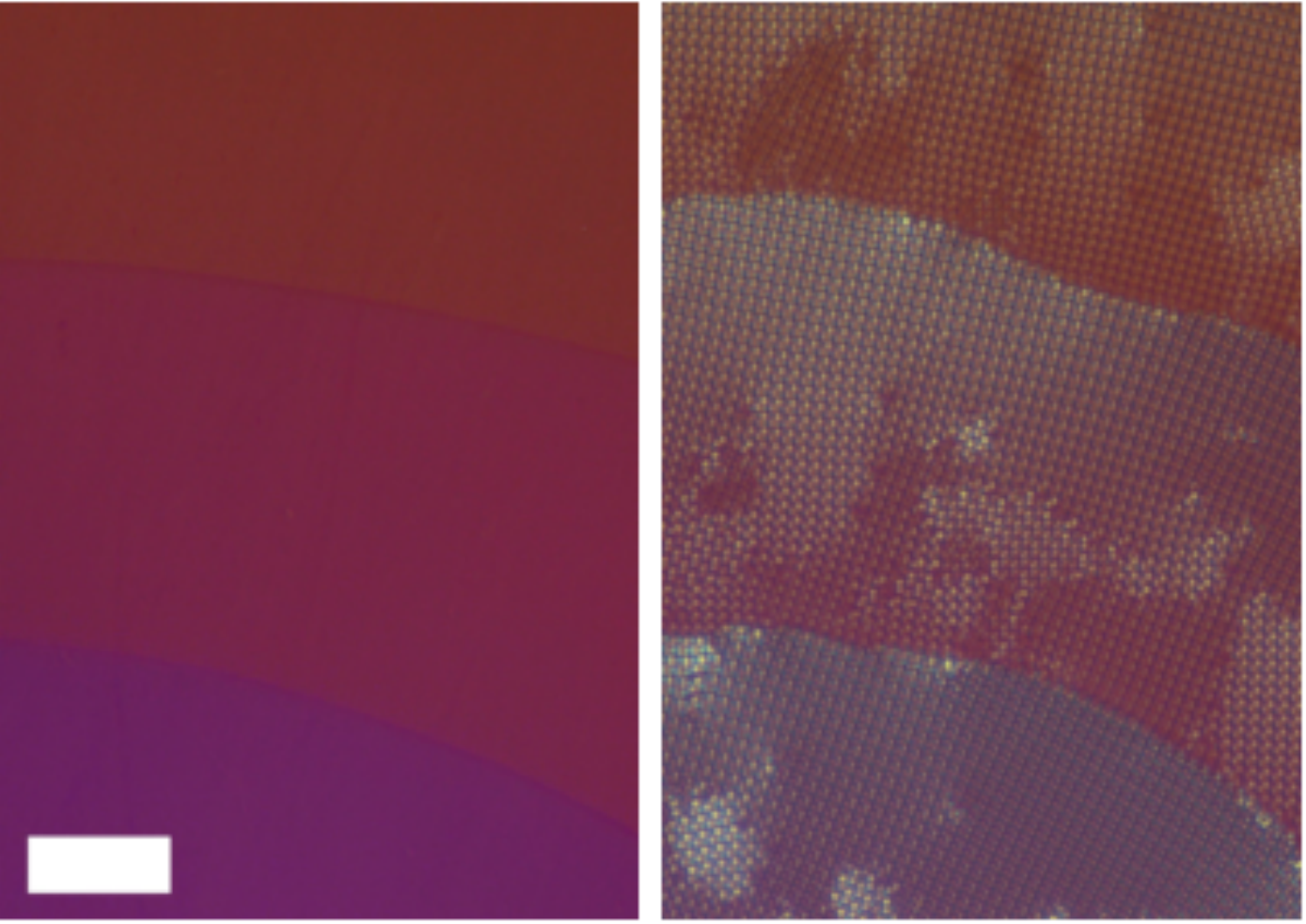}
\caption{ (Left) Planar texture of a cholesteric phase obtained from mixing 4'-pentyl-4-biphenyl-carbonitrile (5CB) and cholesteryl oleate in a planar-aligned cell. Grandjean zones  correspond to a slight gradient of thickness resulting in a discrete change in the number of $\pi$-rotations of the director. (Right) Under large enough AC voltages ($\sim$1 kHZ), typical square grid patterns are observed. Here, a 9V electric field is applied across a 20 $\mu$m thick cell. Images are captured with polarizing optical microscopy under slightly uncrossed polarizers. The scale bar is 50 $\mu$m. }
\label{figEField}
\end{figure}

The HH instability was first observed in cholesteric liquid crystal cells \cite{gerritsma_electric-field-induced_1971,gerritsma_periodic_1971} with strong planar anchoring, where the director $\bm n$ of the molecules is aligned tangent to the top and bottom walls. In this geometry, the cholesteric pitch axis, perpendicular to the nematic director  $\bm n$ and along which the director twists, has a uniform orientation perpendicular to the parallel walls. The application of an electric field \cite{gerritsma_periodic_1971,rondelez_deformation_1971} or a magnetic field \cite{scheffer_electric_1972,rondelez_distorsions_1972} parallel to this helix gives rise to  square-grid patterns above a certain threshold value (Fig.~\ref{figEField}). Here, the driving force of the instability is a gain in dielectric or diamagnetic energy when the cholesteric helix begins to distort. In this geometry, the HH instability occurs only in materials with  positive (nematic)  diamagnetic susceptibility anisotropy $\chi_a$ or dielectric anisotropy $\epsilon_a$ so that the director aligns along the field, antagonizing the helix. The case of AC electric fields is, however, more complex since the presence of conductivity and space charges can also lead to frequency-dependent instabilities for both signs of dielectric anisotropy $\epsilon_a$ \cite{hurault1,rondelez_electrohydrodynamic_1972}.

\subsubsection{Original model\label{originalHelfrich-Hurault}}
\label{original}

In the magnetic case, the threshold and the wavelength of the patterns can be  computed easily at the onset of undulations with two assumptions: (1) the distortions are small, and  (2) the instability wavelength is much larger than the cell thickness $a$, which is itself much larger than the cholesteric pitch $P_0=2\pi/ q_0$ \cite{ishikawa_defects_2002}. Recall that the continuous twist of the director field is described as a pseudolayered structure of $P_0/2$ periodicity [Fig.~\ref{fig:cholHH}(a)]. In the Lubensky-de Gennes coarse-grained approach, the elastic free energy density $f_e$ of a distortion from the planar texture is related to the displacement $u({\bf x})$ of the  pseudolayers along the  $z$-axis, corresponding to the direction of the initial helix \cite{dgennes-prost,lubensky_hydrodynamics_1972}, yielding a free energy density of the same form as Eq.~\ref{ener-displ}, but now with $K$ rewritten as $\bar{K}=3K_3/8$ and $B$ rewritten as $\bar{B}=K_2 q_0^2$. $\bar{K}$ and $\bar{B}$ are the effective elastic moduli related to the Frank-Oseen  elastic constants of twist $K_2$ and bend  $K_3$ of Eq.~\ref{Fmaster}.
In the cholesteric, a distortion $u({\bf x})$ leads to a tilt of the pitch axis $\bm{N}$ from $\hat z$.  To lowest order we get $\tan^2\theta = (\nabla_\perp u)^2$ where $\nabla_\perp = \hat x\partial_x + \hat y\partial_y$.  For small distortions we have
\begin{equation}
\theta\approx \sqrt{\left(\frac{\partial u}{\partial x}\right)^2+\left(\frac{\partial u}{\partial y}\right)^2},
\end{equation}
with a concomitant change in the magnetic energy density of:
\begin{equation}
f_m=-\frac{1}{2}\mu_0 \bar{\chi}_a H^2\theta^2,
 \label{ener-mag}
\end{equation}
where $\mu_0$ is the vacuum permeability and $\bar{\chi}_a= \chi_a/2$ accounts for the continuous twist of the director over a pitch.

The HH model considers a simple undulation pattern along one direction ($\hat x$ here) and compatible with infinitely strong anchoring at the bounding surfaces $z=\pm a/2$ [Fig.~\ref{fig:cholHH}(b)]:
\begin{equation}
u({\bf x})=u_0 \cos \left( \frac{\pi z}{a} \right) \sin (q x).
\end{equation}
The total free energy of such an undulation in a cell of volume $V$ can be computed from Eqs.~\eqref{ener-displ}-\eqref{ener-mag}:
\begin{equation}
F_t=\left(\frac{\bar{B}\pi^2}{a^2}+\bar{K} q^4-\mu_0 \bar{\chi}_a H^2 q^2\right) \frac{V}{8} u_0^2.
\label{ener-total}
\end{equation}
At low fields, undulations are unfavored. An instability occurs for a critical field $H_c$ when the sign of the minimum of $F_t$ (with respect to $q$) changes from positive to negative:
\begin{equation}
H_c^2=\frac{2\sqrt{\bar{K}\bar{B}}\pi}{a\mu_0 \bar{\chi}_a}
=\frac{\sqrt{6K_2K_3}\pi q_0}{a\mu_0 \chi_a}.
\end{equation}

\noindent This  first-order approach also allows to compute the wave vector amplitude $q_c$ at the threshold:
\begin{equation}
q_c^2=\frac{\pi}{a}\sqrt{\frac{\bar{B}}{\bar{K}}}=\frac{2\pi q_0}{a}\sqrt{\frac{2K_2}{3K_3}}.
\label{Helfrich-Huraultqc}
\end{equation}

The main predictions of the wavelengths and the threshold of the HH model were satisfactorily checked experimentally soon after the development of the theory. However, the original model was found to be limited for some experimental situations and thus was amended incrementally over time.  In the following we will show that this basic free energy balance is recapitulated in layered systems subject to stresses both internal and external.  In doing so, we gather all of these effects under the Helfrich-Hurault umbrella.

\subsubsection{Further theoretical refinements and experiments}

Further examination of the HH phenomenon soon showed that undulations along a single direction were rarely observed at the threshold except in large pitch systems, where $a\sim P_0$ \cite{hurault2}. Well-defined square grid patterns were observed for large ratios of $a/P_0$. Delrieu extended the HH theory to examine two-dimensional distortions [Fig.~\ref{fig:cholHH}(c)] and showed that the square lattice was indeed the periodic structure of lowest energy at the onset of the undulations \cite{delrieu_comparison_1974}.

The model outlined in the last section is also too rough to describe the evolution of the patterns above $H_c$. The total free energy scales as the square of the undulation amplitude $u_0^2$ in Eq.~\eqref{ener-total}. Therefore, it is necessary to compute $F_t$  with higher order terms included in the strain to get a consistent undulation amplitude. These terms provide a better description of the compression term in Eq.~\eqref{ener-displ}, accounting for the tilt of the pseudolayers.  In terms of the phase field, a rotationally-invariant strain is $e=[1-(\nabla\phi)^2]/2$ \cite{Kamien2009}. In two dimensions this gives to next-to-leading order:

\begin{equation}
f_e=\frac{\bar{B}}{2}
 \left[\frac{\partial u}{\partial z}-\frac{1}{2}\left(\frac{\partial u}{\partial x} \right) ^2\right]^2+\frac{\bar{K}}{2} \left( \frac{\partial^2 u}{\partial x^2}\right)^2,
 \label{correcteq}
\end{equation}
which yields, after minimization of the free energy $F_t$:

\begin{equation}
u_0=\frac{8}{3}\sqrt{\frac{\bar{K}}{\bar{B}}\left(\frac{H^2}{H_c^2}-1\right)}.
\label{amplitude-undulation}
\end{equation}

However, the exact shape of the experimental patterns was not scrutinized in the 1970s because of a lack of appropriate experimental techniques. It was only later, in a different cell geometry,  that Ishikawa and O. Lavrentovich closely examined an undulation pattern developing along a single direction \cite{ishikawa_undulations_2001}. The two-dimensional system consisted of cholesteric stripes formed in a cell with homeotropic (perpendicular) anchoring of the liquid crystal director, generating a fingerprint texture. The periodic stripes were horizontally sandwiched between parallel spacers in the cell, and a magnetic field was applied in the plane of the cell, perpendicular to the stripes, allowing direct examination of the patterns above the HH instability. The study emphasized the neglected role of anchoring on the bounding substrates, where distortions could still be observed. A finite anchoring yields amplitude undulations much larger than the value predicted by Eq.~\eqref{amplitude-undulation}, as well as a reduced threshold value. This result was later confirmed for the square lattice of the original geometry by Senyuk \textit{et al.}, who used fluorescence confocal polarizing microscopy (FCPM) to image, in three-dimensions, the distorted pseudolayers under an electric field \cite{senyuk_undulations_2006}. We expound upon the influence of anchoring and other surface energies on the HH mechanism when we discuss liquid crystal shells.

The powerful FCPM technique was also employed to analyze  the evolution of the patterns generated by the HH instability, in detail and with increasing fields. It confirmed that the hypothesis of a single Fourier mode in the plane was valid only in a small range above the threshold. When  the field increases, the sinusoidal profile of the square grid pattern gradually changes to a sawtooth one, as predicted by Singer \cite{singer_layer_1993,singer_buckling_2000}. The study also showed that other thresholds are present at higher fields, since the two-dimensional, square grid  pattern  was destabilized in favor of a 1D structure of parallel walls at about twice the first threshold.

The cholesteric mesophase is the  system in which the HH effect was first discovered and theorized. Later, cholesteric systems also enabled subtle experiments for further fundamental studies of the instability. Indeed, the resulting patterns have the advantages of being easily controlled with an external field and of being very regular and stable. This last point even suggested possible applications of these systems, such as the design of switchable two-dimensional, diffractive gratings \cite{senyuk_switchable_2005,ryabchun_cholesteric_2018}. However, a cholesteric phase strained by an external field is not the only scenario leading to an HH instability. In the very first studies, it was already noted that cholesterics exhibit  square-grid patterns transiently under temperature changes or mechanical deformation  \cite{gerritsma_periodic_1971} in the absence of external fields. Moreover, the mechanical-strain-induced HH instability is observed in many other lamellar or columnar systems, including smectic liquid crystals.

\subsection{Mechanical layer strain in smectics\label{History-Sm}}

In smectics, molecules align and arrange into equally-spaced parallel planes, creating molecule-thick layers measurable as a one-dimensional density modulation. The buckling instability of smectic phases was identified shortly after the HH effect was observed in cholesteric phases, but now with pseudolayers replaced by {\sl actual} layers. Unexpected laser light scattering was observed in a smectic-A system, in which the nematic director is parallel to the smectic layer normal.  The system was presumed to be well-oriented, with the smectic layers parallel to the bounding, homeotropic glass substrates, where the molecules anchored perpendicularly to the bounding surfaces. Yet, the scattered pattern observed was well-defined, indicating the presence of periodic structures in the cell. The intensity of the scattered light was shown to be extremely sensitive to the strain of the sample, strongly increasing with dilation but decreasing under compression \cite{clark_light_1973,delaye_buckling_1973}.

 \subsubsection{Mechanically induced Helfrich-Hurault effect}
  The presence of a periodic pattern in strained smectic-A samples was explained by considering displacements of layers with ground state spacing $a$ of the form:
\begin{equation}
u(x,z)=\alpha z+ u_0 \cos \left(\frac{\pi z}{a} \right) \sin (q x)
\end{equation}
were $\alpha = \delta a/a \ll 1$ is the global applied strain \cite{clark_straininduced_1973}. Eq~\eqref{correcteq}  still describes the elastic free energy density of the smectic-A phase, where the modulus $\bar{K}=K$ is now the splay modulus of the director and  $\bar{B}=B$ the bulk compression modulus. Together, they traditionally define the smectic penetration depth, $\lambda=\sqrt{K/B}$, a length usually comparable to the molecular size. Expanding in $\alpha$ gives an expression similar to Eq.~\eqref{ener-total} for the total elastic energy:

\begin{equation}
F_t=\left(\frac{B\pi^2}{a^2}+K q^4-B\alpha q^2\right) \frac{V}{8} u_0^2,
\label{ener-strain}
\end{equation}
showing the formal analogy between a uniform strain in a layered system and the application of an external field. Following the analysis in the previous section, $\alpha$ plays the role of $H^2$, and so the threshold strain is $\alpha_c=2\pi\lambda/a$, above which undulations ensue \cite{clark_straininduced_1973,singer_layer_1993,singer_buckling_2000, napoli_mechanically_2009}. Note that $\alpha>0$ for the analogy to hold -- compression does not lead to buckling in this system. Since $\lambda$ is a molecular length scale, the instability  appears for very small changes of spacing, $\delta a   \approx 2 \pi\lambda$ and with wavevector amplitude $q_c^2=\pi/a\lambda$, in concert with  Eq.~\eqref{Helfrich-Huraultqc}.

This analysis holds for lamellar phases under dilation, including cholesterics, but a quantitative difference may be present. A thermotropic smectic-A phase or a short-period, lyotropic lamellar phase is a much stiffer material than large pitched cholesteric phases, such as the ones studied by Senyuk \textit{et al.} with FCPM \cite{senyuk_undulations_2006}. This implies that, for sample cells of comparable thicknesses, the pattern wavelengths are much smaller in a lamellar phase, but also that the sinusoidal profile of the undulation is rapidly destabilized above the HH threshold. Indeed, smectic-A layers are often considered to be almost incompressible ($B \rightarrow \infty $), as shown by the ubiquitous presence of topological defects called focal conic domains in disordered samples \cite{Friedel_mesomorphes}. These macroscopic structures consist of  curved but parallel layers whose common focal surfaces are degenerated into three-dimensional curves, an ellipse and a conjugate hyperbola \cite{bouligand_recherches_1972}. However, in experiments, the dilation of layers is not expected to be absent from the bulk, but rather confined in curves or, eventually, surface discontinuities \cite{bidaux_statistical_1973, blanc_curvature_1999}.
Because of this, in smectic-A samples, the simple undulation pattern can only be optically observed just above $\alpha_c$, provided thick enough samples are used. Increasing the strain slightly above $\approx 1.7 \alpha_c$ induces focal lines \cite{rosenblatt_parabolic_1977,clark_elastic_1982}. Rosenblatt \textit{et al.} have described an ideal four-fold grid pattern in terms of ordered assemblies of geometrical stacks of parallel layers, introducing parabolic focal conic defects and their corresponding domains \cite{rosenblatt_parabolic_1977}. Such a structure almost satisfies the homeotropic anchoring at the bounding plates while the distortions from dilation remain confined in the line defects. While the ideal square-grid pattern is rarely obtained in smectics with a simple strain [a polygonal structure is often observed,  \cite{rosenblatt_parabolic_1977}] it should be noted that the simultaneous application of a shear flow may help the formation of very long-range, ordered square lattices of parabolic, focal conic domains \cite{oswald_undulation_1982,chatterjee_formation_2012}.

\subsubsection{The role of dislocations and disclinations}\label{SecHistTopology}

\begin{figure}[!ht]
\includegraphics[width=.95\columnwidth]{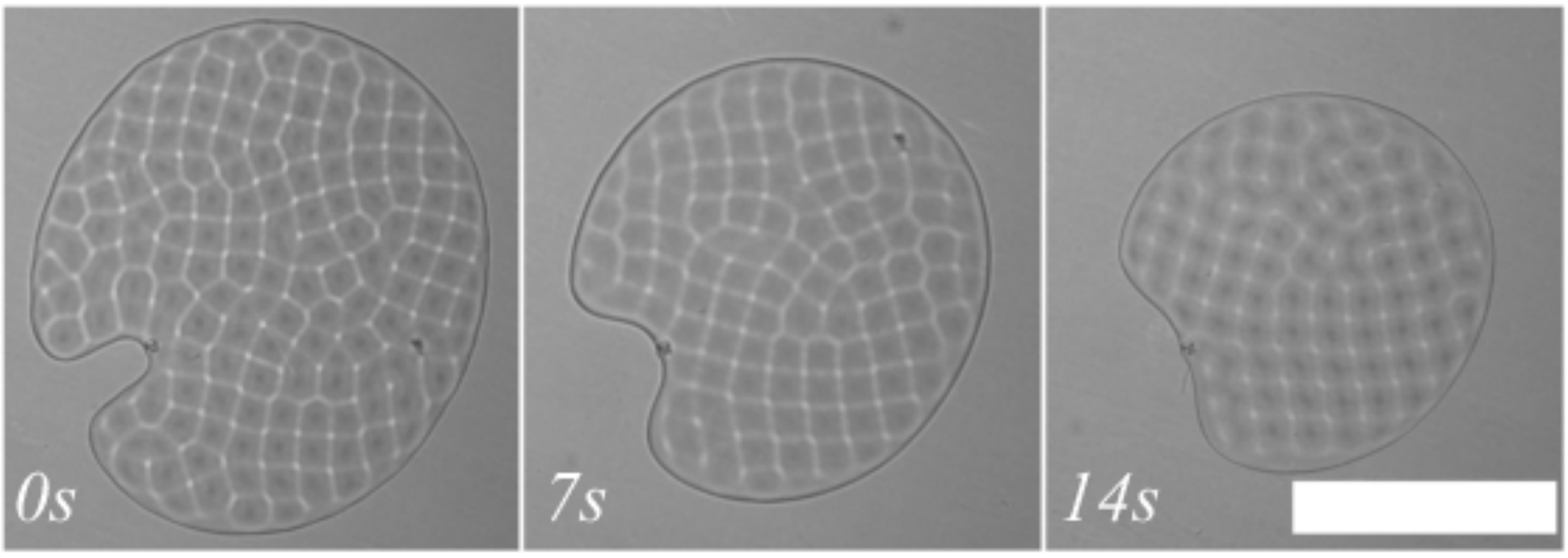}
\caption{Relaxation of a dilated region displaying the square grid pattern through the climb of an edge dislocation loop during a compression-dilation sequence. The cholesteric phase was obtained from mixing 4'-pentyl-4-biphenyl-carbonitrile (5CB) with the chiral dopant (S)-4-cyano-4'-(2-methylbutyl)biphenyl (CB15, 2.8 wt-\%). Images are obtained from bright field optical microscopy. The scale bar is 200 $\mu$m.}
\label{relaxdilat}
\end{figure}

Although these results all support the analogy between electromagnetic field-induced and mechanically-induced HH effects, a major difference exists in the temporal evolution of the textures. Field-induced patterns are caused by a gain in energy accompanying the reorientation of the layers and are stable. On the contrary, after a uniform strain, the planar texture remains most favorable and can be achieved if layers can be added to the slab. Mechanically-induced textures are therefore transient or metastable, as was emphasized by Clark, Meyer, and Delaye in 1973 \cite{delaye_buckling_1973,clark_straininduced_1973}. An efficient mechanism to relax the strain was expected to be the climb of edge dislocations, which are unavoidably present in a smectic slab \cite{bartolino_plasticity_1977, ribotta_mechanical_1977}. Note that a smectic-A wedge cell with a tiny angle on the order of $10^{-3}$ rad already gives rise to a linear density of about one dislocation per micron. This mechanism is difficult to observe directly in smectic-A systems. It can, however, be studied in the vicinity of the smectic-A to smectic-C transition \cite{blanc_defect_2004} and is easily observed in cholesteric phases due to their larger, micron-scale periodicities, as shown in Fig.~\ref{relaxdilat}.  We note that, technically, cholesterics do not have dislocations but $\chi$-disclinations since they do not have a density modulation \cite{dgennes-prost}.

Finally, we point out that buckling instabilities are not only found in lamellar systems, but also in other modulated phases such as columnar phases \cite{livolant_liquid_1986,oswald_nonlinear_1996}. We expound more upon the HH instability in a broad range of materials in the last section of this review.

\section{Liquid crystal shells}\label{secLCShell}

\begin{figure*}[ht]
\centerline{\includegraphics[width=0.8\textwidth]{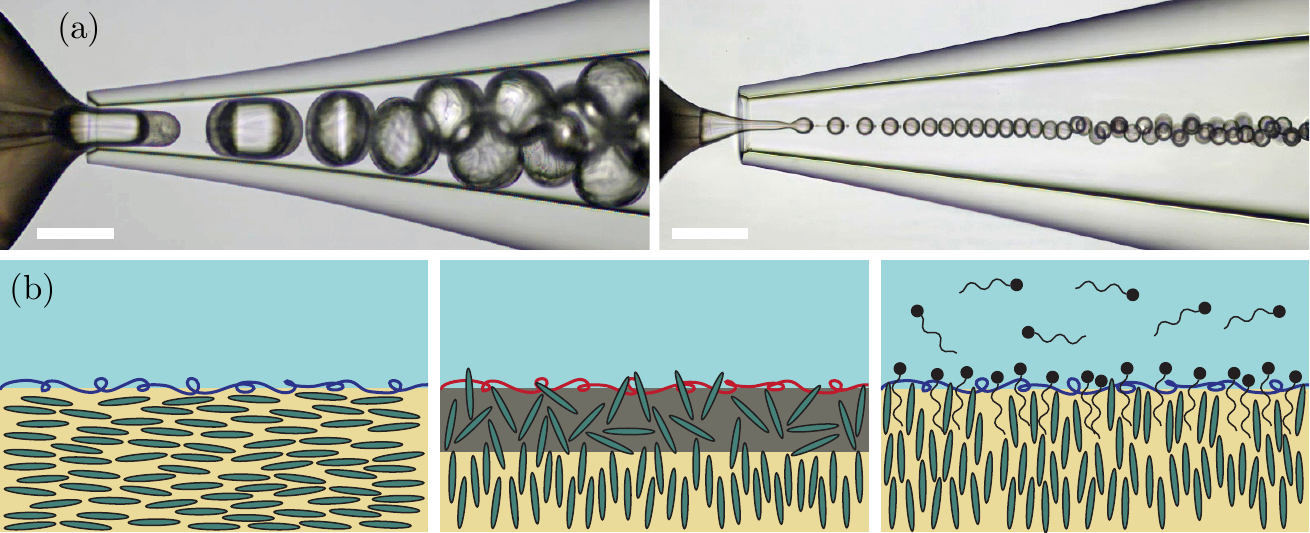}}
\caption{\label{ExperimentsWithShells} (a) Microphotographs of capillary devices used to produce shells. The two photographs show the typical maximal (left) and minimal (right) sizes one can attain with a given device geometry. Scale bar is 200 {$\mu$m}. (b) Diagrams for techniques used to change the anchoring at a water / liquid crystal interface. In the presence of PVA, the anchoring is strongly planar (left). The system transitions to a perpendicular orientation when its temperature is brought a couple tenths of a degree Celsius below the clearing point of the bulk 5CB, resulting from the presence of an interfacial melted layer of 5CB and PVA (center). Homeotropic anchoring can also be regulated by adding surfactants to the aqueous phase (right). The amount of adsorbed surfactant to the water-liquid crystal interface determines the homeotropic anchoring strength.}
\end{figure*} 

In the previous section, we reviewed the history of the HH instability in both cholesteric and smectic liquid crystals confined between glass plates. All of the previous examples have been in systems with \textit{solid} boundaries. However, in sensing applications and in bio-materials, liquid crystal systems with periodic ground states are often in contact with fluid (liquid or gas) phases. The boundary conditions are then deformable, resulting in an interplay between bulk and surface energies that gives rise to more complex dynamics and ground states. 

Liquid crystal shells are attractive systems for investigating the effect of fluid interfaces on the HH instability, due to shell thickness tunability and fine control over the system's boundary conditions through a wide-array of techniques, ranging from tuning the system temperature to altering the system chemistry \cite{fernandez-nieves-shell, Teresa_TopologicalTransformations, Teresa_FrustratedOrder, sm-shell-tll-2, sm-shell-jl, Teresa_Review, Teresa_DefectCoalescence, Sec_DefectTrajectories, sm-shell-tll-1, Lagerwall_DefectSmecticShells, Koning_BivalentInhomogeneousShells, Lagerwall_TuningConfigurations, Alex_waltz, tll-copar-clc-planar-shell, Alex_ElasticInteractions, dePablo_MesoscaleCholShells, Lisa_ChangeStripes, Lagerwall_BCP}. Shells are water-in-liquid-crystal-in-water double emulsions, where a thin liquid crystal layer is confined between an inner water droplet and a continuous water phase, produced in microfluidic devices made of nested glass capillaries [Fig.~\ref{ExperimentsWithShells}(a)]. In these devices, a water-in-liquid-crystal compound jet is sheared by an outer aqueous solution, leading to its breakup into water droplets that are encapsulated by liquid crystal \cite{Utada_DoubleEmulsions, fernandez-nieves-shell}. This technique enables both the production of highly monodisperse samples and independent control over the size of the inner and outer diameters. The thickness and curvature of the shells can be selected by adjusting flow rates during microfluidic production. The shell thickness can also be varied post-production through osmotic swelling or de-swelling, accomplished by changing the concentration of a solute, such as salt or sugar, in the surrounding aqueous phases \cite{Teresa_FrustratedOrder, Sec_DefectTrajectories, Daeyeon_OsmSwell, Alex_waltz, Lisa_ChangeStripes}. Osmotically swelling the liquid crystal shells is useful for observing the temporal evolution of thickness- or curvature-dependent phenomena \cite{Teresa_FrustratedOrder,Alex_waltz,lagerwall-shells,durey-skyrmion,tran-acsnano}. 

Furthermore, the anchoring at the inner and outer water-liquid crystal interfaces of the shell can be set independently. In the simplest case, shells of 4'-pentyl-4-biphenyl-carbonitrile (5CB) in contact with pure water have matching planar boundary conditions on both the inner and outer shell surfaces. The planar anchoring is degenerate, which means that the director is free to rotate on the surfaces. The planar anchoring strength can be increased with the introduction of polyvinyl alcohol (PVA) in the aqueous phases. This polymer surfactant also increases the shell stability by decreasing the water-liquid crystal interfacial tension and by inducing a repulsive force -- a disjoining pressure -- when the inner and outer interfaces get closer. [Fig.~\ref{ExperimentsWithShells}(b), left]. The increased shell stability allows for the shell anchoring conditions to be dynamically and gradually tuned with simple modifications to the system, mainly through two mechanisms. 

The first method involves quasi-statically bringing the system temperature a few tenths of a degree Celsius below the clearing point of the bulk 5CB. The shells undergo a series of anchoring transitions as the temperature rises, from matching planar anchoring on the inner and outer shell surfaces, to hybrid anchoring, and then to matching homeotropic anchoring, before fully transitioning to the isotropic phase \cite{GD_AnchoringTransitions}. This behavior has been linked to the PVA polymer at the shell interfaces, which renders the liquid crystal more disordered near the interfaces compared to the bulk. The shell interfaces then favor the nucleation of the isotropic phase. The melted layer and the bulk nematic create a new, low-anchoring-strength interface accounting for the changes in anchoring observed in the shell with increasing temperature [Fig.~\ref{ExperimentsWithShells}(b), middle].

\begin{figure}
\includegraphics[width=0.48\textwidth]{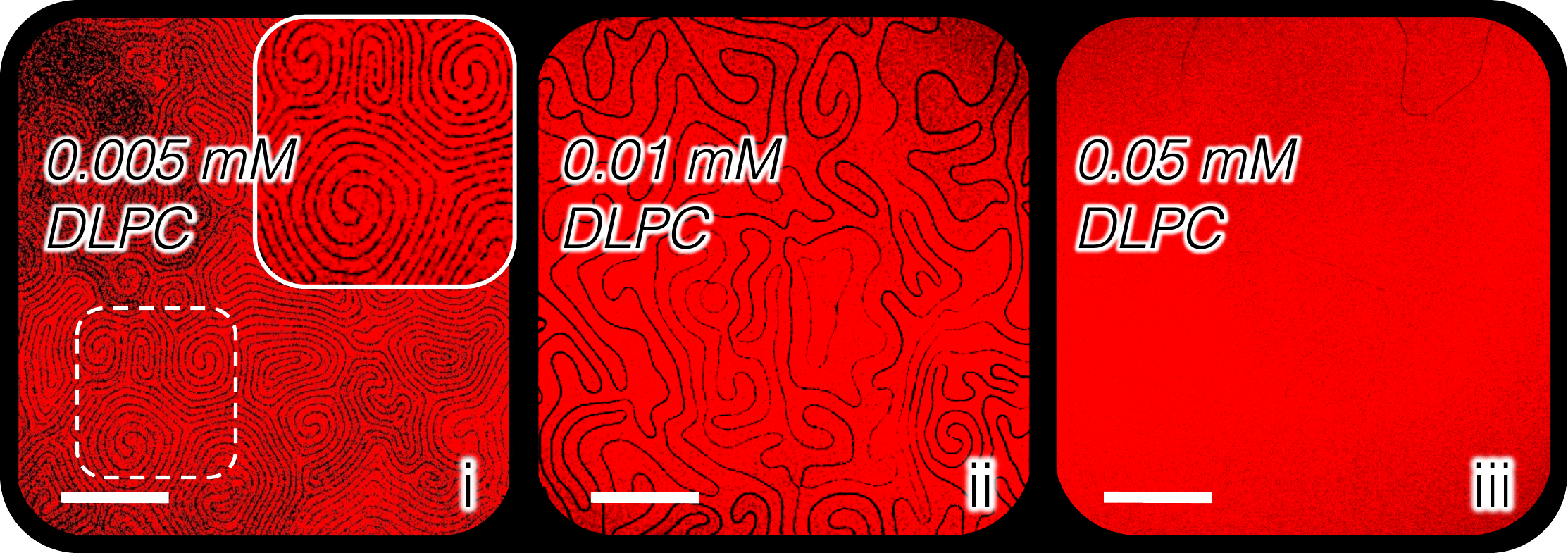}
\caption{\label{CholestericSurfactantSegregation} Laser scanning confocal micrographs of the lipid surfactant 1,2-dilauroyl-sn-glycero-3- phosphocholin (DLPC), labeled with 1 mol\% Texas Red 1,2-dihexadecanoyl-sn-glycero-3- phosphoethanolamine, triethylammonium salt (TR-DHPE) demonstrates the cross-communication of the liquid crystal and the adsorbed surfactant. The surfactant causes homeotropic anchoring, inducing stripe patterns in the cholesteric. The cholesteric subsequently patterns the surfactant, causing them to segregate into stripes at the cholesteric-water interface. As the surfactant concentration increases from i to iii, surface stripes become wider and more disordered (ii) until regions where the cholesteric twist violates the homeotropic anchoring condition are forced away from the surface, as a result of the lipids saturating the interface (iii). Reproduced from \cite{Lisa_ParticleFingerprints}.}
\end{figure}

The second technique relies on the dissolution of surfactants in the water phases. As small amphiphilic molecules adsorb on the shells' interfaces, their aliphatic tails force the liquid crystal molecules to reorient, perpendicular to the boundary, as illustrated in the right-most panel of Fig.~\ref{ExperimentsWithShells}(b) \cite{drzaic1995liquid, poulin-weitz, Lagerwall_InterfacesShells, Lagerwall_SurfactantsAnchoring}. This yields homeotropic boundary conditions with a tunable anchoring strength that increases with the surfactant surface coverage \cite{Abbott_first, Abbott_SurfactantStructure, Abbott_ActiveAnchoring, Abbott_InterfacesReview, Abbott_electrolytes, dePablo_AmphiphilePhaseTransition}. For a cholesteric twisting along a water-liquid crystal interface, it has been shown that surfactants localize in the homeotropic regions and are excluded from planar regions (Fig.~\ref{CholestericSurfactantSegregation}) \cite{Lisa_ParticleFingerprints}. This cross-communication between the bulk and the surface results in patterned chemical heterogeneity at the cholesteric interface and could manifest in other liquid crystal phases in which the bulk competes with the surface anchoring. Responsive surfactants enable further control of surfactant adsorption and conformation at the interface with means beyond the surfactant concentration, such as through temperature, pH, and UV light intensity \cite{Kwon:2016vl,Lagerwall_CTAB,LightResponsive1,LightResponsive2}.

The flexibility of the shell system thus lends itself to investigating the role of surface tension, anchoring, and boundary curvature on the HH instability. In the following sections, the outlined techniques are employed to investigate undulating instabilities in cholesteric and smectic shells.

\section{Cholesteric shells \label{sec:Cholshell}}

Since the classic HH instability was first discovered in cholesterics, we begin by examining cholesteric shells made of 5CB doped with a chiral dopant, (S)-4-cyano-4'-(2-methylbutyl)biphenyl (CB15). In the following, we review how undulations can develop in the cholesteric pseudolayers in response to topological frustration, as well as changes in the liquid crystal anchoring. We also review how undulations occur not only within the bulk, but also at the interface itself. Cholesteric shells demonstrate how fluid boundaries play a significant role in the HH instability, while also illustrating that the instability is, at its core, a response to local, geometrical frustration.

\subsection{Planar cholesteric shells}\label{secPlanarChShell}

\begin{figure}[ht]
\includegraphics[width=0.48\textwidth]{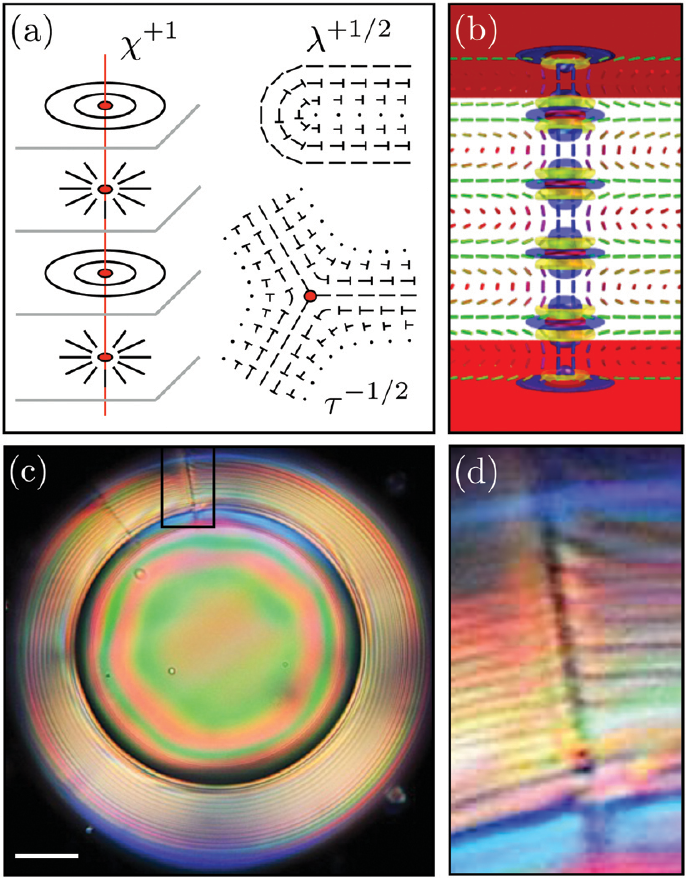}
\caption{\label{PlanarCholShell} \textbf{$\chi^{+1}$ disclination line in a planar cholesteric shell: experiment and simulation.} (a) Schematics of $\chi^{+1}$, $\tau^{-1/2}$ and $\lambda^{+1/2}$ disclinations in cholesterics. (b) A simulated cross section of an $m=+1$ defect. Blue and yellow regions respectively indicate zones of high splay and bend distortion; red indicates director singularities. (c) Side view of a shell with two $m=+1$ defects between crossed polarizers, revealing a visible nonuniform structure of the defect core, which is enlarged in (d). Scale bars are {20} {$\mu$m}. Reproduced from \cite{tll-copar-clc-planar-shell}.}
\end{figure} 

Planar anchoring in cholesteric shells frustrates the bulk ordering and induces structures that can be seen as a manifestation of the HH instability, broadly construed.  Why is there frustration when the pitch axis does not lie in the tangent plane of the shell?  The answer is topology!  Since it is the director that lies in the planar shell's tangent plane, the Poincar\'e-Brouwer-Hopf theorem requires that the sum of the indices of the zeros of a line field is equal to the Euler character of the shell \cite{Poincare_theorem, Brouwer_theorem,Hopf_theorem}.  Zeros of the line field are topological defects -- places where the local orientation is undefined, while the index of the zero is its signed winding. For a sphere, the Euler character is $2$, and so the net winding of the defects on the shell surface must be $2\times 2\pi$, manifesting as four $+1/2$ defects,  two $+1$ defects, or one $+2$ defect, two $+1/2$ defects and a $+1$ defect, or three $+1$ defects and one $-1$ defect, {\sl etc}. Although the necessity of a minimum number of defects can be thought of as \textit{topological} frustration that arises from the system's \textit{global} curvature, the defects can also be viewed as manifesting from \textit{local} incompatibilities, \textit{i.e.}, as \textit{geometrical} frustration.
Moving inward from the shell surface along its normal is equivalent to moving along the cholesteric pitch axis, by definition.  Thus there is a slightly smaller sphere just below the outer surface which also has planar anchoring and thereby must also have these defects (note that the global rotation of the director field does not contribute to the defect charge). If the pitch axis remains radial from the outer to the inner surface of the shell, then the shell would consist of a series of concentric spheres each with two-dimensional defects. From the three-dimensional perspective, these defects are not independent and would connect up into line defects with net winding $4\pi$. This is seen in planar nematic shells, where the shell thickness controls the amount and winding number of defects \cite{fernandez-nieves-shell,Teresa_FrustratedOrder,Koning_BivalentInhomogeneousShells,Vitelli_NematicShells,Vitelli_TLL}.   Recall, however, that in three dimensions, integer-winding defect lines are not topologically stable: they can ``escape into the third dimension'' \cite{Meyer_escape}.  Of course, this deformation has an associated bend energy (and possibly twist) and so for thin shells, this does not happen.  However, as the shells thicken,
the director goes smoothly from being horizontal (parallel to the tangent plane of the sphere) in the periphery of the defect to being vertical at the core. The only singularities left in the system after this escape are point defects, or ``boojums,'' that have been ``pushed away'' to the shell surfaces \cite{Volovik_boojums, Lavrentovich_WordsAndWorlds}.

However, cholesteric defects are considerably more complex than nematic defects. While a nematic is characterized by a single director field, $\bm{n}$, an unfrustrated cholesteric is properly described at large scales by three, mutually-orthogonal, line fields: the director $\bm{n}$, the pitch axis $\hat{\mathbf{P}}$, and their cross product $\notn\equiv {\bm{n}}\times\hat{\mathbf{P}}$.  Winding defects are now characterized by both their strength and by the axis around which they rotate.  Adopting the notation by Friedel and Kleman \cite{Friedel_dislocations}, defects where the pitch axis and $\notn$ rotate around the director are labeled  $\lambda$. On the other hand, defects where $\notn$ and the director axis rotate around the pitch axis are labeled $\chi$.  Finally, defects in both the director and pitch axis, where the two rotate about $\notn$ are labeled  $\tau$.  Examples of each of these defects are illustrated in Fig.~\ref{PlanarCholShell}(a). Though similar in their algebra to defects in biaxial nematics \cite{Mermin_Defects}, the existence of pseudolayers spoils a precise correspondence \cite{Kamien_GeomChol}.  However, just as in biaxial nematics, defects cannot escape into the third dimension: as a defect in the director attempts to escape, a new defect in either $\hat{\mathbf{P}}$ or $\notn$ appears.
In the na\"ive mapping between cholesteric pseudolayers and smectics, the $\chi$ defects correspond to dislocations, while the $\lambda$ and $\tau$ defects are the standard disclinations. It should be noted that while the $\lambda$ defects do not have a singularity in the director field, they have a singularity in the cholesteric structure since the pitch axis is undefined.

\begin{figure*}[ht]
\centerline{\includegraphics[width=\textwidth]{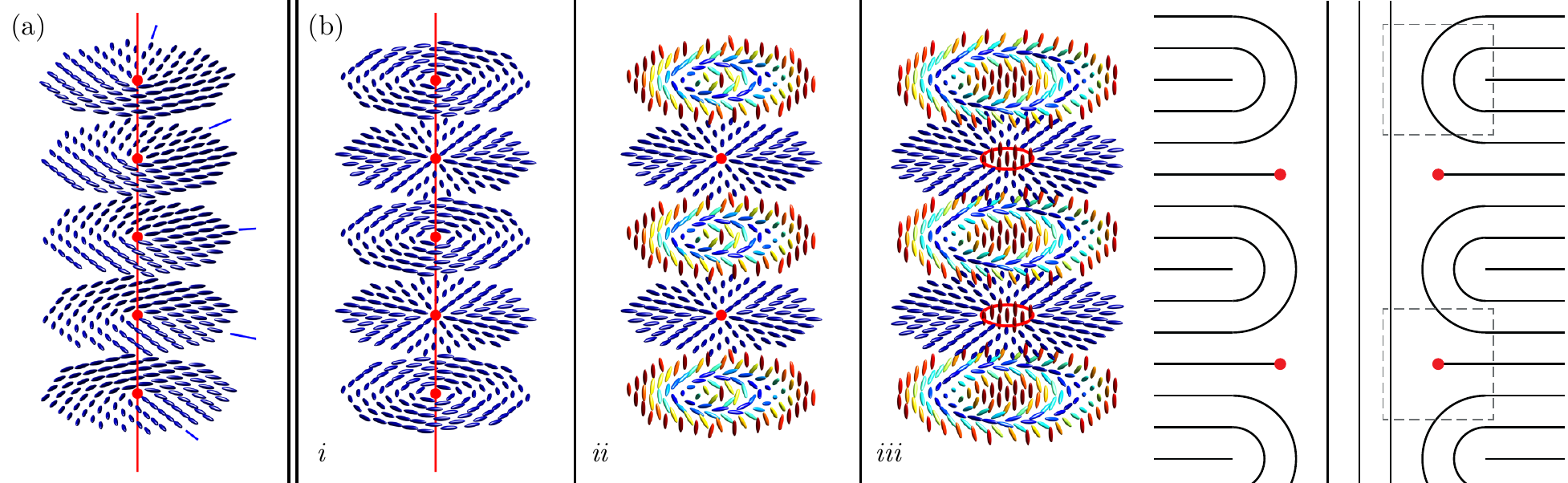}}
\caption{\label{ChiLine} (a) Schematic of the structure of a $\chi^{+1/2}$ line, which consists of a smoothly twisted $m=+1/2$ defect line. The red line denotes the singularity, and the red dots mark the intersection of that line with the represented cross sections of the director field. Blue lines are drawn to guide the eye towards the rotation of the $m=+1/2$ defects. (b) Schematics of the $\chi^{+1}$ line. i. ``Textbook'' version of the $\chi^{+1}$ line seen as alternating bend and splay $m=+1$ director defects along a vertical singular line. ii. Semi-escaped version of i., in which the line singularity ``escapes'' in between the splay defects, transforming regions of high bend into $\lambda^{+1}$ defects. iii. The line in ii. is further relaxed. In the left panel, the $m=+1$ splay point defects are relaxed into $m=+1/2$ loops, and the core of the $\lambda^{+1}$ defects also expand. This relaxation creates a column of vertically aligned nematic in the center of the defect. This is apparent in the right panel, which depicts the director as black lines within a vertical cross section. In the right panel, connecting the horizontal layers of the far field with the vertical layers in the center frustrates the system, generating undulations reminiscent of the Helfrich-Hurault instability. The undulations produce periodic defects, highlighted by dashed boxes (top box: $\lambda^{+1/2}$, bottom box: $\tau^{-1/2}$).}
\end{figure*}

To illustrate a $\chi$ defect, it is useful to view them as line disclinations within a three-dimensional nematic, but with an added modulation along their length due to the cholesteric twist. Consider any point defect with $m \neq 1$ in a two-dimensional nematic: locally rotating the director by a constant angle at every point of the plane will simply induce a global rotation of the defect. Thus, a $\chi$ line disclination with $m \neq 1$ in a cholesteric can be pictured as a two-dimensional point defect extended in the third direction, which is then smoothly twisted [Fig.~\ref{ChiLine}(a)]. 

However, we see the possibility of a more complex $\chi$ defect within a cholesteric shell that has (degenerate) planar anchoring on the inner and outer boundary. A cross-polarized micrograph of a shell with this morphology is in Fig.~\ref{PlanarCholShell}(c). The pseudolayers form concentric spheres with the smallest and largest corresponding to the shell boundaries. The signature of the pseudolayers is visible as a series of concentric dark rings, spaced apart by half of the pitch. The twist axis lies along the radial direction, since it is perpendicular to those layers. One can imagine that defects in cholesteric shells are radially-oriented, singular lines spanning the shell thickness. Structures that seem like radial lines are visible in Fig.~\ref{PlanarCholShell}(c). However, at higher resolution, the defects appear to be more complex than a simple line, with periodic distortions along their length. We can imagine the defect within the shell as a charge $+1$ $\chi$ disclination running from the inner to the outer surface, locally depicted in Fig.~\ref{ChiLine}(b)-i. Compared to $\chi$ disclinations with $m \neq 1$, rotating the director of a $+1$ $\chi$ disclination produces an alternating pattern of pure splay and pure bend defects separated by a quarter pitch. Were we to trace out a surface of constant director orientation, we would find something with the topology of a helicoid -- a dislocation in the pseudolayers, as promised.  However, the defects deform -- in the plane perpendicular to the disclination, the director field attempts to unwind.

Though in the nematic, escaping into the third dimension could lower the amount of elastic distortion in the system and remove any singularities in the director, this is not possible in a cholesteric. The cholesteric's triad of line fields prevents a full escape of the line singularity. The singularity can only escape in alternating regions with a periodicity set by the pitch. Regions of high splay retain director discontinuities at their centers, while the regions of high bend in between are escaped. By escaping, these bend regions become $\lambda^{+1}$ defects. At the core of a $\lambda^{+1}$ defect, the director is vertical (\textit{i.e.,} radial in the reference frame of the shell), and moving away from the core, the director twists smoothly in all directions, becoming points of double twist. On both interfaces of the shell, the semi-escaped $\chi^{+1}$ line terminates with a boojum as in the nematic. This semi-escaped $\chi^{+1}$ line is shown in Fig.~\ref{ChiLine}(b)-ii.

Moreover, the singularity in Fig.~\ref{ChiLine}(b)-ii can relax further into the structure in Fig.~\ref{ChiLine}(b)-iii, to reduce the overall amount of elastic distortion. $+1$ splay defects can open up into looped $+1/2$ disclinations. Inside those defect rings, the director field is uniformly vertical. The vertically-oriented director field at the core of the $\lambda^{+1}$ defects similarly expands. The line singularity in Fig.~\ref{ChiLine}(b)-i is replaced by a vertically aligned director field (\textit{i.e.,} radially aligned in the reference frame of the shell), as depicted in Fig.~\ref{PlanarCholShell}(b) \cite{Sec_frustration, Alex_waltz, tll-copar-clc-planar-shell}. 

With the singularity in Fig.~\ref{ChiLine}(b)-iii being the most energetically favorable, we can imagine how the defects form in experimental systems. Looking at a vertical cross-section of the relaxed, semi-escaped $\chi^{+1}$ line, there is a clear incompatibility in the director orientation between the center of the singularity and the director field far from it. To connect the vertical director lines at the center of the relaxed, semi-escaped $\chi^{+1}$ line with the concentric planar layers that constitute the rest of the shell, undulations along the singularity can result (Fig.~\ref{ChiLine}(b)-iii, right). The ``crests'' of the undulations can generate $\lambda^{+1/2}$ disclinations, while the ``valleys'' can create $\tau^{-1/2}$ disclinations, reminiscent of alternating $\lambda^{\pm1/2}$ defects often seen in cholesterics \cite{Kamien_GeomChol}. As this system has rotational invariance around the axis of the original $\chi^{+1}$ line, the $\tau^{-1/2}$ and $\lambda^{+1/2}$ are looped defects, as illustrated in Fig.~\ref{ChiLine}(b), left \cite{Sec_frustration, Alex_waltz, tll-copar-clc-planar-shell}.

The semi-escaped singularities in cholesteric shells can be viewed through the lens of the HH instability. The mismatch between the vertical director field lines and the far-field, horizontal layers embodies geometrical frustration that is topologically-induced. As in the classical HH systems, the frustration is relieved through periodic elastic distortions that can generate a regular array of defects. However, unlike the original HH analysis, the undulation wavelength in planar cholesteric shells is set by the pitch --- the pseudolayer periodicity. This difference arises from how the geometrical frustration in planar cholesteric shells is induced by the system's global curvature, rather than by an external field. Defects in planar cholesteric shells reveal how topology, \textit{i.e.,} global curvature, can give rise to local, geometrical frustration in layered liquid crystal systems. That the frustration in planar cholesteric shells is relieved through periodic distortions demonstrates the ubiquity of the HH instability, interpreted in this broad sense of relieving layer strain through an undulation with its own periodicity. In this case, the ``undulation'' is a periodic array of defects.

\subsection{Homeotropic cholesteric shells} \label{HomeotropicChShell}

\begin{figure}[!ht]
\centering
  \includegraphics[width=0.48\textwidth]{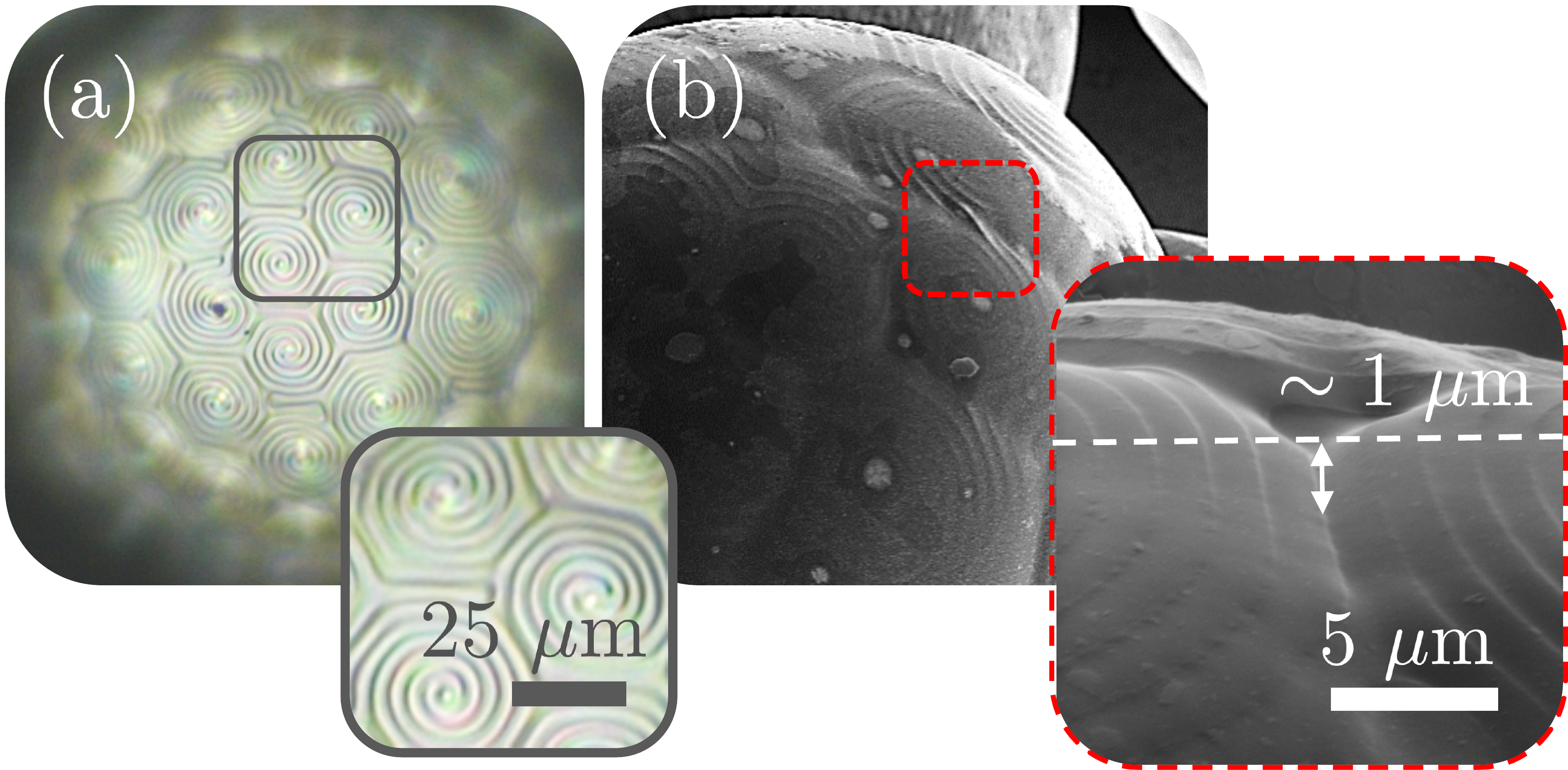}
  \caption{\label{HomeotropicShell} (a) Polarizing micrograph of a cholesteric shell with homeotropic anchoring, due to the presence of a surfactant in the surrounding aqueous solution. (b) A polymerized and dried cholesteric shell with homeotropic anchoring accentuates interfacial deformations due to the underlying focal conic domains. Scanning electron micrographs courtesy of Daeseok Kim.}
\end{figure}

Beyond applied external fields and topological frustration, the competition between the interface and the bulk can also trigger the HH instability, exemplified by cholesteric shells with homeotropic anchoring, shown in Fig.~\ref{HomeotropicShell}. Homeotropic anchoring conditions are particularly frustrating for cholesterics, as the  anchoring  \textit{always} favors an untwisted configuration of molecules and is incompatible with the pseudolayer structure preferred by the bulk. This incompatibility   induces defect structures (arrays of disclination lines), much like the ones shown in Fig.~\ref{ChiLine}. However, unlike the case of  planar anchoring discussed in the previous subsection, the homeotropic cholesteric shell typically has defects tiling the entire surface -- not just at a few, topologically-required points, evidenced by the micrograph in Fig.~\ref{HomeotropicShell}(a). Indeed, the anchoring incompatibility is an example of local frustration. Additionally, the interface itself may locally undulate and deform in response to these defects, to further accommodate the anchoring conditions, shown in the scanning electron micrograph in Fig.~\ref{HomeotropicShell}(b). In this case, the surface tension $\sigma$ must necessarily play a role in establishing the shape of the fluid interface.

Consider the energy contributions of the boundary. A fluid interface introduces both an anchoring energy and a surface tension $\sigma$ that will generally compete with the bulk free energy. These boundary effects may be significant, distorting the interface and modulating the ordering within the layered system \cite{freeCholAnchor}. Assuming a simple model of the interface as a height field\footnote{Note that $h(x,y)$ is a Lagrangian displacement variable of the surface while $u(x,y,z)$ is the Eulerian displacement of the layers. The difference matters at nonlinear order, in principle \cite{KLallthat}.
} $h \equiv h(x,y)$, a general surface energy at a fluid interface would have the form
\begin{equation}
f_{\mathrm{s}}=\int \mathrm{d}^2 \mathbf{x}\,\sqrt{1+(\nabla h)^2}\left[\sigma+A(\hat{{\boldsymbol\nu}}, {\bm{n}}) \right], \label{eq:FEsurface}
\end{equation}
where $\sigma$ is a surface tension and $A(\hat{\boldsymbol\nu},\bm{n})$ is an anchoring strength that will depend on the orientation between the interface normal $\hat{\boldsymbol\nu}$ and the nematic director $\bm{n}$ at the surface. $\bm{n}$ is perpendicular to the pseudolayer normal ${\bm{N}}$ in cholesteric phases, but Eq. \ref{eq:FEsurface} holds generally for all lamellar liquid crystals ($\bm{n}$ can be parallel or at an angle to ${\bm{N}}$ for smectic-A or other smectic phases, respectively).  
\begin{figure}[!ht]  
\includegraphics[height=2.5in]{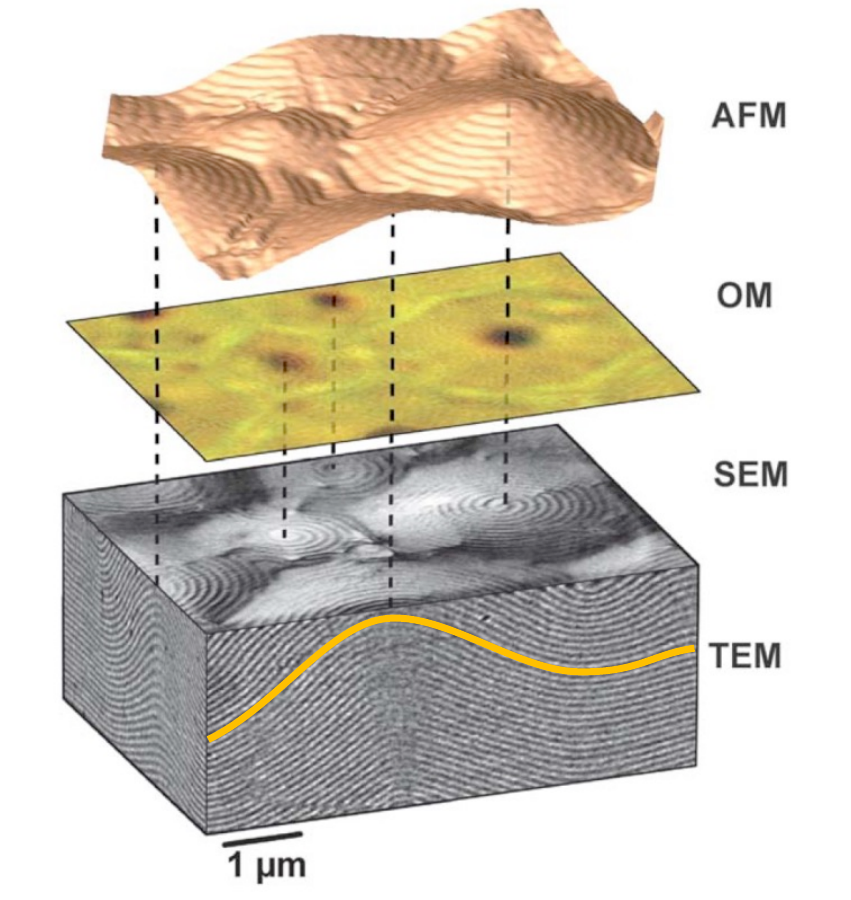}
\caption{  \label{fig:fcdprofile} An undulated cholesteric fluid interface, showing how the incompatible homeotropic anchoring at the surface forces the bulk layers to undulate (orange line) and turn upward, forming focal conic domain ``hills''. Reproduced from \cite{mitov}.}
\end{figure}

The anchoring term must be invariant under ${\bm{n}}\rightarrow - {\bm{n}}$, so we can write $A(\hat{\boldsymbol\nu},{\bm{n}}) = W[1-(\hat{\boldsymbol\nu}\cdot {\bm{n}})^2]/2$, with an anchoring strength $W>0$ for homeotropic alignment and $W<0$ for degenerate planar alignment \cite{rapinipapoular}. Note that this anchoring energy can compete with the periodic ordering of the bulk. This is necessarily the case for cholesterics \cite{freeCholAnchor}:  if the cholesteric pitch axis is oriented in any direction away from the surface normal, the twist of the cholesteric competes with the boundary condition of that surface, whether planar or homeotropic.   Indeed, when $W>0$, there is no configuration that is compatible with a periodic cholesteric and the surface would tend to unwind the cholesteric, competing against the ground state pitch. The anchoring therefore takes the role of  an applied, electromagnetic field, but here, the reorientation of the director occurs only at the surface, instead of throughout the entire system. Just as in the classic system, presented in Sec. \ref{History}, anchoring can also trigger the HH-instability, inducing undulations in the cholesteric pseudolayers. These reorientations undulate the layers just underneath the cholesteric surface, as indicated by the orange line in Fig.~\ref{fig:fcdprofile}.

The onset of undulations is not surprising when the magnitudes of anchoring, surface tension, and bulk elastic energies of typical systems are considered. For example, in common cyanobiphenyl-type liquid crystals with chiral dopants that induce micron-scale pitches, the nematic-isotropic or aqueous interface has anchoring strength $W \sim10^2\text{-}10^5~kT/\mu\mathrm{m}^2  $ \cite{nemisointerface} and surface tension $\sigma \sim 10^5\text{-}10^6~kT/\mu\mathrm{m}^2$ \cite{surfacetensions}.  The  bulk elasticity terms have magnitudes $K_{1,2,3}\sim10^3~kT/\mu\mathrm{m}$ \cite{Frankconstants} and so when the liquid crystal is forced to have defects (with cores on the scale of $1\text{-}10~nm$) to accommodate a frustrating boundary condition, the defects can contribute energy per unit area on the order of $K_i/(10~nm) \sim 10^5~kT/\mu\mathrm{m}^2$. Therefore, for the cholesterics considered here, all of these energetic contributions can compete with one another. 

In cholesterics, the ratio $\sigma/W$ between the interface surface tension $\sigma$ and the homeotropic anchoring strength $W$ determines whether one finds a smooth $(\sigma/W \gg 1)$ or cusped $(\sigma/W \ll 1)$ interface shape \cite{freeCholAnchor}. Moreover, depending on how the cholesteric rearranges near the interface, the interface shape will change to accommodate any defect structures. For example, for an array of disclination lines, the interface may buckle into a wrinkled shape. These considerations also come up near the interface between a cholesteric and an isotropic phase, which favors homeotropic alignment, as discussed in Sec.~\ref{secLCShell} \cite{wettingch, GD_AnchoringTransitions}.

When bulk layer distortions become large and one is far above the threshold for the undulation instability, more complex states emerge. Secondary instabilities are possible, where undulations develop on top of the original undulations. For cholesterics, layers may undulate in two orthogonal directions, creating an array of ``focal conic'' domains \cite{senyuk_undulations_2006, freeCholFCD}, seen also in the classic smectic system detailed in Sec.~\ref{History-Sm}. In extreme cases, such as with very strong incompatible anchoring, the layer structure will strongly distort or break up entirely, yielding intricate defect structures \cite{knots,defectarray1}.
 
 \begin{figure}[!ht]  
\includegraphics[height=2in]{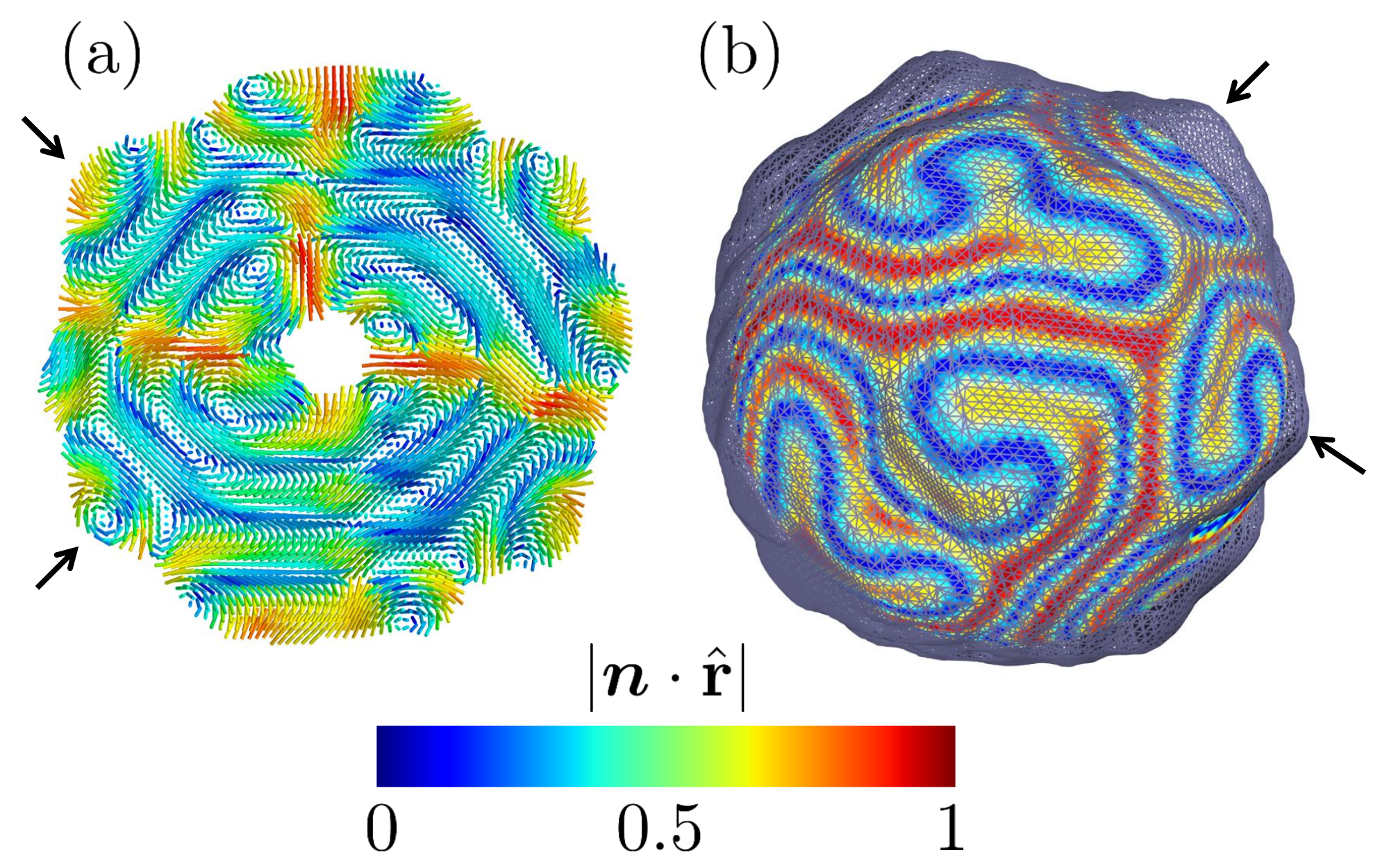}
\caption{(a) Cross section through a cholesteric shell with a free, isotropic-cholesteric interface. The layers are distorted near the boundary and concentric in the bulk. Arrows indicate ``hills'' formed by focal conic domains. The color indicates the director ${\bm{n}}$ orientation relative to the radial direction $\hat{\mathbf{r}}$. The pitch $P_0$ to shell thickness $t$ ratio is $t/P_0 \approx 2$. (b) The distorted, outer interface is shown, with the director distribution just underneath the surface colored by $|{\bm{n}} \cdot \hat{\mathbf{r}}|$. Reproduced from \cite{MOL_LT_PRR}. \label{fig:bumpyspheresim}}
\end{figure}

Multi-scale simulation methods are often employed to capture the interplay between the anchoring energy, the bulk elasticity, and the interfacial surface energy \cite{cholwrinkle, MOL_LT_PRR, Lisa_ChangeStripes}. An example is shown in Fig.~\ref{fig:bumpyspheresim}. We simulate a cholesteric liquid crystal near coexistence between a cholesteric phase (with a pitch, $P_0$) and an isotropic phase. By initializing a shell of the cholesteric inside a bulk isotropic phase, it is possible to generate isotropic-cholesteric, fluid interfaces. As previously mentioned in Sec.~\ref{secLCShell}, these interfaces have a weakly homeotropic anchoring for the cholesteric, creating an anchoring incompatible with the concentric spherical layer arrangement in the droplet bulk. We see in Fig.~\ref{fig:bumpyspheresim} that there is layer reorientation and formation of focal conic domain ``hills'' at the shell surface. The parameters and details of the simulation are described in \cite{MOL_LT_PRR}. These focal conic domain hills are also visible in the cholesteric surface relief, shown in Fig.~\ref{fig:fcdprofile}. Accounting for a deformable boundary and surface tension in the HH instability allows us to capture the interfacial deformations seen in homeotropic cholesteric shells. 

\subsection{Anchoring transitions}\label{AnchTransitionsSec}

\begin{figure}[!ht]
\centering
  \includegraphics[width=0.4\textwidth]{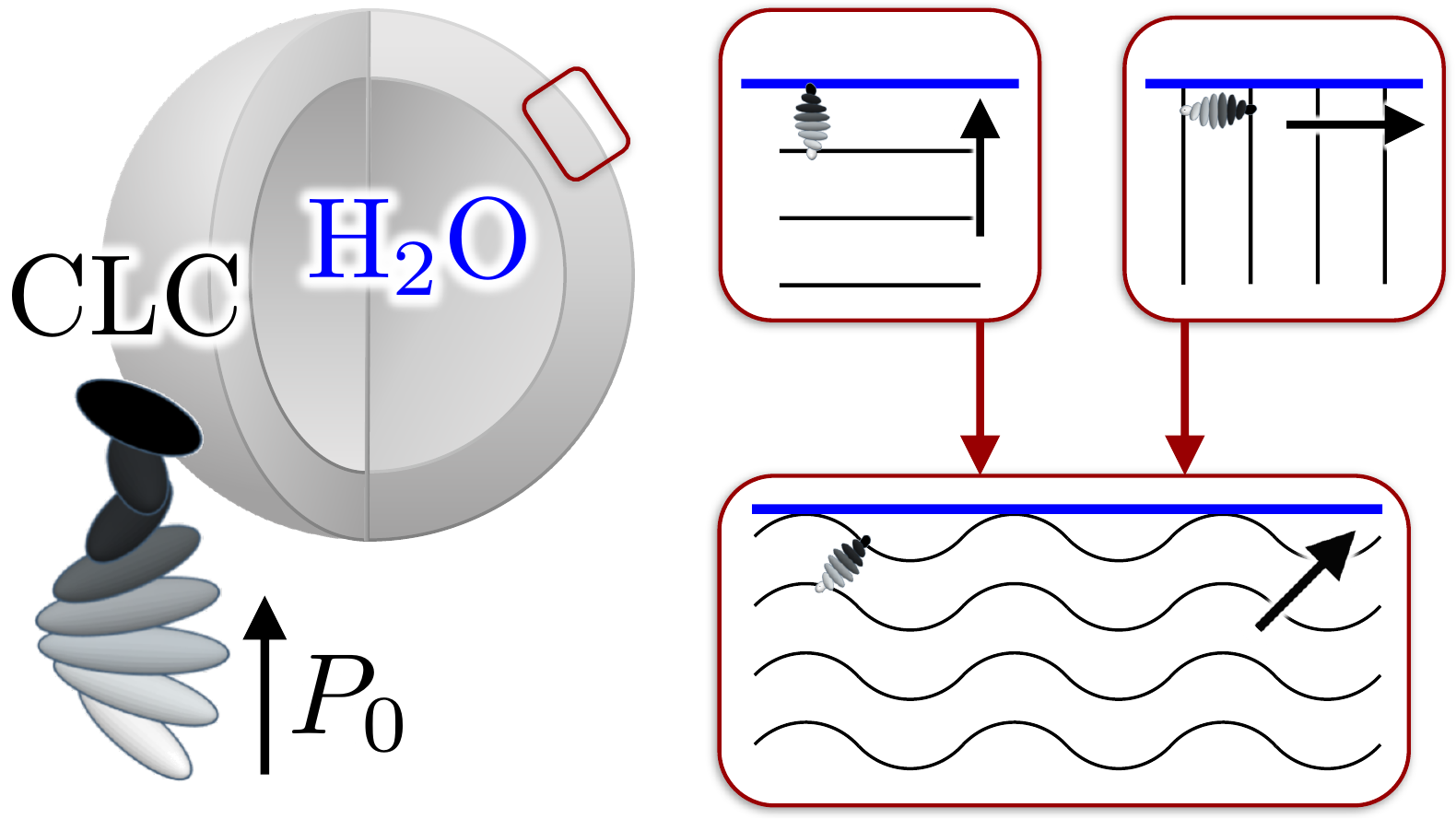}
  \caption{\label{HH-AnchTrans-Schematic} Schematic of a cholesteric liquid crystal shell. The red insets illustrate how changing the anchoring at the shell interface alters the pitch axis orientation, which can lead to a HH-like instability in the bulk (bottom right). Reproduced from \cite{MOL_LT_PRR}.}
\end{figure}

The HH instability can also describe transient states that arise from transitions between the planar and homeotropic structures detailed in the preceding subsections. The changing anchoring is analogous to the application of an external field, but with molecular realignment occurring only at the confining surfaces. As in the classical HH instability, transitioning from one type of anchoring to another at an interface causes the cholesteric pseudolayers to reorganize in order to accommodate the new boundary condition, leading to frustration in the system. As described in the previous subsection, the frustration in the layers can be relieved by an HH-like, undulation instability, as illustrated for a cholesteric shell in Fig.~\ref{HH-AnchTrans-Schematic} \cite{MOL_LT_PRR}. In this subsection, we focus on modeling the onset of the HH instability triggered by anchoring transitions. 

\begin{figure}[!ht]
\centering
  \includegraphics[width=0.42\textwidth]{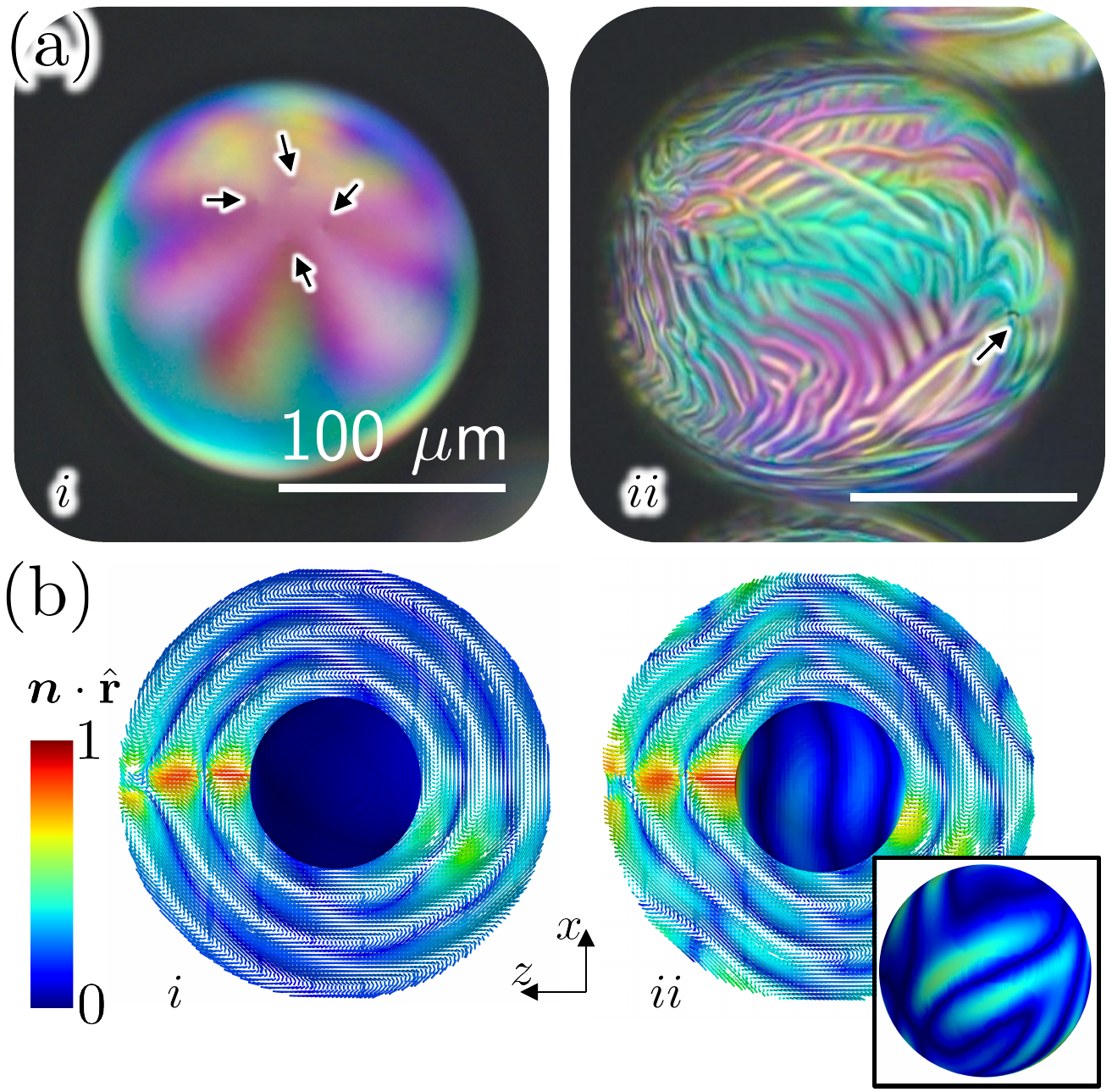}
  \caption{\label{HH-AnchTrans-P2H} (a) An initially planar cholesteric shell (i) has 4 topological defects with charges totaling +2. After the shell is introduced to a solution including 10 mM sodium dodecyl sulfate, the anchoring at the outer shell surface transitions from planar to homeotropic, tilting the pitch axis away from the radial direction. Large stripes are generated at the shell interface with a periodicity around 10 $\mu$m, twice the pitch (ii). Arrows indicate defect locations. (b) An initially planar cholesteric shell (i) with a 1.2 $\mu$m pitch, a 6.6 $\mu$m diameter and a 2.1 $\mu$m thickness is minimized under moderate homeotropic anchoring conditions ($\sim 2 \times 10^{-4}$ J/m$^2$). After $t = 5000$ minimization steps, the cholesteric pseudolayers undulate (ii) and generate stripes, shown in the inset. Adapted from \cite{MOL_LT_PRR}.}
\end{figure}

\begin{figure*}
\centering
  \includegraphics[width=0.8\textwidth]{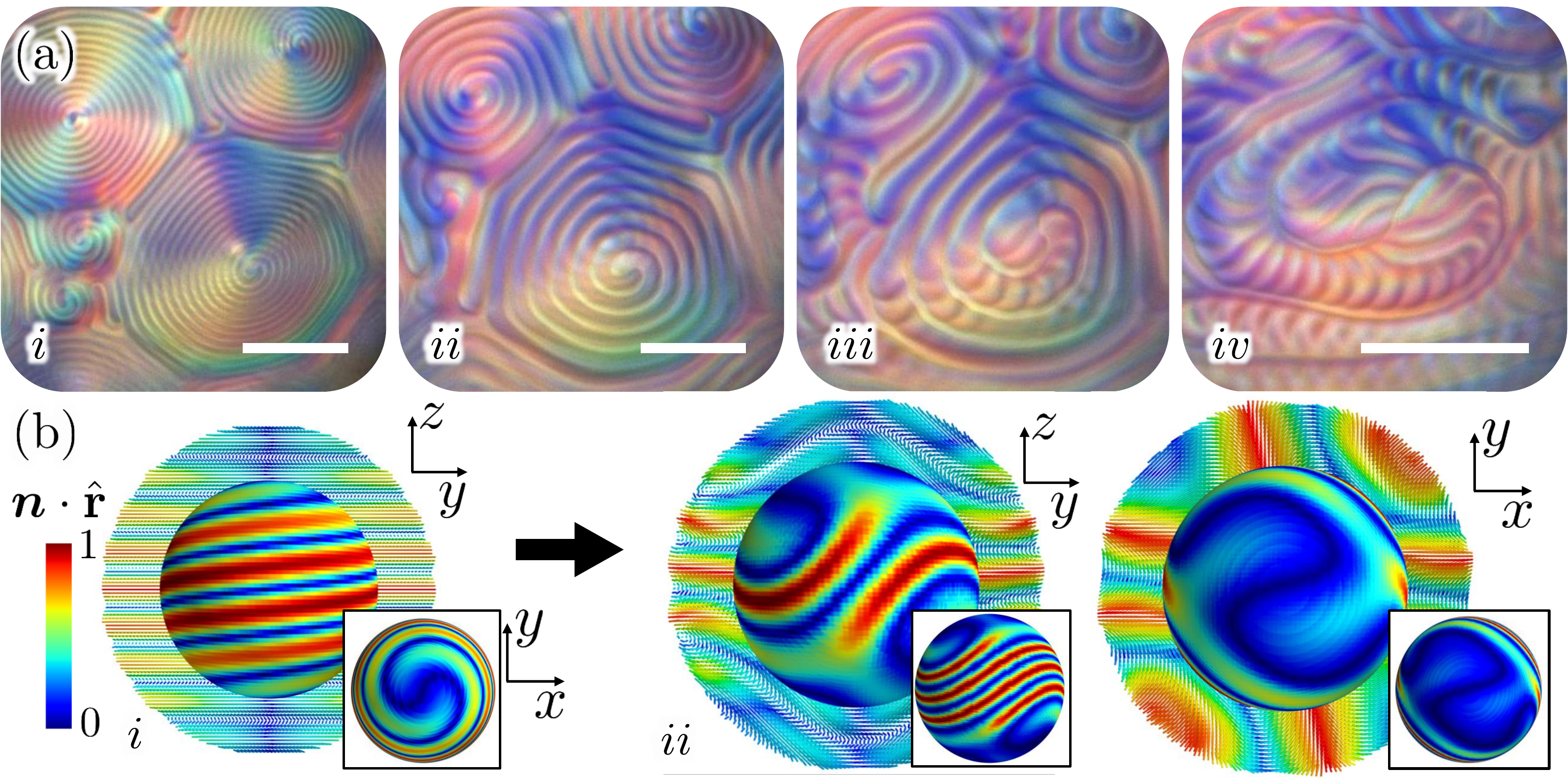}
  \caption{\label{Fig-Onset} (a) A thick cholesteric shell in an aqueous solution with 7 mM sodium dodecyl sulfate, 1 wt \% polyvinyl alcohol, and 0.1 M sodium chloride has a focal conic domain texture initially. The pitch is 5 $\mu$m. The shell is transferred to another, similar aqueous solution, but without sodium dodecyl sulfate, and the texture evolves over time (i-iv). As the outer interface loses homeotropic anchoring strength with surfactant removal to the surrounding solution, the planar anchoring stripes widen (i-ii). When the stripes widen to the point of becoming greater than around twice the pitch ($\sim10$ $\mu$m), the planar stripes fill in with perpendicular stripes of a second periodicity that is also around twice the pitch. Scale bars are 25 $\mu$m. (b) An initially homeotropic cholesteric shell (i) with a 0.18 $\mu$m pitch, a 0.84 $\mu$m diameter and a 0.18 $\mu$m thickness has a pitch axis oriented along the $\hat{z}$-axis. The shell is minimized under planar anchoring conditions ($\sim 2 \times 10^{-4}$ J/m$^2$), resulting in a local energy minimum in which the stripes are partially unwound (ii), with a side view on the left and a top view on the right. Undulations are visible at the poles, where the stripes have unwound. Adapted from \cite{MOL_LT_PRR}.}
\end{figure*}

As detailed in Sec.~\ref{secLCShell}, the anchoring on a cholesteric shell can be tuned experimentally by the addition or removal of surfactant in the surrounding aqueous phases. For the planar to homeotropic anchoring transition, in which surfactant is added to the outer aqueous solution, stripes with a $2P_0$ periodicity cover the cholesteric shell surface without forming a distinguishable pattern (Fig.~\ref{HH-AnchTrans-P2H}). Defects in the nematic director are still present in the system but do not influence the conformation of the stripes beyond their termination at said defects, seen in Fig.~\ref{HH-AnchTrans-P2H}(a)-ii. Similar stripe patterns are captured in Landau-de Gennes simulations of an initially planar cholesteric shell set to minimize under homeotropic anchoring conditions [Fig.~\ref{HH-AnchTrans-P2H}(b)]. Large, transient stripes are formed on the shell surfaces in the beginning of the minimization, similar to experimental observations. Cross sections of the simulated shell reveal that the origin of the large stripes are undulations of the initially concentric, cholesteric pseudolayers [Fig.\ref{HH-AnchTrans-P2H}(b)-ii]. Furthermore, layer undulations are greatest in cross sections that intersect with the radial director defect, indicating that defects are energetically preferred sites for pitch axis and, consequently, cholesteric layer rearrangement.

The transition to planar anchoring similarly produces large surface stripes, where surfactant is removed from the outer aqueous solution. However, unlike for the transition to homeotropic anchoring, the composition of stripe instabilities for planar transitions is dictated by the initial shell patterning, seen in Fig.~\ref{Fig-Onset}(a). As surfactant leaves the interface, weakening the homeotropic anchoring, the planar stripes of the focal conic domain widen until they reach a width $\sim 2P_0$, after which the planar stripes are filled by orthogonal stripes that have a $2P_0$ periodicity. The overall double spiralled structure of the initial focal conic domain is preserved [Fig.~\ref{Fig-Onset}(a)-iv].

\begin{figure}[!ht]
\centering
  \includegraphics[width=0.48\textwidth]{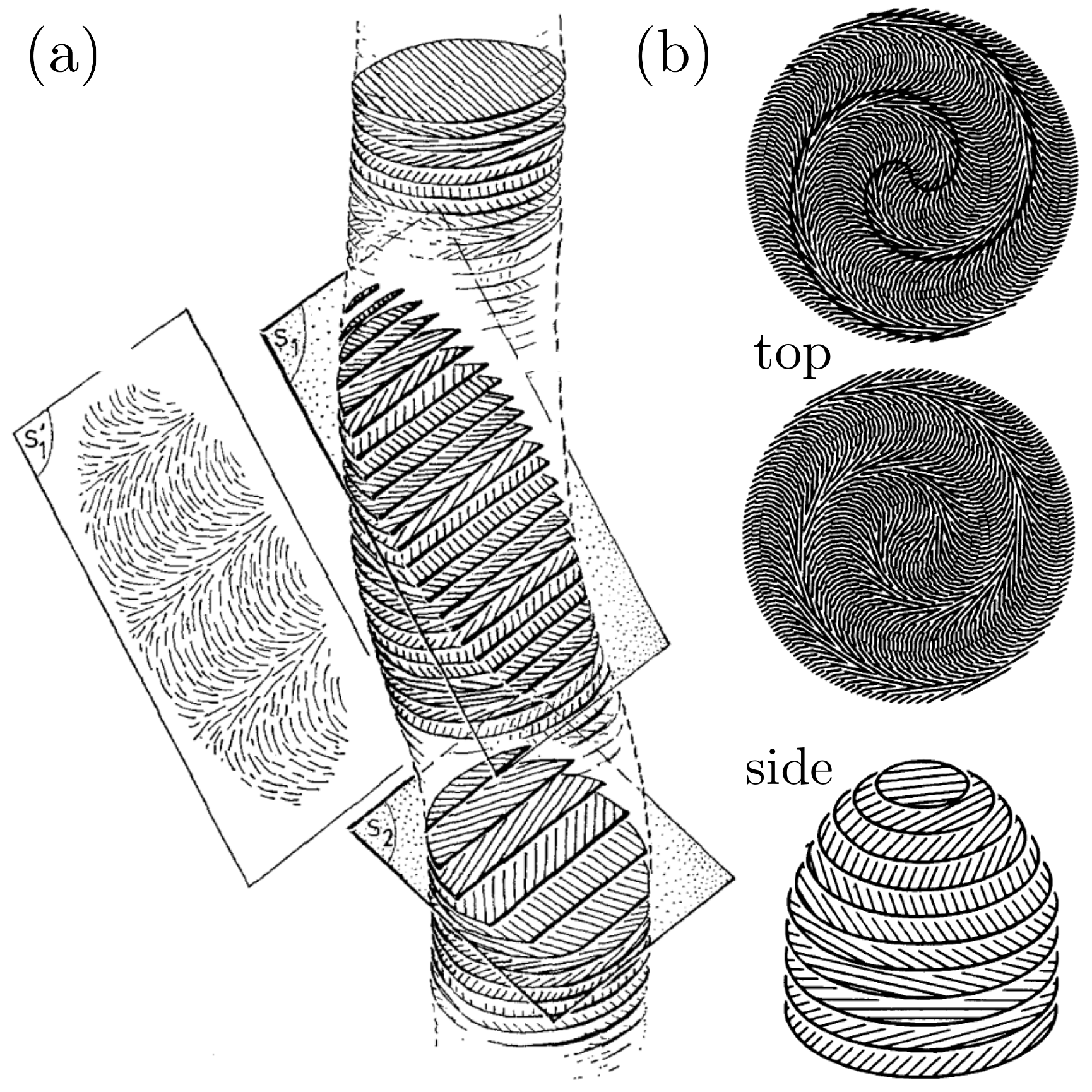}
  \caption{\label{BouligandArches} (a) Schematic of a cholesteric liquid crystal, with its pitch axis oriented vertically. Planes S$_1$ and S$_2$ slice into the cholesteric at an angle to the pitch axis. A curving, periodic texture, called Bouligand arches, is apparent on the surfaces of the planes. Plane S$^{'}_1$ simplifies the pattern on plane S$_1$, highlighting how Bouligand arches reveal cholesteric ordering of the sliced material. Reproduced from \cite{bouligarches}. (b) A cholesteric with a vertical pitch axis is cut into a hill-like shape (side view). Viewing the hill from the top reveals Bouligand arches that follow a double spiral pattern (top). The double spiral pattern is emphasized by black lines. Reproduced from \cite{bouligand-livolant-spher}.}
\end{figure}

We note that the curvature and composition of the orthogonal stripes in the planar transition is reminiscent of Bouligand arches, illustrated in Fig.~\ref{BouligandArches}(a). Bouligand's 1968 work on the chromosomes of dinoflagellates attributed bands of bow-shaped lines found in thin sections of chromosomes to the chiral ordering of filaments in the chromosomes \cite{bouligarches}. The arches that fill in the striped texture of chromosomes are a result of viewing them on a surface that cuts the cholesteric at an angle from the pitch axis.

Indeed, anchoring transitions force the pitch axis to tilt at an angle to the interface, as illustrated in Fig.~\ref{HH-AnchTrans-Schematic}. It is therefore plausible that the structure of the stripe instability is influenced by Bouligand's geometrical arguments. Specifically, the micrograph of Fig.~\ref{Fig-Onset}(a)-iv is evocative of the 1984 study by Bouligand and Livolant of cholesteric spherulites \cite{bouligand-livolant-spher}. Fig.~\ref{BouligandArches} reproduces their illustration that describes the origin of double spiralled structures seen in their experiments. A cholesteric with a vertical, unfrustrated pitch axis is drawn with an angled view in Fig.~\ref{BouligandArches}(b) and is cut into the shape of a hill. Viewing this hill from the top [Fig.~\ref{BouligandArches}(b), top] uncovers a double spiral pattern that is filled in by Bouligand arches.

Although this geometrical model hints at the bulk cholesteric arrangement, this description does not account for the periodicity of the orthogonal stripes that appear to follow an arch-like pattern. As with the homeotropic transition, the organization of the cholesteric layers can also be examined through Landau-de Gennes simulations \cite{MOL_LT_PRR}. Fig.~\ref{Fig-Onset}(b) depicts a cholesteric shell with a pitch axis oriented along the $z$-axis. The focal conic domains are slightly stretched at the poles, resulting in greater regions of planar anchoring that are marked in blue by the ${\bm{n}} \cdot \hat{\mathbf{r}}$ color map. Minimizing this shell under planar anchoring conditions causes the stretched focal conic domains to unwind, generating undulating, orthogonal stripes in regions where the planar anchoring is increased, similar to experimental observations, shown in Fig.~\ref{Fig-Onset}(a). Cross sections of the shell after minimization [Fig.~\ref{Fig-Onset}(b)-ii] reveal that the orthogonal stripes arise from undulation of the underlying cholesteric layers.

We can build a HH-type model of the planar anchoring transition in cholesteric shells by estimating the energy scales associated with imposing an anchoring that induces a tilt in the existing cholesteric pseudolayers on a local patch of the emulsion surface. As detailed in \cite{MOL_LT_PRR}, the free energy of the cholesteric pseudolayers in a small, flat area of the shell surface can be written in the form given by Eq.~\eqref{correcteq}.  Any antagonistic anchoring would tend to reorient the pseudolayers. The associated anchoring energy would have the form of Eq.~\eqref{eq:FEsurface}.  This anchoring energy induces an undulatory instability (a modulation of $u$ in a direction perpendicular to the layers) whenever the anchoring strength $|W|> \pi \sqrt{\bar{K} \bar{B}}$ [see Eq.~\eqref{correcteq}]. Moreover, the wavevector associated with the modulation is $q_c = (\bar{B}/\bar{K})^{1/4} (\pi/2 \ell)^{1/2}$, with $\ell$ being the size of the deformation region near the droplet surface (typically on the order of the pitch). For the cholesteric shells shown in Fig.~\ref{Fig-Onset}(a), these heuristic arguments give reasonable estimates for both the critical $|W| \approx 10^{-5}~\mathrm{J}/\mathrm{m}^2$ and the modulation wavelength $2\pi/q_c \sim 10~\mu\mathrm{m}$ \cite{MOL_LT_PRR}. 
 
For both the planar and homeotropic transitions, the anchoring-induced, HH instability arises from local geometrical frustration between the bulk layer arrangement and the prescribed molecular orientation at the interface. Yet, the conformation of the resultant stripes differs between the two anchoring transitions. For the transition to homeotropic anchoring, the stripes are disordered, with the topologically-required nematic defects serving as favorable sites for initial pitch axis reorientation. For the homeotropic transition, the pitch axis is initially radial and tilts to become tangent to the interface. Since all directions away from radial are equivalent, the onset of the stripe instability is disordered. For planar transitions, pitch axis reorientation occurs first at the pitch defects, evidenced by the unwinding of focal conic domains, where the pitch axis begins to tilt towards radial. Unlike the homeotropic transition, pitch axis reorientation is more constrained. The shortest path for the initial position of the pitch axis to tilt is along the plane orthogonal to the interface that includes the pitch axis. This constraint results in the onset of stripes being orthogonal to, and thus ordered by, the starting stripe pattern, set by the initially tangent pitch axis. Note that the presence of topological defects is not necessary for the anchoring-induced instability to occur. Although the defects generated by \textit{topological} frustration influence the conformation of the stripe instability, the root cause of the stripe instability remains a local, \textit{geometrical} incompatibility between the bulk cholesteric layers and the anchoring condition.

\section{Smectic shells \label{secsmecticshell}}
In the previous section, we introduced the concept of the HH instability in the context of cholesteric shells, where the instability arises as a way of reconciling antagonistic boundary conditions. This is just one of the multiple ways in which geometrical frustration can perturb the structure of a layered system embedded in a spherical shell. Local curvature and global topological constraints can also induce strain in the layers and set off an undulation instability, exemplified by smectic shells with planar boundary conditions.

\subsection{Planar smectic shells in experiments}\label{SmecticShells}
 
\begin{figure}[!ht]
\centering
  \includegraphics[width=0.48\textwidth]{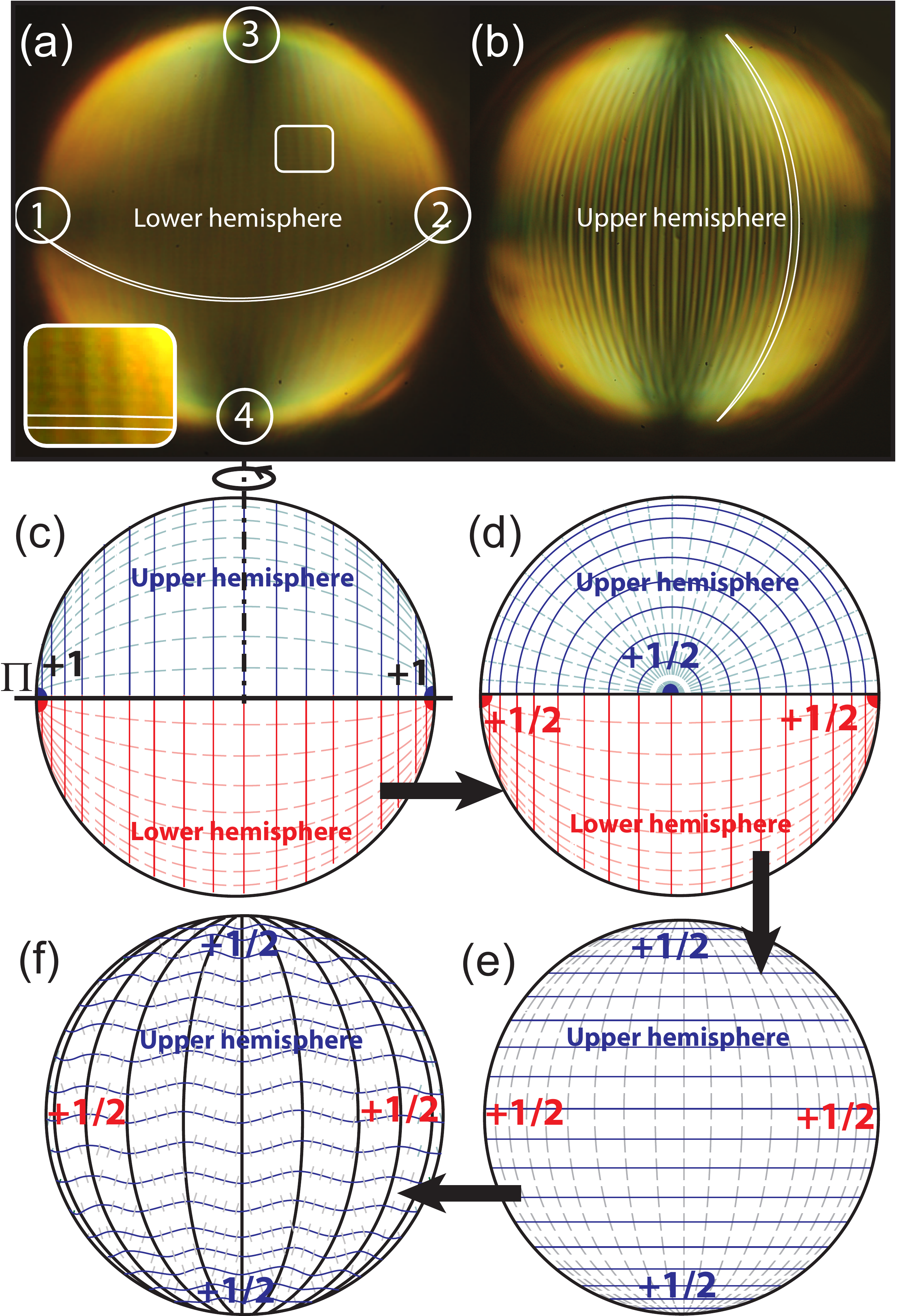}
  \caption{\label{fig:smecticshell} Cross-polarized images of a smectic shell of radius $R=98\text{ }\mu$m and mean thickness $h=1.96\text{ }\mu$m. (a) \& (b) The birefringent texture at the bottom of the shell (a) is different from the birefringent texture at the top of the shell (b) due to the different thickness of these two regions. The inset reveals stripes on the lower hemisphere after image enhancement. Two stripes are outlined at the bottom of the inset to guide the eye. (c) For a two-dimensional smectic shell, theory predicts a configuration with two $s=+1$ defects organized in a bipolar fashion. (d) This configuration is energetically equivalent to any other one that results from splitting the bipolar shell into two halves by a plane $\Pi$ that contain the two $s=+1$ defects, and then, rotating one half with respect to the other one by an angle that can have any value. (e) All the configurations resulting from this transformation have four $s = +1/2$ defects lying on a great circle. (f) This smooth texture is only a first order description of the configuration observed experimentally, where a periodic modulation of the smectic layers is observed. }
\end{figure}

The first study of smectic shells involved bringing planar nematic shells close to the nematic/smectic phase transition temperature, where the elastic ratio $K_3/K_1$ diverges \cite{sm-shell-tll-2, Lagerwall_DefectSmecticShells}. This operation entails the formation of a bend-free state in which the nematic defects relocate to the equator. At the transition, a periodic pattern forms on the shell surface. In Fig.~\ref{fig:smecticshell}(a) and (b) we show cross-polarized micrographs of the lower and upper hemispheres of the same shell. We see  four $+1/2$ defects required by topology and inherited from the nematic state, as described in Sec.~\ref{secPlanarChShell}.  Here the four defects are equally spaced along the equator. Additionally, two sets of longitudinal stripes divide the shell into crescent domains. The first set of stripes connects defects $\small\circled{3}$ and $\small\circled{4}$ by semi-circles that run along the upper hemisphere of the shell, while the second set of stripes connects defects $\small\circled{1}$ and $\small\circled{2}$ by semi-circles that run along the lower hemisphere of the shell. The first set of stripes is visible on the upper hemisphere [see the highlighted crescent domain in Fig.~\ref{fig:smecticshell}(b)]. The second set of stripes is also visible in Fig.~\ref{fig:smecticshell}(a), especially in the top half of the photograph [see the inset of Fig.~\ref{fig:smecticshell}(a)]. This second set of stripes is faint because the bottom part of the shell is thinner than the top. The two set of lines in each hemisphere are orthogonal to each other.

This stripe texture results from an intricate interplay between the curvature of shell, the local energetic constraint of equally spaced layers, global topological constraints, and anchoring conditions.   To understand this,  first consider the limit of vanishing shell thickness where there is no frustration of the smectic layers between the inner and outer surface, as shown in Fig. ~\ref{fig:smecticshell}(c).  The condition of equal spacing results in the layers becoming lines of latitude  \cite{blancEPJE}. The director aligns along the lines of longitude, tracing out geodesics \cite{ccc1,Kamien2009}, depicted as dashed lines in Fig.~\ref{fig:smecticshell}(c-f).  In this situation, there are two $+1$ defects at the two poles.  However, each $+1$ defect can be split in half, and the upper and lower hemispheres can be rotated independently, as shown in Fig.~\ref{fig:smecticshell}(c) and (d) \cite{blancEPJE, Bowick_SphericalNematics, Bates2008}.
The four $+1/2$ defects resulting from this simple surgery sit on a great circle of the sphere [Fig.~\ref{fig:smecticshell}(e)]. The energy difference between the state with two $+1$ defects and those with four $+1/2$ defects comes from the defect core energies and is negligible for large system sizes. While there is a single state for two $+1$ defects, there is an infinite number of states with four $+1/2$ defects. Thus, generically, we expect to see four $+1/2$ defects in the smectic shell, lying along a great circle.  Further minimization of the director energy yields a rotation angle of $\pi/2$, as depicted in Fig.~\ref{fig:smecticshell}(d).  This configuration maximizes the distances between the $+1/2$ defects.

In experiments, however, the shells are three-dimensional and have a thickness that leads to a frustration between the surface anchoring and the layer spacing. Such frustration involves creating either dislocations, layer dilation or anchoring violation, due to the different curvatures of the inner and outer boundaries \cite{sm-shell-tll-1}. In a configuration without dislocations, imposing planar anchoring at the boundaries necessarily implies layer dilation. Again, this frustration is precisely the type that leads to the HH instability --  the smooth texture sketched in Fig.~\ref{fig:smecticshell}(e) is disturbed by the presence of a set of periodic lines and the rapid variation of ${\bm{n}}$ across these lines. By examining the birefringent texture of the experimental shells under rotation, it has been observed that ${\bm{n}}$ is tilted by an almost constant angle $\pm\phi$ [by a few degrees for the shell in Fig.~\ref{fig:smecticshell}(a)] with respect to the two-dimensional director field sketched in Fig.~\ref{fig:smecticshell}(e). Since ${\bm{n}}$ is tilted in opposite directions in two adjacent crescent domains, the visible lines  that separate them roughly correspond to symmetric curvature walls \cite{blanc_curvature_1999}. The sawtooth periodic undulation of the smectic layers schematically represented in Fig.~\ref{fig:smecticshell}(f) is yet another HH instability pattern observed at large strains and is connected to the three-dimensional nature of the shells.

A zero-strain, smectic texture is possible in thick smectic shells provided that the director tilts away from the outer shell surface, incurring an anchoring penalty (see \cite{sm-shell-tll-1}). A first approach to relax this additional surface energy has been developed by Manyuhina and Bowick \cite{sm-shell-ovm}. They examined the influence of a finite anchoring strength $W$ on a \emph{nematic} shell texture with large bending modulus $K_3 \gg K_1$ that is expected to mimic the smectic behavior.  Within the frame of nematic elasticity, they adopted a perturbative approach for thick shells, starting from the ideal two-dimensional structure in Fig.~\ref{fig:smecticshell}(d), while imposing infinitely strong anchoring at the shell inner surface, as well as allowing the director to tilt with respect to the tangent plane and to vary along the shell thickness. The authors proposed a plausible criterion for the onset of director tilting  which should occur when the shell mean curvature $\kappa = 1/R$ gets larger than $W/K_3 $. Moreover, they showed that the axisymmetric texture is unstable beyond this same threshold, where a spontaneous herringbone texture develops.  

This first approach can be complemented with geometrical considerations based on the elasticity of smectic layers, more in line with the HH model. Indeed, the experimental results suggest that, in shells with strong planar anchoring, the strain associated to layer dilation $\gamma$ is released by undulations of the smectic layers, related to a mechanical HH instability \cite{sm-shell-tll-2}.

\subsection{Strain from boundary curvature}\label{Theory-BoundaryCurvature}

\begin{figure} [h]  
\includegraphics[width=2.8in]{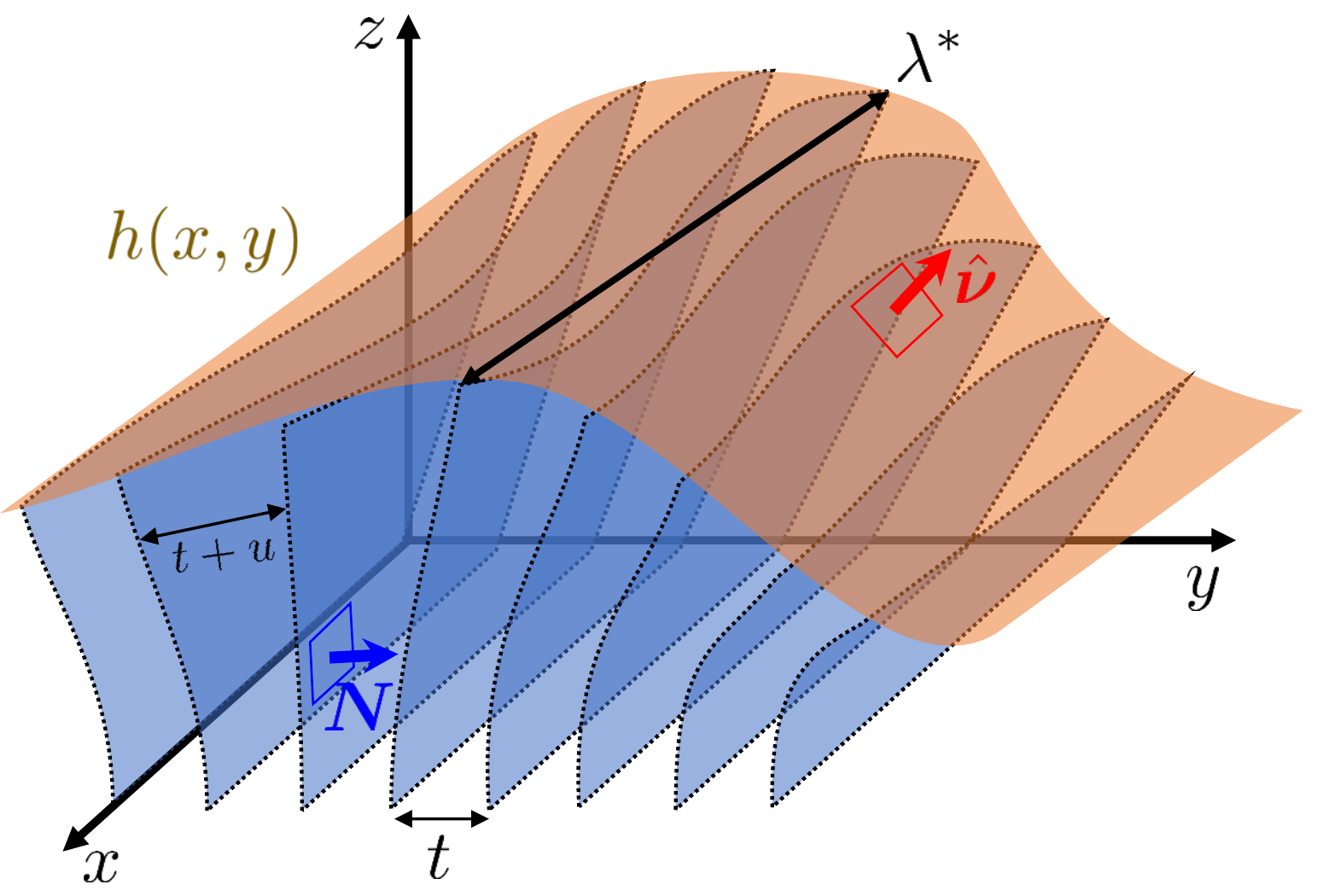}
\caption{ \label{fig:theorysetup} A schematic of a free-interface-induced instability in which a curved, deformable interface (orange) described by a height field $h(x,y)$ induces undulations with characteristic wavelength $\lambda^{\mathrm{*}}$ in the blue layered system. At the interface, the layer normals ${\bm{N}}$ prefer to be perpendicular to the interface normal $\hat{{\boldsymbol\nu}}$. As described in the main text, a curved interface like this will dilate or compress the layers relative to their preferred spacing $t$. The resultant strain may be relieved via layer buckling in a perpendicular direction. }
\end{figure}

Before delving into the specifics of the HH instability in smectic shells, let us take a step back and consider more generally how boundary curvature can strain smectic layers. Consider an interface with some spatially varying surface normal $\hat{\boldsymbol\nu} \equiv \hat{\boldsymbol\nu}(x,y)$, written in terms of the height field $h$ as $\hat{\boldsymbol\nu}=(-\partial_x h,-\partial_y h,1)/\sqrt{1+(\nabla h)^2}$, as depicted schematically in Fig.~\ref{fig:theorysetup}. For illustrative purposes, consider a simple surface shape:  $h(x,y) = d+(2\kappa)^{-1}[\sqrt{1- (2 \kappa  y)^2}-1]$, where $\kappa$ is the mean curvature of the surface and $d$ the film thickness at $y=0$. For $\lvert y \rvert \ll 1/\kappa$ the surface has a parabolic profile $h\approx  d- \kappa y^2$ along the $y$ direction.  So, near the maximum of the parabola, we expand in powers of $y$ and consider the interaction between the surface and  the smectic layers in the bulk.

Using the phase field $\phi \equiv \phi(\mathbf{x})$, the layered structure is recovered by solving $\phi= na$ for $\mathbf{x}$, where $n\in\mathbb{Z}$ labels the layer and $a$ is the layer spacing [see Fig.~\ref{fig:theorysetup}].  Suppose that, in an unperturbed configuration, the layers are stacked along the $y$-direction so that  $\phi=y$.  The layer normal, then, is ${\bm{N}}=\nabla \phi/||\nabla \phi||=\hat{\mathbf{y}}$. If we have perturbations in the layer spacing, this may be captured by a small deformation: $u=y-\phi$. In this case, the layers are still roughly stacked along the $y$-direction, as sketched in Fig.~\ref{fig:theorysetup}, but with deviations described by $\delta \phi$.  Then, assuming planar boundary conditions at the interface that prefer an orientation ${\bm{N}} \perp \hat{\bm{\nu}}$, the surface free energy $f_s$  at the interface  for small $\delta \phi$ and small $\nabla h$ is given by 
\begin{eqnarray}
f_{\mathrm{s}}&\approx&\frac{W}{2} \int \mathrm{d}x\,\mathrm{d}y\, \left.(\hat{{\boldsymbol\nu}}\cdot {\bm{N}})^2\right|_{z=d} \nonumber \\ &\approx& \frac{W}{2}\int  \mathrm{d}x\,\mathrm{d}y\, \left. [\partial_z(\delta \phi)-\partial_y h]^2\right|_{z=d}, \label{eq:anchoring}
\end{eqnarray}
where $W$\ is an anchoring strength. We have substituted $\hat{\bm{\nu}}\approx \hat{\mathbf{z}}-\nabla h$ for the interface normal and  ${\bm{N}} \approx \hat{\mathbf{y}}+\nabla(\delta \phi)$ for the layer normal. This surface free energy is minimized for $\delta \phi(y,z)=-2\kappa yz$, representing a layer \textit{dilation} with increasing $z$.  Therefore, at the top edge of the film, the layer spacing experiences a dilating strain, $\gamma \approx 2  \kappa d$ (relative to the spacing on the bottom of the film), with $d$ being the film thickness at $y=0$. This dilation will  be energetically costly due to the  bulk layer compression elasticity.  The system may relieve this energetic cost in a variety of ways, including disrupting the layer structure via dislocations or developing layer undulations, as illustrated for a generic curved surface in Fig.~\ref{fig:theorysetup}. Here, one sees layers mostly stacked along that $\hat{\mathbf{y}}$ direction, but undulating along $\hat{\mathbf{x}}$ to relieve the strain imposed by the curvature of the interface.   

The details of the layer relaxation are generally complex, since the undulations will coexist with defects, and the details of their interactions are subtle. Analogous issues are seen in smectic systems confined to wedge geometries \cite{bartolino_wedge_1977}. Yet, we can make a basic estimate of the critical strain $\gamma^*$ (applied along the layer normal ${\bm{N}}$) required to induce an undulation.

First, note that the layer compression and bending moduli $K$ and $B$, respectively, combine to yield a characteristic length $\lambda=\sqrt{K/B}$, which governs the size of deformations. This length scale is again the smectic penetration depth, first introduced in Sec.~\ref{History-Sm}. Second, the undulation instability occurs when the layer strain $\gamma$ exceeds a critical value $ \gamma^* \approx 2 \pi \lambda/\ell$ (or, equivalently, if the layer stress exceeds $B \gamma^*$), with $\ell$ being a characteristic sample size in the direction of the applied strain. In the case of our simple  example of an interface $h(x,y) = -  \kappa y^2$ with the layer normals along the $\hat{\mathbf{y}}$ direction and the dilation induced by an interface curvature, $\ell$ would be the extent of the bent region in the $\hat{\mathbf{y}}$ direction. However, the critical strain would also depend on the anchoring strength $W$ and would generally have a complicated form.  Alternatively, if the layers are arranged such that ${\bm{N}} \parallel \hat{\boldsymbol\nu}$ and are dilated by a strain along that same direction (as in the classic instability shown in Fig.~\ref{fig:cholHH}), then $\ell$ would be the film thickness $d$  and $\gamma^*=2 \pi \lambda/d$, as expected.  Furthermore, depending on the nature of the mechanical deformation, there may be some modifications to $\gamma^*$ \cite{napoli_mechanically_2009}. For instance, the surface tension at a fluid interface may modify $\lambda$, introducing an additional length $\lambda \rightarrow \lambda+\lambda_{\mathrm{s}}$,  with $\lambda_{\mathrm{s}}\sim \sigma/B$, with $\sigma$ being the surface tension \cite{dbcp-3}. Nevertheless, the basic scaling $\gamma^* \sim \lambda/\ell$ is predictive in a wide range of cases in which this mechanical instability is observed.
 
Note that the critical strain $\gamma^*$ may be connected to the usual HH critical field $H_c$, since the strain $\gamma$ introduces an energy penalty due to the compression term proportional to $B$. The coefficient $|\chi_a| H^2$ is completely analogous to the stress $\gamma B$ \cite{onuki,delaye_buckling_1973}. The critical field then is directly related to $\gamma^*$ as 
\begin{equation} 
|\chi_a|H_c^2 = \gamma^*B=\frac{2 \pi K}{\lambda \ell},
 \end{equation} 
 which reduces to the $\gamma^* = 2 \pi  \lambda/\ell$ result. The connection to the usual HH scenario, described in Sec.~\ref{originalHelfrich-Hurault}, also allows us to extract the characteristic wavelength $\lambda^*$ of the undulations, given by
\begin{equation}
\lambda^*= 2 \sqrt{ \pi \lambda \ell},
\label{curv-critic-lambda}
\end{equation}
which is consistent with Eq.~\eqref{Helfrich-Huraultqc}. The instability has the same character in smectics and cholesterics \cite{clark_straininduced_1973} and the discussion in  Sec.~\ref{originalHelfrich-Hurault} can be directly mapped to these strain-induced undulations.  
 
The strain $\gamma$ may be imposed externally due to a particular confinement, applied force, or thermal expansion. If the strain occurs near a curved interface, the interface geometry will modify the character of the instability.  For instance, in a smectic with concentric cylindrical layers, a layer dilation induces an instability in which the layers begin to undulate along the cylinder axis. Unlike a flat geometry, the curvature makes the onset of the instability more complex, with the shape of the layer playing an important role \cite{de_gennes_instabilities_1976}. 

Now that we understand how boundary curvature can strain smectic layers enough to trigger the HH instability, we turn back to smectics in a shell geometry. In the following, we consider the simpler case of a cylindrical shell and build on that to qualitatively interpret the spherical shell data.

\subsection{Cylindrical smectic shells}

\begin{figure}[!ht]
\centering
 \includegraphics[width=0.85\columnwidth]{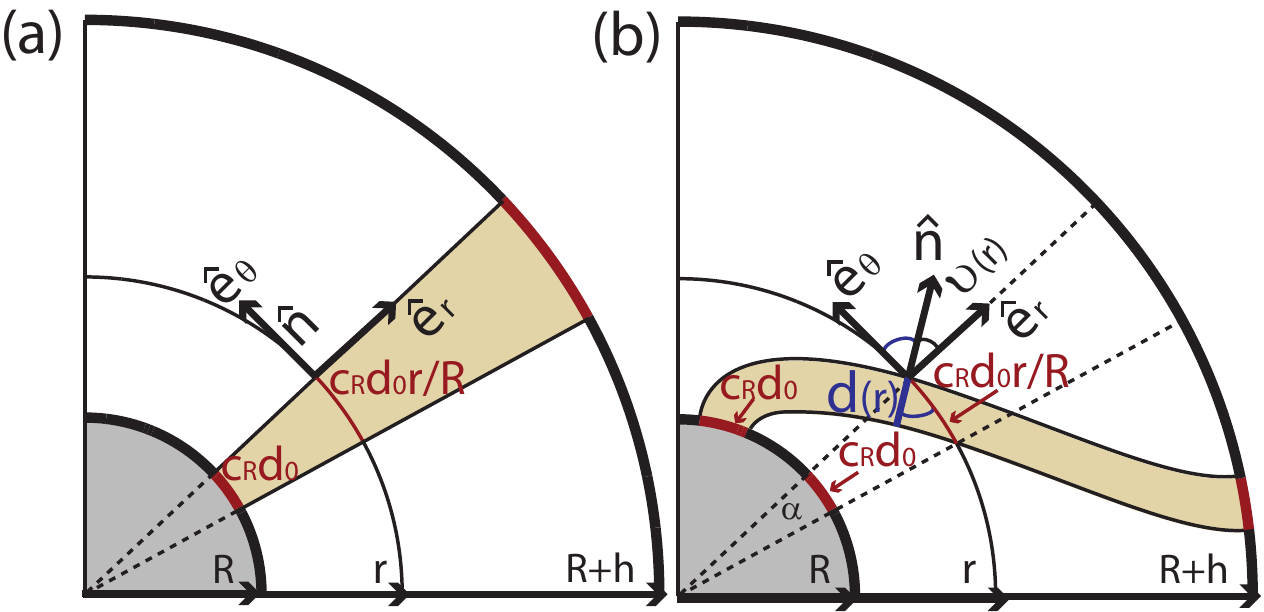}
 \caption{\label{fig:smecticcylinder} Smectic phase confined between two cylinders of radii $R$ and $R+h$. The anchoring is planar on the confining boundaries. (a) In absence of dislocations, a radial texture necessarily induces some dilation in the thickness of a smectic layer (shaded region). (b) The absence of dilation in the bulk implies that the ``orthoradial'' thickness of a layer $d$ varies with the radial coordinate $r$. However, some dilation is necessarily ejected to the outer surface.}
\end{figure}

Consider a smectic slab confined between two cylinders of radii $R$ and $R_0=R+h$, with $h\ll R$ and with strong planar anchoring, \emph{i.e.} $ {\bm{n}}$ lies parallel to the inner and outer cylinders, along $\hat{\mathbf{e}}_\theta$ [Fig.~\ref{fig:smecticcylinder}(a)]. The layers spacing is $a$, and the appropriate smectic free energy is given by Eq.~\eqref{EqFS}, with $\bm{N}$ the layer normal which we call $\bm{n}$ in this section and  $1-|\nabla \phi|^{-1}=1-d/a$, where $d$ is the layer thickness and $a$ the equilibrium layer spacing. If the layers were dilated but not curved [$\mathbf{\nabla}\cdot {\bm{n}}=0$, as schematically represented in Fig.~\ref{fig:smecticcylinder}(a)], their thickness would increase as $d(r)=r c_Ra/R$, where $c_Ra$ is the layer thickness at the inner boundary. Note that the constant $c_R$ is close to 1 and can be chosen to minimize the energy for the bend-free state: for $h\ll R$, we have $c_R\approx 1-h/2R$ with a smectic free energy per unit length of $f_e\approx \pi Bh^3/12R$.

\begin{figure}[!ht]
\includegraphics[width=0.9\columnwidth]{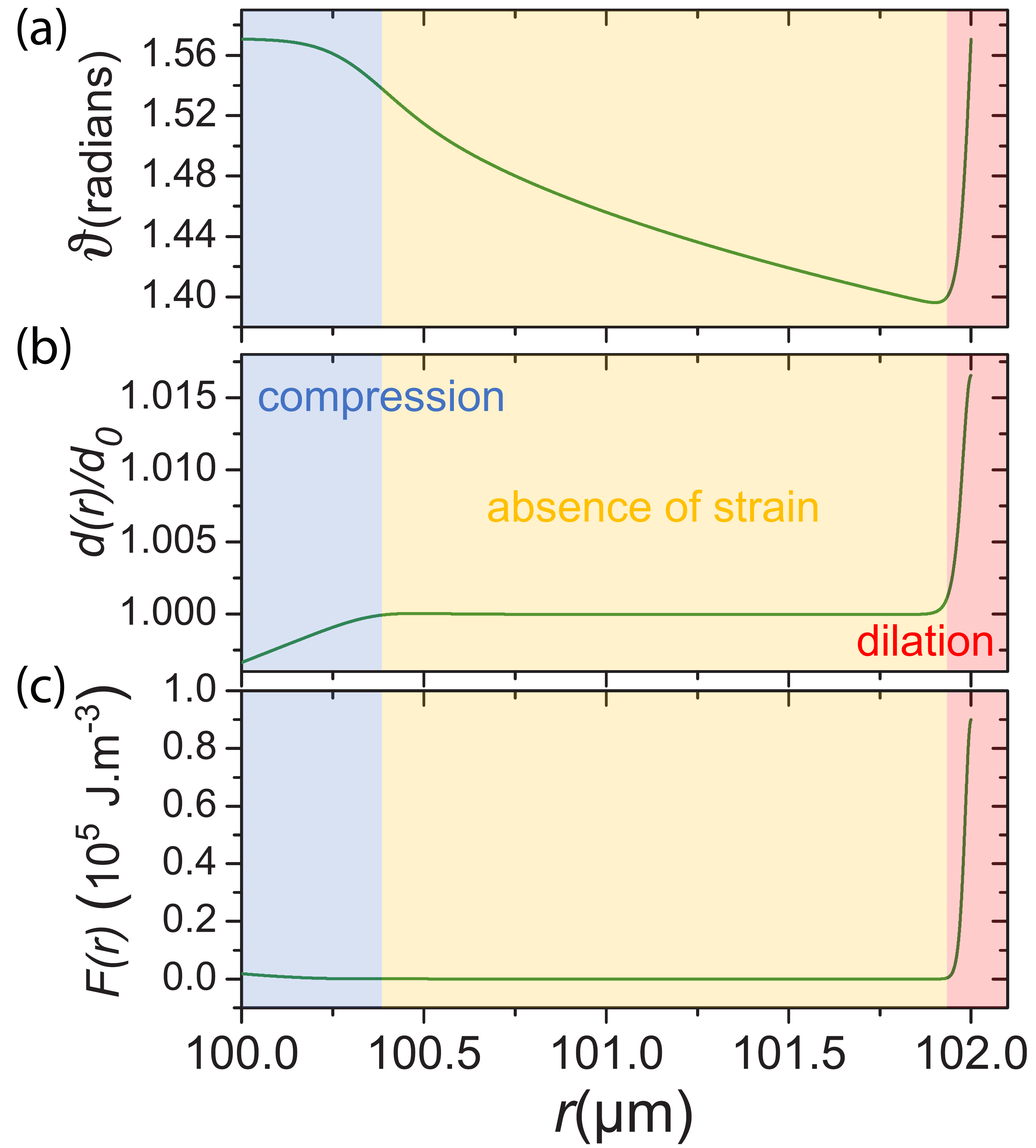}
\caption{\label{fig:shellplots} (a) Tilt, (b) relative dilation, and (c) free energy density of a smectic layer as a function of the radial coordinate $r$ for a shell of 2 $\mu$m thickness and inner radius $R=100\text{ }\mu$m, using $\lambda=3$ nm.}
\end{figure}

Even with an infinite anchoring strength, the elastic energy decreases when we consider a more general scenario, where the smectic layers are allowed to curve into an ``S''-shape, as depicted in Fig.~\ref{fig:smecticcylinder}(b). Treating the system as two-dimensional with no variation along the cylinder, we consider the axially-symmetric director field ${\bm{n}}(r,\theta)=\cos \vartheta(r)\hat{\mathbf{e}}_r+\sin \vartheta(r) \hat{\mathbf{e}}_\theta$, where $\vartheta(r)$ is the tilt of the director and the normal of the layers, with respect to the unit radial vector $\hat{\mathbf{e}}_r$, [Fig.~\ref{fig:smecticcylinder}(b)]. The width of the layers is $d(r)=\sin \vartheta (r) r c_Ra/R $ [Fig.~\ref{fig:smecticcylinder}(b)] and the free energy density is

\begin{equation}
f_e=\frac{B}{2}\left(1-\frac{c_Rr\sin \vartheta(r)}{R}\right)^2+\frac{K}{2} (\mathbf{\nabla}\cdot
\bm{n})^2.
\end{equation}

\noindent The Euler-Lagrange equation reads

\begin{equation}
\frac{d^2 \vartheta}{dr^2}=\frac{c_R^2r^4-R^2\lambda^2-\frac{c_R r^3R}{\sin \vartheta}}{
r^2R^2\lambda^2\tan \vartheta}-\frac{d\vartheta}{dr}\left(\frac{1}{r}+\frac{d\vartheta}{dr}\frac{1}{\tan
\vartheta}\right), \label{eqcylind}
\end{equation}

\noindent where we impose the boundary conditions $\vartheta(R)=\vartheta(R+h)=\pi/2$. The tilt angle $\vartheta(r)$ and the shape of layers can then be obtained by numerically solving Eq.~\eqref{eqcylind} using standard two point boundary value methods and optimizing the resulting elastic energy $f_e(c_R)$ as a function of $c_R$. Fig.~\ref{fig:shellplots}(a) shows the numerical solution, $\vartheta(r)$, for $R=100\text{ }\mu$m and $h=2\text{ }\mu$m, which are typical values for the shell radius and thickness of the experimental shells. The relative dilation of the layers $d/a$ and the free energy density $f_e(r)$ of the ground state configuration are shown in Fig.~\ref{fig:shellplots}. Three different regions can be distinguished: the layers are slightly compressed in a thin inner region, highlighted in blue, while dilation is mostly confined at the outer surface, highlighted in red.  The dilation is nearly vanishing in the yellow region, between the two boundary layers.

\begin{figure}[!ht]
\centering
 \includegraphics[width=0.85\columnwidth]{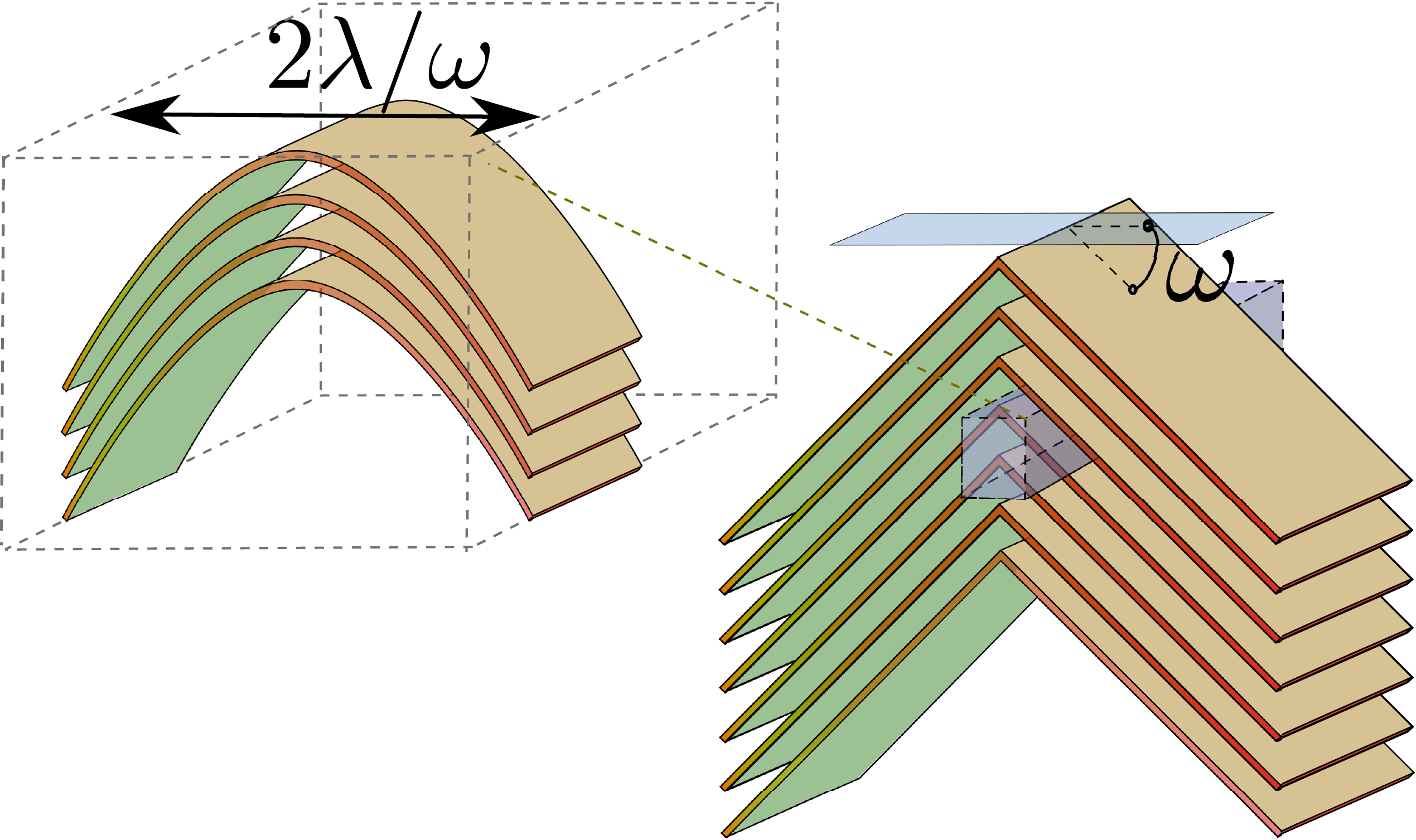}
 \caption{\label{fig:wallbend} Curvature walls separate domains in which $\bm{n}$ displays opposite orientations $\pm \omega$. If $\omega$ is not too high, the smectic layers undergo a continuous bend across the wall.}
\end{figure}

The fact that dilation is expelled from the bulk is a well known phenomenon for layered systems \cite{blanc_curvature_1999}. Indeed, comparing the two terms of Eq.~\eqref{EqFS} reveals that dilation can be present only in regions where the curvature of the layers ($\mathbf{\nabla}\cdot\bm{n}$) is of order $\lambda^{-1}$, \emph{i.e.} where the layers rapidly reorient, or near the common focal surfaces of a set of equidistant layers. That is why the macroscopic textures of layered systems can be described by an extended geometrical description, based on a combination of  domains with equidistant layers separated by curvature walls of varying shapes, as illustrated in Fig.~\ref{fig:wallbend} \cite{blanc_curvature_1999}. If the ``mis-orientation'' angle  $2\omega$ of a wall is not too high, the layers remain continuous. The width of a wall is thereby $2\lambda/\omega$, and its free energy per unit area is
\begin{equation}
f_w\approx \frac{2K}{\lambda} (\tan \omega-\omega)\approx \frac{2K\omega^3}{3\lambda}. \label{enerwall}
\end{equation}
Here, the geometrical construction driven by a strong planar anchoring at the inner surface yields $\sin \vartheta(r)=R/r$ in the bulk and a tilt $\omega_0=\pi/2-\vartheta(R+h)\approx\sqrt{2h/R}$ at the outer cylinder. The dilation is thus expelled in a curvature wall that contains most of the elastic energy per unit length: 

\begin{equation}
f_e =\frac{4K\pi h}{3\lambda} \sqrt{\frac{2h}{R}}. \label{eqenercylind}
\end{equation}

\begin{figure}[!ht]
\centering
  \includegraphics[width=0.85\columnwidth]{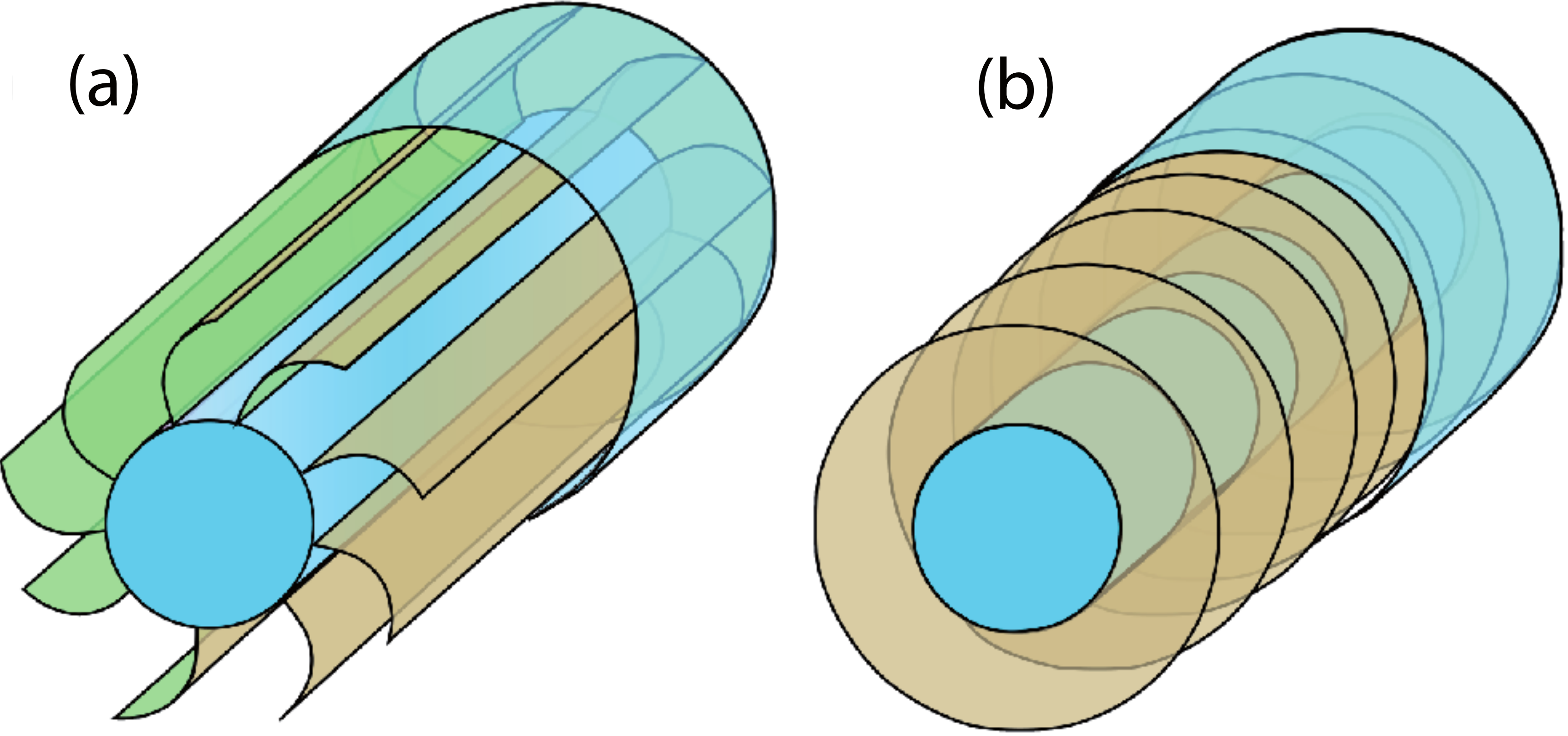}
 \caption{\label{fig:3Ddistortion}
 In three dimensions, the distortion of the smectic layers is efficiently relaxed by an overall rotation of the layers, allowable when the anchoring is planar degenerate.}
\end{figure}

With this two-dimensional approach, dilation is confined to the neighborhood of the outer cylinder. However, three-dimensional distortions of the director field are expected to further lower the resulting elastic energy. For example, with degenerate planar anchoring on both cylinders, the elastic energy can be entirely relaxed when the director is oriented along the other principal curvature direction, where curvature is null (Fig. \ref{fig:3Ddistortion}).

\subsection{Spherical smectic shells}\label{secSm-Secondary}

A similar geometrical frustration is present in spherical smectic shells of finite thickness, but contrary to the cylindrical case considered in the previous section, the elastic energy cannot be globally relaxed. Unlike a cylinder which only bends in one direction, the two principal curvatures on a sphere are nonvanishing (and equal).  Thus the geometrical strain remains, regardless of the layer orientation. 
Let us first examine the vanishing thickness limit shown in Fig.~\ref{fig:smecticshell}(c). Such an ideal smectic sphere has two +1 defects located at the north and south poles, and the surface director is given by ${\bm{n}}=\hat{\mathbf{e}}_\theta$, written with the usual spherical coordinates $r,\theta,\varphi$ and the corresponding unit vectors $\hat{\mathbf{e}}_r,\hat{\mathbf{e}}_\theta,\hat{\mathbf{e}}_\varphi$. For shells of finite thickness, a geometrical construction from the inner sphere of radius $R$, similar to that shown in Fig.~\ref{fig:smecticcylinder}(b) for the cylindrical case, is also possible, giving rise to  a director field $\bm{n}=\sin \vartheta(r) \hat{\mathbf{e}}_\theta\text{ } \pm \text{ }\cos \vartheta(r) \hat{\mathbf{e}}_r $ where $\sin \vartheta(r)=R/r$. The resulting angular misfit at the outer surface of a  shell of thickness $h$ is still $\omega_0=\arccos R/(R+h)\approx \sqrt{2h/R}$, but now uniform all over the spherical system. Therefore, the resulting half-curvature wall located at the outer sphere has an elastic energy:

\begin{equation}
f_e \approx \frac{4\pi R^2K}{3\lambda} w_o^3. \label{eqenersphere}
\end{equation}

\begin{figure}[!ht]
\centering
  \includegraphics[width=\columnwidth]{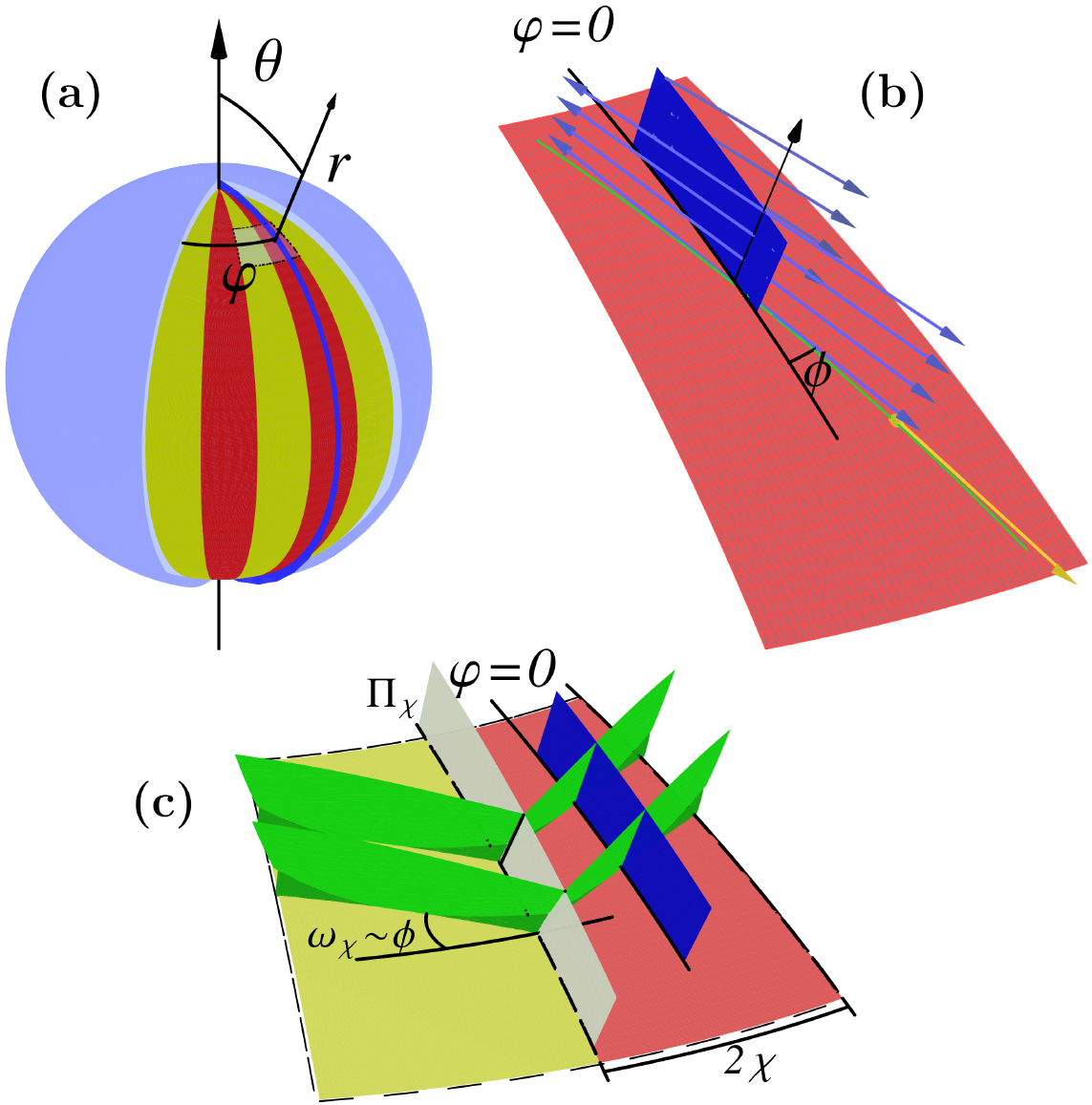}
 \caption{\label{figconstruction}
 (a) Spherical coordinates used to describe the smectic textures of a shell. (b) The director field $\bm{n}(r,\theta,\varphi=0)$ defined on the $\varphi=0$ plane is used to construct the director field in its vicinity (blue lines). The region of the shell near the inner surface can be complemented with a set of lines (yellow line) tangent to each great circle parallel to $\bm{n}(r=R,\theta,\varphi=0)$.  (c) This defines parallel smectic layers (light and dark green surfaces) in a limited neighborhood $-\chi<\varphi<\chi$  but the process can be iterated periodically along the azimuthal direction (as shown here for the region $-3\chi<\varphi<-\chi$)  at the cost of additional curvature walls $\Pi_\chi$ of tilt mis-orientation $\sim 2\phi$.}
\end{figure}

\noindent Breaking of rotational invariance around the $z$-axis can strongly reduce this elastic energy. A simple geometrical construction, which shares many common features with experimental shells, is detailed in Fig. \ref{figconstruction}.

 We first consider the following director field, defined in the shell region  $R\leq r\leq R+h$ of the half plane $\varphi=0$:
\begin{equation}
\bm{n}(r,\theta,\varphi=0) = \frac{ R\cos \phi}{r}\,\hat{ \mathbf{e}}_\theta+\sqrt{1-\frac{ R^2\cos^2 \phi}{r^2}} \,\hat{\mathbf{e}}_\varphi
\label{eqfield}
\end{equation}
where $\phi$ is nonvanishing. For the given azimuthal angle $\varphi=0$, this field ensures a strong planar anchoring on inner and outer spheres.  It is also compatible with smectic layers since it defines a set of non-intersecting straight lines, the common normals of a set of parallel layers, in the neighborhood of $\varphi=0$.  The straight lines representing the smectic director field built in this way are defined almost everywhere in the shell, excluding in a region near the inner sphere. In this region, the set of straight normals can be complemented with half-lines tangent to great circles of the sphere, that are themselves tangent to the director $\bm{n}(R,\theta,0)$ [Fig.~\ref{figconstruction}(b)]. Since great circles are geodesics of the sphere, this construction ensures that the resulting layers are everywhere perpendicular to the inner sphere \cite{blancEPJE}. However, topology comes into play: this geometric construction cannot be extended to the whole sphere without additional defects. For instance, the great circles will intersect at some distance from $\varphi=0$. The process is therefore limited in the azimuthal direction up to angles  $\varphi=\pm\chi$, but it can be iterated periodically along $\varphi$ with additional curvature walls $\Pi_\chi$ separating the crescent domains, as shown in Fig.~\ref{figconstruction}(a) and \ref{figconstruction}(c).

In this texture,  the tilt of the layers at the outer surface is much smaller than the uniform tilt $\omega_0$ of  the rotational invariant texture. By construction, the tilt is indeed zero at $\varphi=0$ and slowly increases when departing from this azimuthal angle. The dilation in the curvature wall at the outer sphere is thereby strongly reduced. We can straightforwardly compute $w_e(\theta,\varphi)\approx |\varphi|\sin \theta/\sin \phi$ to first order, which gives a maximal extension $\chi\approx \omega_0 \sin \phi$ for the crescent domains.

The gain in elastic energy at the outer wall is counterbalanced by an energy cost of the $\pi/\chi$ additional curvature walls, $\Pi_\chi$, located at the crescent boundaries. However, numerical computations of the energy $F_T$  resulting from the combination of the energies at the outer sphere and in the $\Pi_\chi$ walls show a large net gain in elastic energy. A rough computation can also be obtained by considering that the mis-orientation angle of the $\Pi_\chi$ wall is $2\omega_\chi\approx2\phi$. Using Eq.~\eqref{enerwall} for wall density energy and $w_e(\theta,\varphi)\approx |\varphi|\sin \theta/\sin \phi$ one obtains:

\begin{equation}
F_T \approx \frac{2\pi^2K}{3\lambda \chi}\phi^3 R h+ \frac{\pi^2 K\chi^3}{16\lambda \sin^3 \phi}R^2  \label{eqenercombined}
\end{equation}

which reduces to
\begin{equation}
F_T \approx \frac{2\pi^2K}{3\lambda \omega_0}\phi^2 R h+ \frac{\pi^2 K\omega_0^3}{16\lambda }R^2  \label{eqenercombinedreduced}
\end{equation}
when $\chi\approx \omega_0 \sin \phi$ and $\phi\ll 1$. For small enough values of $\phi$, the combined energy is much lower than that of the global solution Eq.~\eqref{eqenersphere}.

\begin{figure*}[!ht]
\centering
  \includegraphics[width=1\textwidth]{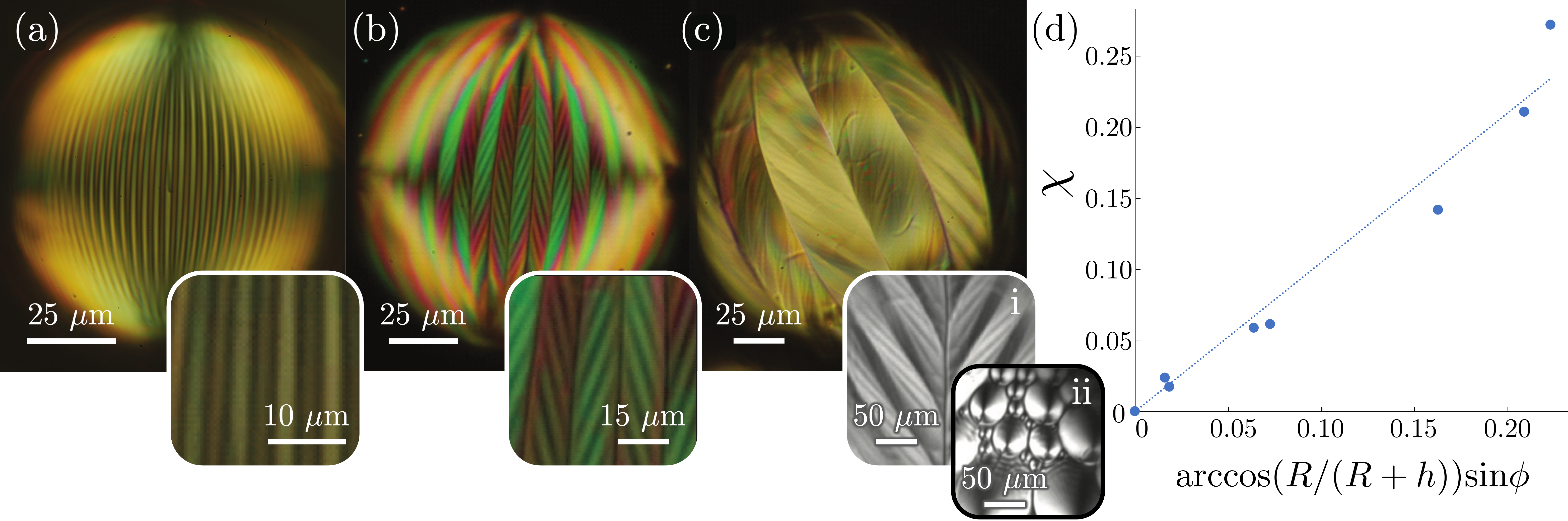}
  \caption{\label{fig:smecticthickness}
Effect of the shell thickness in the smectic texture. (a-c) Cross-polarized images of smectic shells with a normalized mean thickness $h/R= 0.020$, $h/R= 0.036$, and $h/R=0.145$, respectively. Polarizing optical microscopy; the focus is set on the top of the shell (the thickest region). The insets highlight (a) primary curvature walls ($h/R= 0.020$), (b) secondary curvature walls ($h/R= 0.036$), (c)-i tertiary curvature walls ($h/R= 0.145$), and (c)-ii focal conics ($h/R \approx 1$) in smectic shells. (d) Experimental data showing a linear dependence between $\chi$ and $\arccos (R/(R+h))\sin \phi$. The slope of the fitting line is $1.05$, consistent with the geometrical construction in Fig.~\ref{figconstruction}, where $\chi\approx\arccos (R/(R+h)) \sin \phi$.}
\end{figure*}

Note that these considerations are appropriate for studying the ground state of a thick smectic shell for which $h/R > \gamma^*$ is much above the critical strain to induce the initial mechanical HH instability. It would also be interesting to study very thin shells and to observe the onset of the  instability. The wavelength $\lambda^*$ of the initial   instability should be given by Eq.~\eqref{curv-critic-lambda}. For example, at the equator, the relevant length scale over which the mechanical deformation occurs is just the shell circumference $2\pi R$, so we would expect $\lambda^* \approx 5~\mu\mathrm{m}$ for the $\lambda \approx 30~\mathrm{\AA}$ layer spacing of an 4'-octyl-4-biphenylcarbonitrile (8CB) smectic liquid crystal confined to a shell with radius $R=100~\mu\mathrm{m}$, for example.  This is consistent with the spacing of the initial curvature walls (see, e.g., Fig.~\ref{fig:smecticthickness}).

Experimentally, the period and the amplitude of the elastic instability are highly dependent on the shell thickness. As the shell becomes thicker, the angular period decreases, the wavelength of undulations increases, and fewer domains are observed. This effect can be qualitatively seen in Fig.~\ref{fig:smecticthickness}(a)-(c), where the number of crescent domains dividing a shell decreases as the normalized thickness, $h/R$, increases. Additionally, the tilt angle $\phi$ inside the domains concomitantly increases with the thickness. According to the geometrical considerations described above, the width of the domains $2\chi$ should be related to the tilt angle $\phi$ as $\chi\approx \omega_0 \sin \phi$, where $\omega_0=\arccos (R/(R+h))$ a linear dependence seen in experiments [Fig.~\ref{fig:smecticthickness}(d)].

In addition to the quantitative changes observed for the amplitude and period of the instability, further increase of the shell thickness entails deeper structural changes. In very thin shells, only the primary curvature walls discussed in the last section can be distinguished and the modulation is simple [Fig.~\ref{fig:smecticthickness}(a)]. However, in thicker shells, primary curvature walls of large tilt angle ($\phi > 10^\circ$) are filled in with secondary curvature walls of a few degrees of tilt to form a herringbone texture [Fig.~\ref{fig:smecticthickness}(b)], and the secondary curvature walls can be further patterned by tertiary curvature walls with increasing shell thickness [Fig.~\ref{fig:smecticthickness}(c)]. Observations of the light extinction between crossed polarizers show that each set of walls are roughly perpendicular to the average orientation of the modulated layers.  In very thick shells, this hierarchical organization is broken and is replaced by a complex texture made of focal conic domains [Fig.~\ref{fig:smecticthickness}(c)-ii], reminiscent of the ones observed in large single spherical droplets with planar anchoring \cite{Fournier_spherulites,BlancPRE2001,blancEPJE}. 

The appearance of the secondary and tertiary patterns in thick shells can be qualitatively understood in the geometrical framework examined in the previous section. After the first instability, the tilt $\omega_e$ of the layers at the outer spheres has strongly decreased but is almost nowhere null. The layers are roughly tilted with an angle $\pm \phi_1$ with respect to the latitude lines. Iterating the process with smaller angles $ \phi_2$ allows for the decrease of $\omega_e$ once again, at the cost of additional curvature walls of smaller energy. The dilation that was localized only at the outer sphere in the rotational invariant construction is then strongly reduced, while a part of it is redistributed in the whole shell in the form of mis-orientation walls.

The smectic layers, antagonized by the system's spherical geometry, undulate to maintain their preferred spacing, patterning the shell with curvature walls. The wavelength of the undulations increases and fills in with hierarchical undulations with increasing shell thickness, similar in spirit to the undulations observed in planar anchoring transitions of cholesteric shells. The incompatibility of the shell curvature with the smectic layers and the emergent, periodic textures that result exemplify how geometrical frustration is at the core of the HH instability.

\section{Other mechanisms to the HH instability}

The liquid crystal shells examined in this review underwent the HH instability due to frustration from topological constraints, changes in anchoring conditions, and boundary curvature. Other sources of frustration have also been found in other systems, including changes in layer spacing due to phase transitions and sample thicknesses incompatible with the layer spacing, which can be described by the classic strain mechanism of HH. Before reviewing other phenomena that fall into the HH umbrella, we note that there are other possible contributions to the HH instability in lamellar systems.   

\begin{figure}[ht!]
\includegraphics[width=0.48\textwidth]{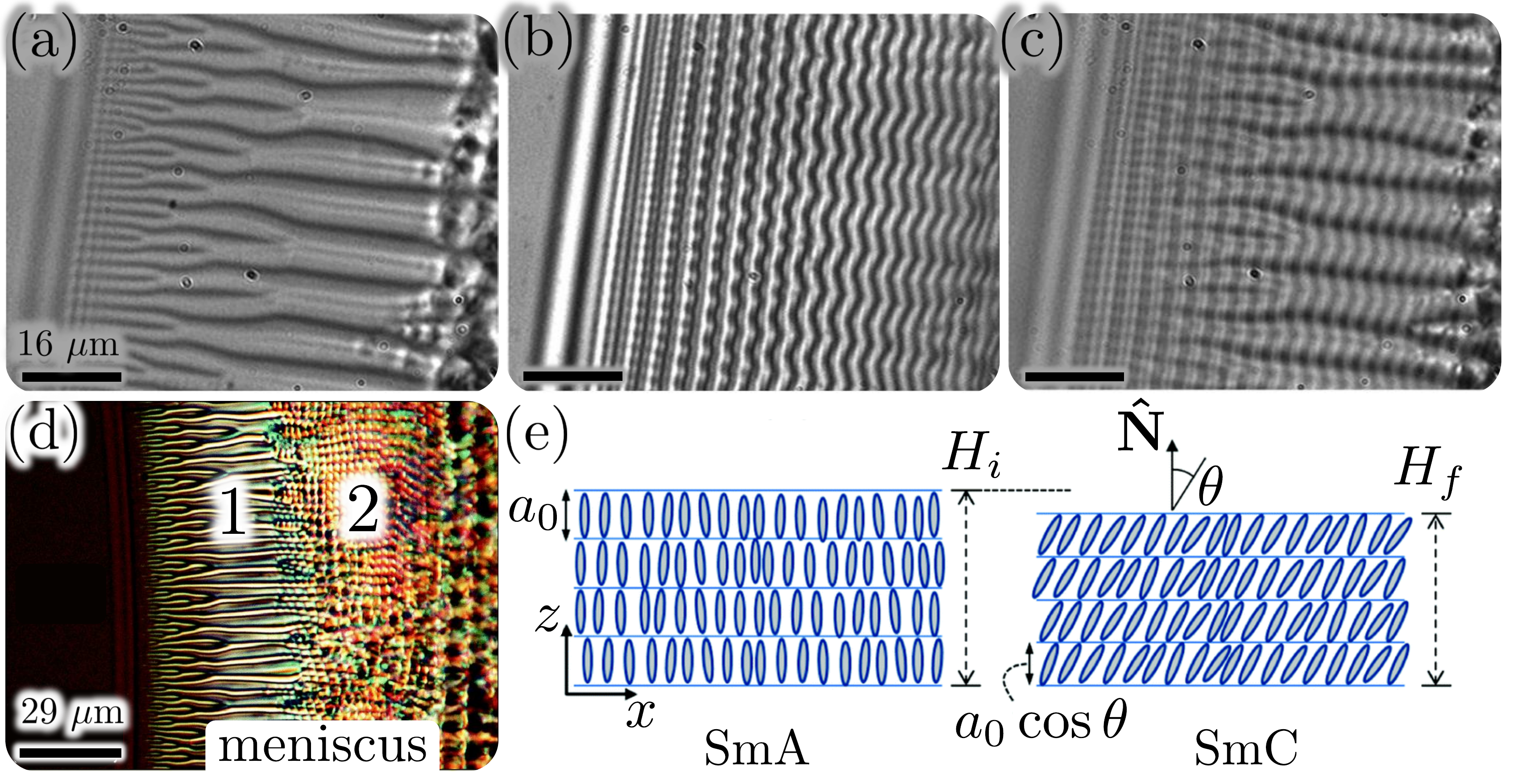}
\caption{\label{ThinSmFilm} (a) Bright-field (transmitted light) optical micrographs of the meniscus of a smectic-C* film (compound SCE-9, $T = 25$ $^{\circ}$C). (b) Interferogram of the sample at the same location as in (a), obtained with phase shifting interferometry (PSI), reveals distorted interference fringes. (c) Superposition of the micrographs in (a) and (b), show how the $\mathbf{c}$-director splay distortions seen in bright-field correspond to the interface undulations captured by the interferogram. (d) Polarized micrograph of the smectic-C* meniscus reveals two regions: region 1 exhibits radial stripes, shown also in (a) and region 2 shows a two-dimensional structure of focal conic domains. (e) Schematic depicting the HH instability as a possible origin of smectic layer undulations from the smectic-A---smectic-C phase transition (left is before, right is after the transition). After the phase transition, the director tilts with an angle $\theta$, causing the natural layer spacing of $a_0$ to reduce to $a_0\cos{\theta}$. Although the phase transition causes a decrease in the natural layer thickness, the gradient in meniscus thickness fixes in a certain number of dislocations in the system, thereby fixing in a certain number of layers and the thickness. A mechanical stress analogous to a dilation of the smectic-C layers results from this incompatibility, triggering undulations to accommodate the director tilt.}
\end{figure}

A striking example is the work of Loudet \textit{et al.} on smectic-C* films.  Recall that unlike the smectic-A phase, a smectic-C phase has its nematic director canted at a non-zero angle with the layer normal and the projection of the director onto the plane of the layers is referred to as the \textbf{c} director.  Finally, a smectic-C* phase has the same geometry as smectic-C on each layer but, because of intrinsic chirality, the \textbf{c} director rotates from layer-to-layer; like the cholesteric pitch, the period of the \textbf{c} rotation is typically much longer than the period of the smectic layers and will not alter the ensuing discussion \cite{loudet-2017}. A  geometry mismatch occurs at the smectic-A to smectic-C transition where the molecules tilt relative to the smectic layer normal, decreasing the layer thickness.  Indeed in thin films of the smectic-C phase, the meniscus exhibited stripes that appeared to correlate to the interface shape, and Loudet \textit{et al.} hypothesized these structures to be the result of the HH instability \cite{sm-C-free-interf}. A bright-field image in Fig.~\ref{ThinSmFilm}(a) shows the meniscus of a compound in the smectic-C* phase (SCE-9, from Merck, England, at 25$^{\circ}$C).  Here, the stripes are attributed to splay deformations of the \textbf{c} director, induced by frustration from the surface \cite{meyer-pershan}. Note that particles within smectic-C thin films are also found to induce similar structures, due to thickness gradients created by wetting of the inclusions \cite{colloid,harth-stann-2009,stebesmectic}.

Another effect is saddle-splay, though saddle-splay is an oft-neglected term in the Frank free energy because it is a total derivative.  However, when topological defects form they provide boundaries inside the sample.  
Classic studies of the saddle-splay term in nematics use hybrid-anchored nematic thin films with homeotropic and degenerate planar conditions on the two film surfaces. In this case, the saddle-splay contributes to a stripe instability \cite{han-str-1,hybridK24}.  We would expect analogous contributions at, say, the interface of a cholesteric, if the interface prefers a different orientation of the layers than the bulk.  
The saddle-splay contribution in the case of uniformly spaced  smectic layers is proportional to the Gaussian curvature of the layers and, according to the Gauss-Bonnet theorem, becomes a purely topological contribution. As such, we do not expect it to be pertinent for small undulation instabilities \cite{ishikawa_defects_2002}, but it certainly contributes when the layered system develops caustic-like cusps and folds \cite{didonna}. The saddle-splay term also plays a role if the nematic order is distorted at a fluid interface. For instance, in the case of thin nematic films with deformable boundaries, the saddle-splay is also involved in the onset of stripe instabilities, which have been the subject of some interest since the early 1990s and remains a topic of interest in the current millennium \cite{han-str-1,han-str-2,han-str-3,han-str-4,han-str-5}.

Finally, an incompatibility of the layer number and the thickness can also trigger undulations. The boundary condition may force the system to have an integer number of layers between the top and bottom of a film. This creates an intrinsic strain on the layers if the film thickness $d$ is not an integer multiple of the preferred layer size $t$. If the sample has a free surface, the surface itself will undulate and the surface tension $\sigma$ will play a role in determining the onset of the instability, as shown for the 1995 study by Williams of a block copolymer system \cite{dbcp-3}. Layer strain induced by the incompatibility of the system thickness with the number of layers has also been simulated in cholesterics \cite{machon-njp}. The induced corrugations on the interface from undulation instabilities are ubiquitous across systems with periodic ground states.

In summary, the interface plays an essential role in undulation instabilities, as it provides a mechanism for applying strains to a layered system through anchoring conditions, surface tension, and boundary curvature, amongst other sources. The instability, in turn, typically modulates the shape of a free and deformable interface, introducing corrugations. These features may be understood by taking into account the basic elastic properties of the layered system (\textit{i.e.}, layer bending and compression), along with the anchoring energy and surface tension at the interface. In any individual case of undulation instabilities, the energetic contributions from the anchoring conditions, the surface tension, and the bulk elasticity must be accounted for. The complex interplay between these various contributions generates an astonishing number of variations on this theme of geometrical-frustration-induced, undulation instabilities.

\section{Helfrich-Hurault: here, there, and everywhere}
 
As seen in the systems we have reviewed thus far, undulation instabilities in smectics and cholesterics are induced by geometrical frustration, with important and often neglected contributions from deformable boundary conditions, interfacial curvature, and surface anchoring conditions. However, similar responses to bulk and surface incompatibilities are also prevalent in other materials with periodic ground states. These same mechanisms can be extended to account for phenomena seen in both biological and other synthetic systems. In the final section of this review, we briefly discuss undulation instabilities across a wide array of materials to demonstrate the ubiquity and utility of the HH mechanism, beyond the traditional smectic and cholesteric phases.

\subsection{Twist-bend nematic phases}

\begin{figure}[ht]
\includegraphics[width=0.45\textwidth]{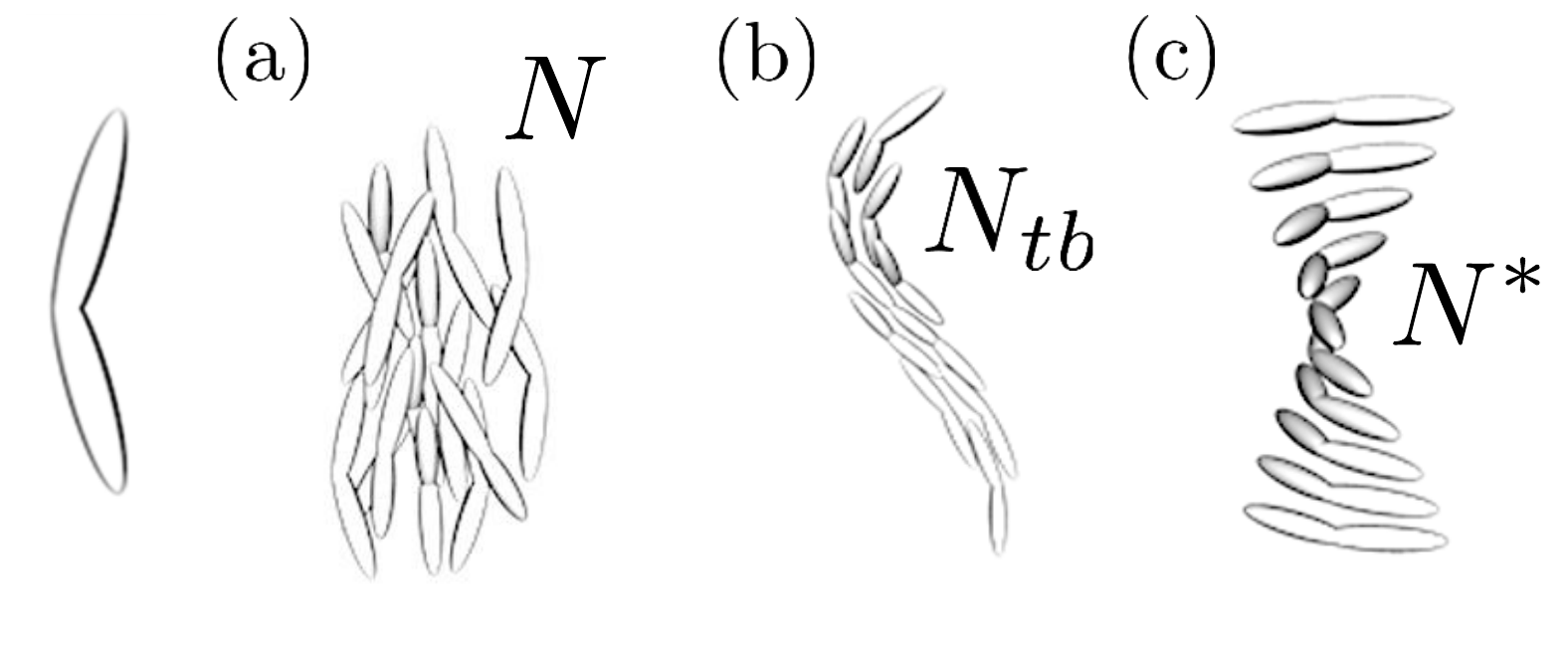}
\caption{\label{Ntb-Schematic} Schematics of bent-core molecules (left-most) forming (a) a nematic, uniaxial liquid crystal ($N$), (b) a twist-bend nematic with an oblique helicoid ($N_{tb}$), and (c) a cholesteric (chiral nematic, $N^*$) liquid crystal with a right helicoid. The pitch for $N_{tb}$ phases is typically on the order of 10 nm. [Reproduced from \cite{mandle2016dependency}.]}
\end{figure}

Liquid crystals phases formed by banana-shaped, bent-core mesogens undergo the HH instability through undulation of their structures in response to mechanical stress, such as applied electric and magnetic fields or a reduction in layer-spacing with decreasing temperature. Depending upon their rigidity and the presence of system chirality, bent-core molecules can form over 50 types of liquid crystal phases, including a wide range of layered liquid crystals, including smectic and cholesteric phases (Fig.~\ref{Ntb-Schematic}) \cite{bentcore-rmp-jakli-etal}. Strains on the periodic structure of these smectic and cholesteric phases will undergo the HH instability, similar to systems discussed previously. However, banana-shaped molecules can also form a twist-bend nematic phase (Fig.~\ref{Ntb-Schematic}(b)), in which the director follows a helicoid at a constant oblique angle with respect to the helical axis, resulting in twist and bend deformations throughout the system. Twist-bend nematics have a nanoscopic, molecular-scaled pitch but can create periodic textures on the micron-scale, depending on the system thickness. We focus here on the HH instability exhibited in twist-bend nematics. 

\begin{figure}[ht]
\includegraphics[width=0.4\textwidth]{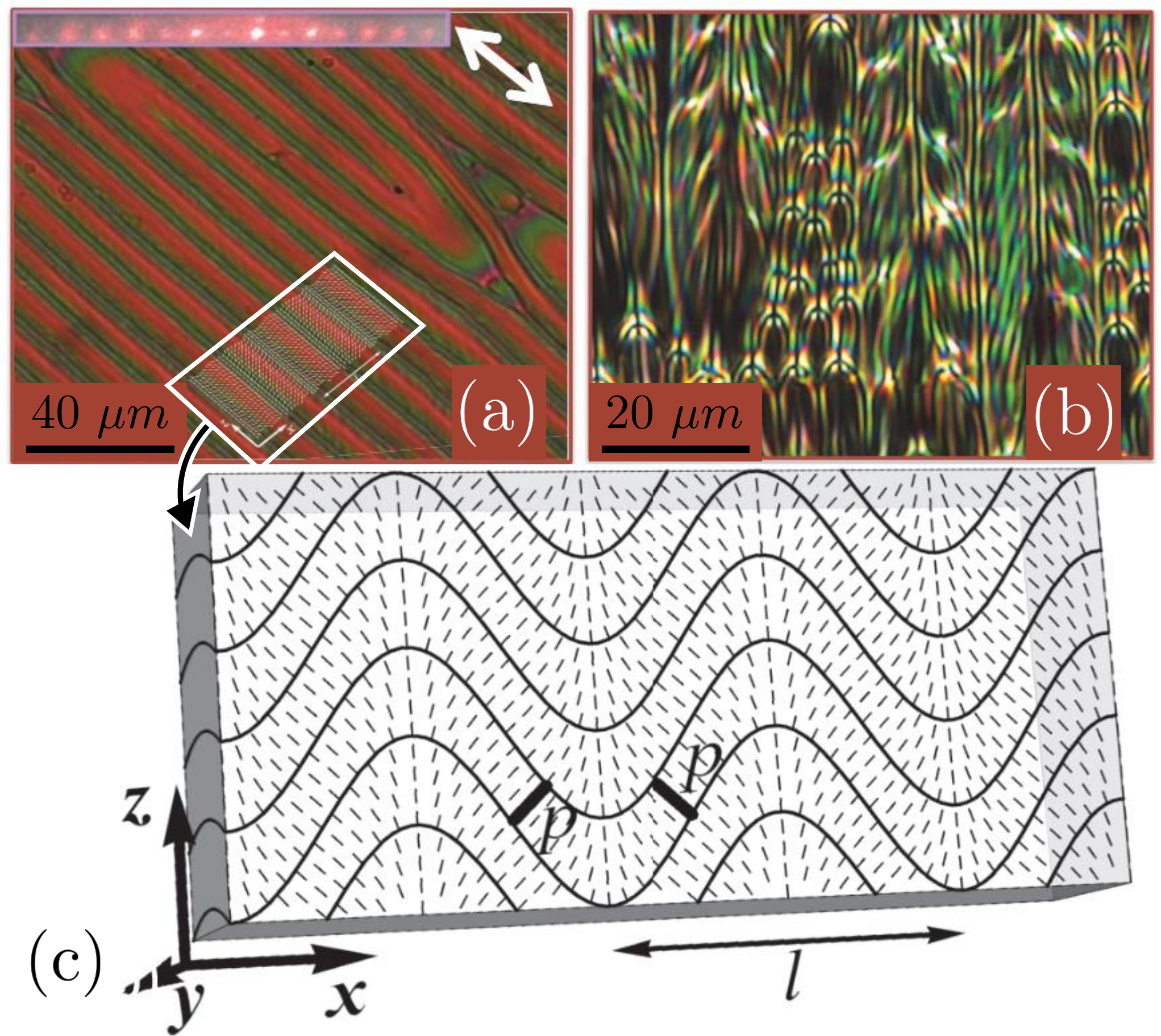}
\caption{\label{TwistBend} Micrographs and schematic of stripes formed by bent-core molecules in the nematic twist-bend phase, sandwiched between a 10-$\mu$m thick, planar cell. The micrographs in (a) and (b) are KA(0.2) and CB7CB, respectively. (a) KA(0.2) has stripes shown through crossed-polarized light microscopy, distinguishable also by the diffraction pattern in the top-most inset. The white arrow indicates the direction of rubbing. The bottom inset depicts the modulation of the helical axis of KA(0.2), made larger in (c). (b) The stripes in CB7CB are more complex, generating arrays of focal conic domains. The period of stripes in (a) and (b) are proportional to the cell thickness. (c) The thickness-dependent stripes in twist-bend phases are well-modeled by the HH model, illustrated in the schematic. The thickness of the pseudolayer, $p$, is the pitch of the conical helix. The direction of the heliconical axes (short lines) undulate in the $x$-direction, with a period $\ell$. [Reproduced from \cite{challa2014twist}.]}
\end{figure}

The model bent-core molecule first studied is 1'',7''-bis(4-cyanobiphenyl-4-yl)heptane (CB7CB). CB7CB within a glass cell treated for planar anchoring can form focal conic domains that are reminiscent of those observed in smectic phases, depicted in Fig.~\ref{TwistBend}(b). Friedel and Grandjean established that the presence of focal conic domains represents a phase with one-dimensional positional ordering \cite{friedel1910observations}. However, x-ray diffraction and deuterium magnetic resonance measurements of CB7CB reveals no density modulation, while suggesting some form of chirality in the system \cite{cestari2011phase}. These findings led Cestari \textit{et al.} to be the first to conclude that CB7CB is a twist-bend nematic. Similar to cholesterics, twist-bend nematics can form a pseudolayer structure defined by the pitch [Fig.~\ref{Ntb-Schematic}(b) and (c)]. 

Both CB7CB and KA(0.2) [another twist-bend nematic material, composed of 20 mol\% 1'',9''-bis(4-cyano-2'-fluorobiphenyl-4'-yl)nonane (CBF9CBF) added to a mixture of five odd-membered liquid crystal dimers with ether linkages containing substituted biphenyl mesogenic groups \cite{ka02}] can generate optically detectable stripes within planar glass cells [Fig.~\ref{TwistBend}(a)]. The stripe periodicity is micron-scaled, at least an order of magnitude larger than the measured pitch of the twist-bend nematic's conical helix. The stripe periodicity also depends on the system thickness, and the stripes are not thermodynamically stable. For samples with dielectric anisotropy both greater than and less than zero, the stripes could be eliminated by applying an electric or magnetic field. Only upon decreasing the temperature of the system afterwards would the stripes return \cite{odl-twistbend-2013,challa2014twist}. That this periodicity is larger than the phase's intrinsic periodicity and that the stripes are not thermodynamically stable are all properties reminiscent of the HH instability in smectics and cholesterics. 

Stripes and focal conic domains dependent upon system thickness or process history are signatures of the HH instability, as exemplified by smectic and cholesteric shells and thin films. Challa \textit{et al.} use a ``coarse-grain'' model of twist-bend phases to describe the optical stripes seen for both CB7CB and KA(0.2). The framework of the HH instability is then applied (Fig.~\ref{TwistBend}(c)) to capture the critical magnetic field strength necessary for stripe elimination, and to estimate the elastic properties of CB7CB and KA(0.2) \cite{challa2014twist}. Notably, the undulations in twist-bend phases are hypothesized to be created by the shrinking of the pseudolayers from decreasing the system temperature, reminiscent of the stripe formation in smectic-C menisci.

Lastly, we note that twist-bend nematics are also the first example of a fluid with local polar order without density modulation, and measurements on structures generated by the HH mechanism confirm this. Pardaev \textit{et al.} performed light scattering on a twist-bend nematic sample exhibiting parabolic focal conic domains that nucleated from the HH instability to detect the existence of this local polar order, evidenced by a second harmonic signal that is absent in the parabolic focal conic domains of smectic-A phases \cite{pardaev2016polNtb}. Again, structures attributed to the HH instability since the 1970s, such as parabolic focal conic domains in smectics and cholesterics, are being found in recent phases, like the twist-bend nematic phase, illustrating the pervasiveness and relevance of this mechanism in ordered systems. 

\subsection{Lyotropic liquid crystals}

\begin{figure} [h]  
\includegraphics[height=2.4in]{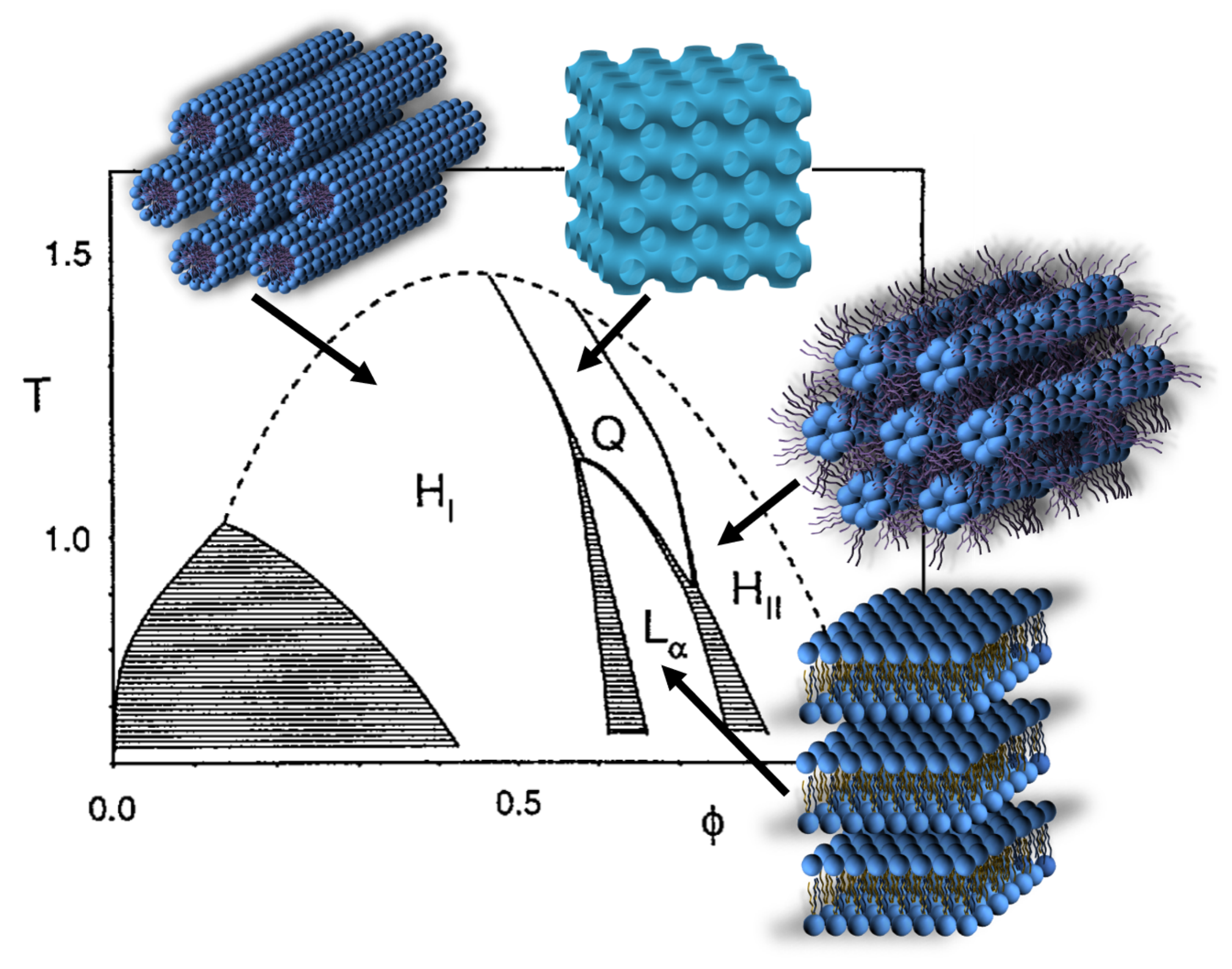}
\caption{Mean-field phase diagram, adapted from \cite{landaulyo}, of amphiphillic molecules in solution. The various ordered phases are indicated. The dark regions  have two-phase coexistence. We will particularly focus on undulational instabilities in the $L_{\alpha}$ lamellar phase and in the hexagonal columnar phase $H_I$.               \label{fig:lyophase}}
\end{figure}

A significant class of materials that also exhibits spatially modulated phases including cylindrical, layered, and foam-like configurations are lyotropic liquid crystals, which are typically collections of amphiphilic molecules in a solvent. The mixtures can involve multiple components, but typically include surfactant molecules and a solvent mixture that may contain salts or organic compounds, such as cyclohexane or alcohols. The thermodynamic phase of these materials is controlled by the concentration of the solute (typically the surfactant molecule), along with the temperature. An exemplary phase diagram is shown in Fig.~\ref{fig:lyophase}. Note that at sufficiently low temperatures $T$, we transition between a series of various ordered phases as we increase the concentration $\phi$ of the amphiphile in solution. The typical sequence of phases starts with a dilute micellar solution at low concentrations, transitioning to a hexagonal arrangement of micellar cylinders at higher concentrations, then to a lamellar arrangement (or a bicontinuous phase $Q$  as shown in Fig.~\ref{fig:lyophase}), until finally transitioning to an inverted micellar cylinder phase at high concentrations. All of these ordered phases are spatially modulated structures with some characteristic length of spacing $\lambda$. As such,  frustration imposed on the system that competes with the spacing $\lambda$ may lead to undulation instabilities.

The aqueous nature of lyotropics allows one to strain the system in a myriad of ways, including via shear flows and doping with nanoparticles, which may in turn be controlled with electric or magnetic fields. Many of these perturbations result in the classic HH instability because the  lamellar phase ($L_{\alpha}$ in Fig.~\ref{fig:lyophase}) is for all intents and purposes equivalent to the layered smectics and cholesterics described previously in this review. For the lyotropic lamellae, shear flow  may be applied to induce layer undulations \cite{lyoshear,lyoshear2}.   At small shear rates, the buckling instability may be directly related to an undulation produced by a dilative strain, with a characteristic length given by $\lambda_c \sim \sqrt{\lambda \ell}$, with $\lambda$ being the lamellar spacing and $\ell$ being the sample thickness \cite{hh-surfactants}.  It is also possible to induce an instability in these smectic-like states via confinement that is incompatible with a particular number of layers, which then reduces the problem to essentially an identical analysis as a smectic liquid crystal in a cell \cite{lyomackintosh}.

\begin{figure} [h]  
\includegraphics[height=1.7in]{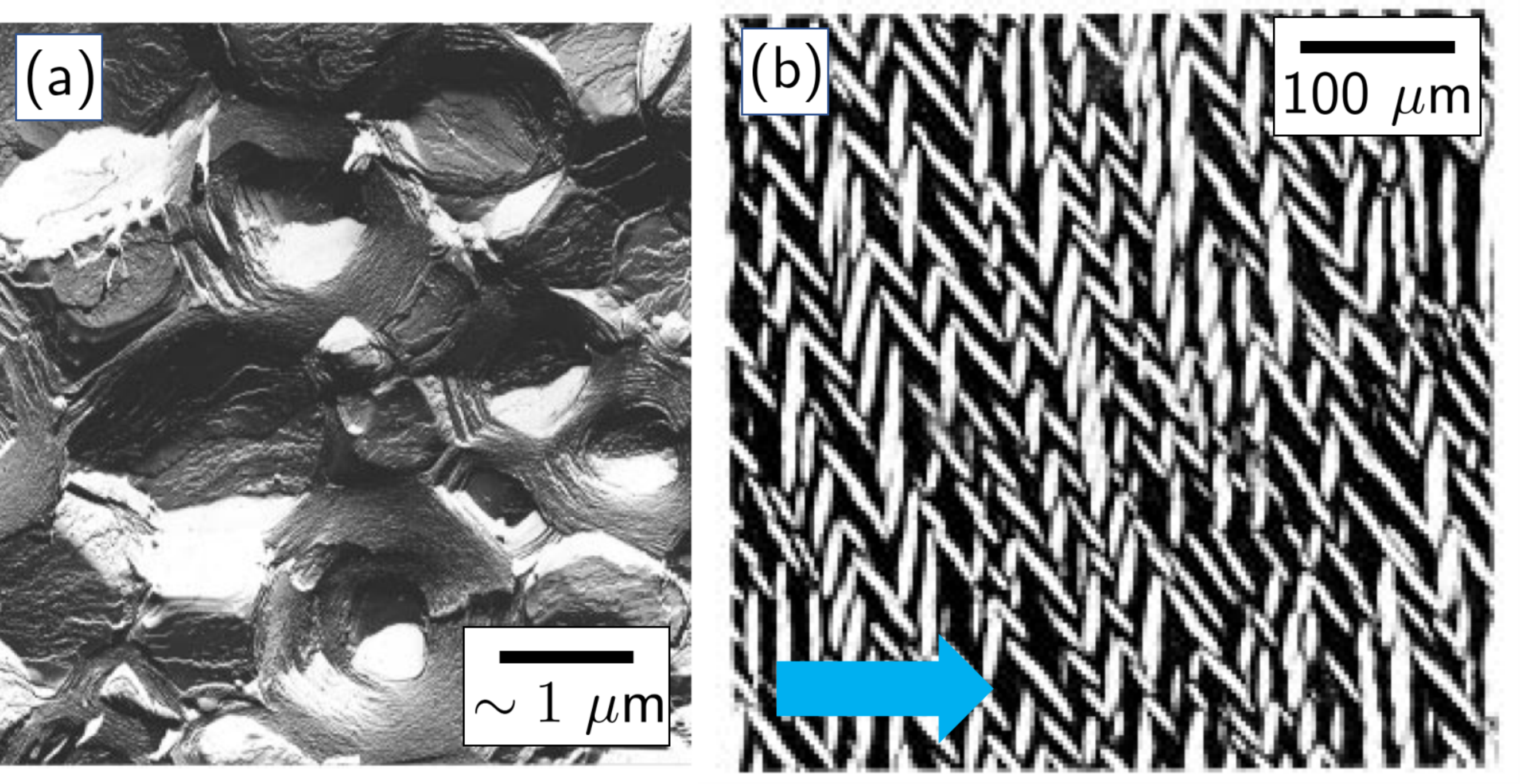}
\caption{ (a) Freeze  fracture electron microscopy section of a lyotropic lamellar phase after an applied shear, taken from \cite{lyoshear3}.  The lamella turn into a dense packing of multilamellar ``vesicles''. (b) Polarized microscopy image, taken from \cite{maglyo}, of a lyotropic columnar phase ($H_I$ in Fig.~\ref{fig:lyophase}) undergoing an instability to a herringbone pattern.  This is achieved by doping the material with magnetic nanoparticles and applying a field $\mathbf{B}$ (blue arrow) which acts to reorient the cylinders.  \label{fig:lyounstable}}
\end{figure}

 Under larger flows, the lamellar phase may break up into a packing of multilamellar vesicles \cite{lyovesicle,lyoshear2,lyoshear3} or analogs of focal conic domains \cite{lyoFCD}. An example of the resultant structure is shown in Fig.~\ref{fig:lyounstable}(a).  Under these more extreme shear conditions, interesting intermediate phases may also form, including a phase in which multilamellar cylinders orient along the shear direction \cite{multicylinder}.  These multilamellar cylindrical structures may, in turn, also exhibit undulatory instabilities via, for example, the alteration of the spacing between lamellae or an induced spontaneous curvature \cite{bundle2}.

The cylindrical phases ($H_I$ and $H_{II}$ in Fig.~\ref{fig:lyophase}) also have interesting ground states that can undergo HH-like instabilities. The characteristic size $\lambda$ between adjacent pairs of cylinders may be frustrated by an applied strain or cylinder reorientations under flow or applied fields. The cylinders may accommodate these strains by undulating or buckling. It is also possible to induce undulatory instabilities in the cylindrical phases by, for example, doping the phase with magnetic particles and then reorienting the phases with an applied magnetic field. At high fields, a herring-bone structure is observed as shown in Fig.~\ref{fig:lyounstable}(b), reminiscent  of the herringbone structures we see in smectic shells, described in Sec.~\ref{SmecticShells}. 

Given the multi-component mixtures involved in forming the lyotropic phases and the complex set of interactions in forming the ground states with an associated characteristic length $\lambda$, it is difficult to model these systems without resorting to a phenomenological description. One possibility is to use molecular dynamics simulations. However, even simple, single lipid bilayers present challenges, even with the rapid advance of computational tools \cite{lipidMD}. To our knowledge, there are no existing detailed, microscopic models of these HH-like instabilities in lyotropic materials.

\subsection{Diblock copolymers \& polymer bundles}

Block copolymers also show ordered phases similar to lyotropic liquid crystals (Fig.~\ref{fig:lyophase}). However, unlike lyotropics, block copolymers typically have a fixed density. Therefore, tuning between different ordered phases is achieved by changing the structure of the constituent polymers themselves, instead of varying the concentrations of system components, as is typically done for lyotropic systems. In this section, we focus primarily on \textit{di}-block copolymers, where two polymers of $A$ and $B$-type monomers are grafted together.   

\begin{figure} [h]  
\includegraphics[height=2.4in]{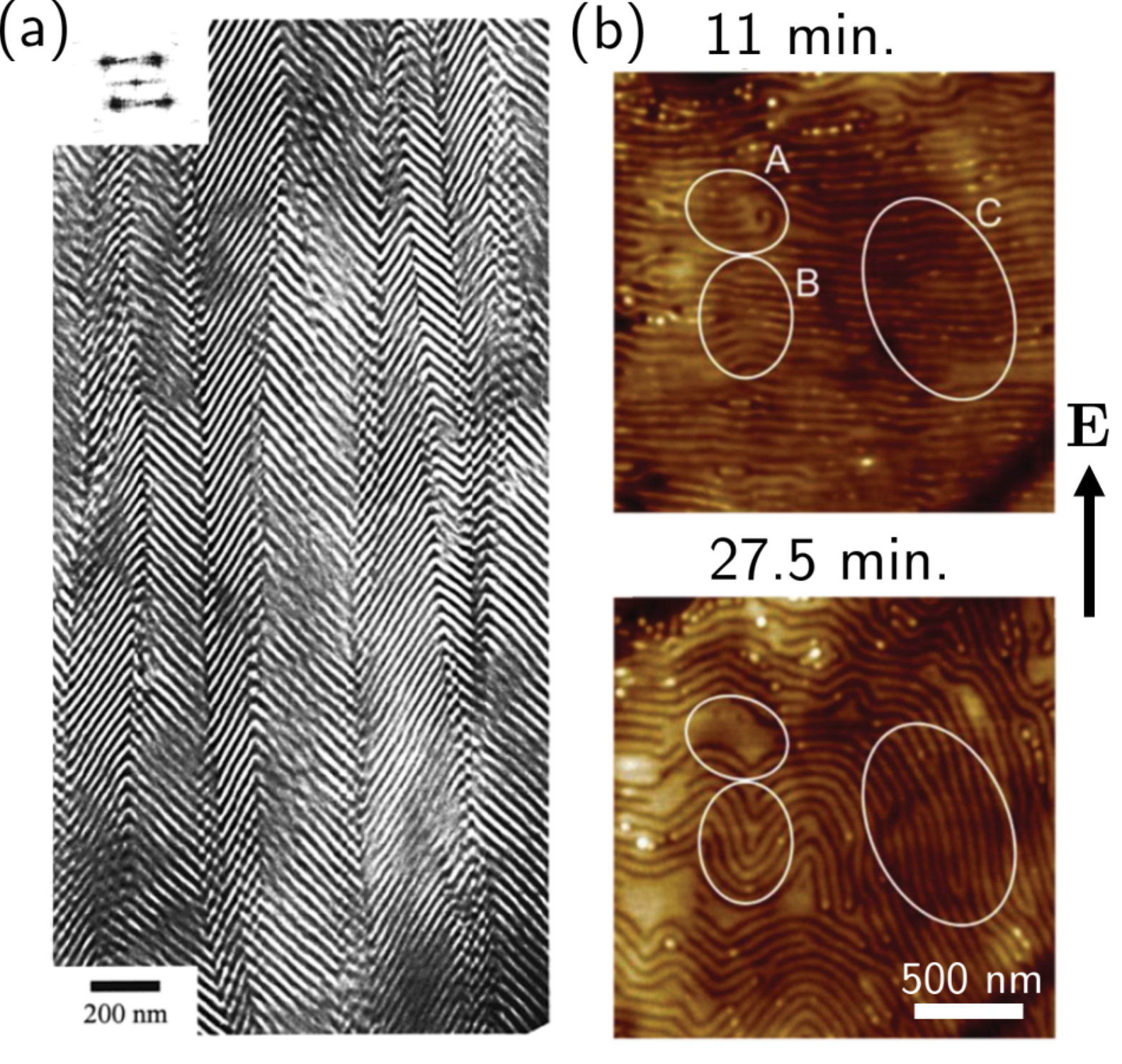}
\caption{(a) A TEM micrograph of a diblock copolymer under a large uniaxial strain (300\%), exhibiting a characteristic herringbone structure.  This structure may develop from an HH instability. Results are taken from \cite{dbcp-2}. (b) Two snapshots of a diblock copolymer lamellar phase shown after the indicated number of minutes under an applied electric field $\mathbf{E}$. The lamellae are initially oriented horizontally, and the applied field reorients the layers. Three regions are indicated, with different reorientation behaviors, including an undulation-like mode, but with a sharp kink that is on the order of the lamellar spacing. Results are taken from \cite{boker}. \label{fig:copolymer}}
\end{figure}

The two $A$ and $B$ segments of the  copolymer typically have some incompatibility, which is captured via a Flory-Huggins term in the free energy: $\chi \int \psi_A(\mathbf{r}) \psi_B(\mathbf{r})\,\mathrm{d}\mathbf{r}$, where $\psi_{A,B}$ are the local volume fractions of $A$ and $B$-type monomers [taken to satisfy $\psi_A(\mathbf{r})+\psi_B(\mathbf{r})=1$].   A self-consistent mean-field analysis of the total free energy does a reasonable job in predicting the observed phases of these materials, which include a lamellar phase, a phase with hexagonally-packed cylinders, and gyroid phases, amongst others \cite{matsenschick,batesfredrickson}. In the weak segregation limit, where the $A$ and $B$ portions only weakly demix, the system is effectively described by a Brazovskii-type free energy \cite{brazovskii,fredricksonhelfand}:
\begin{align}
f_{\mathrm{cp}}&=\frac{1}{2}\int [r+(q-q^*)^2][\psi(\mathbf{q})]^2\mathrm{d}\mathbf{q}+ \frac{\mu}{6}\int [\psi(\mathbf{x})]^3\,\mathrm{d}\mathbf{x}  \nonumber \\
& \qquad {}+\frac{u}{24}\int [\psi(\mathbf{x})]^4\,\mathrm{d}\mathbf{x}, \label{eq:brazo}
\end{align}  
where $\psi(\mathbf{x})$ (and its Fourier transform $\psi(\mathbf{q})$) describes the deviation of the relative $A/B$ monomer density from the well-mixed, disordered phase. The unstable mode $q^*$ is related to the wavelength $\lambda^*$ of the $AB$ domains via $q^* = 2\pi /\lambda^*$. In the strong segregation limit, the $(q-q^*)^2$ term has to be replaced with an appropriate interaction term that couples the Fourier transformed fields $\psi(\mathbf{q})$ at different modes $\mathbf{q}$  \cite{ohtakawasaki}.

The phases of block copolymers are analogous to the lyotropic phases, highlighted in Fig.~\ref{fig:lyophase}. Although the lyotropics typically have more dilute phases, such as a suspension of spherical vesicles, these phases are not achievable in a block copolymer. The major difference between the two systems is that the amphiphile concentration $\phi$ is replaced with the relative density of the $A$ and $B$ monomers $\psi$, which always has some molecular variation due to the block copolymer molecular structure. We focus primarily on the lamellar and cylindrical phases to examine the HH instability in diblock copolymers. For these phases, perturbations of the system away from the ground state can be examined in both the weak- and strong-segregation limits.

The lamellar phases of diblock copolymers exhibit the same undulatory instabilities as discussed for the other lamellar phases. Uniaxial strain applied perpendicular to the lamellae leads to the HH instability, similar to the classic smectic and cholesteric systems \cite{dbcp-1}. At large strains, the copolymer can develop a ``herringbone'' structure, reminiscent of those observed in the smectic shells of Sec.\ref{secSm-Secondary} \cite{dbcp-2}. An important difference, however, is that the phases of diblock copolymers depend on an interaction parameter $\chi$ and can exhibit a strong segregation regime, when the $A$ and $B$ portions of the copolymer are highly repulsive, or a weak segregation regime, when $\chi$ is small. Yet, it is possible to perform a perturbative analysis to examine the HH instability in both regimes. In the strong segregation limit, the approach is the same as for magnetic systems, which we detail later in this section \cite{magsmectic1,magsmectic2}. In the weak segregation limit, a smectic-like free energy can  be derived by perturbing away from a uniform stripe phase $\psi(\mathbf{x})= A \cos(2\pi \hat{\mathbf{d}} \cdot \mathbf{x}/\lambda)$, with $\hat{\mathbf{d}}$ being the direction of the stripes/lamellae. The details of such an analysis are given in \cite{sabetta}. It is also possible to model the HH instability by simulating the relaxation of a system with the free energy in Eq.~\eqref{eq:brazo} under an appropriate perturbation.

For diblock copolymers, a possible perturbation that induces an undulation is a strain from an electric field applied normal to the lamellae. Since the lamellae prefer to lie along the field, the applied field rotates them. The resultant undulations may be phenomenologically described by a smectic-like free energy with an associated HH-like instability \cite{diblockonuki}. We note that such a phenomenological theory has some deviations that are better captured by a self-consistent field theory treatment \cite{matsen1,matsen2}, where the basic prediction $\lambda^* \sim \sqrt{ \lambda \ell}$ holds \cite{matsen2} under certain conditions. However, it is also possible for the block copolymers to develop an instability at a wavelength that is close to the lamellar spacing itself ($\lambda^* \sim \lambda$). The undulations may also occur in two-dimensions, creating a square lattice of deformations that are reminiscent of parabolic focal conic domains, detailed in Sec.~\ref{History-Sm} \cite{dbcp-4,tsoriandelman}.   

In general, the layer reorientation mechanism of diblock copolymer systems under an applied field is complex, and there is a sustained interest in elucidating all of the possible regimes \cite{glasner}. One may observe some of the subtleties in Fig.~\ref{fig:copolymer}(b), where three different regions are identified in a single sample, under the same applied field. Despite the variety in the morphology of the instability, in all cases, we observe a frustration between some applied strain and the equal layer spacing of the ground state, as with the other systems considered in this review.

\begin{figure} [h]  
\includegraphics[height=1.5in]{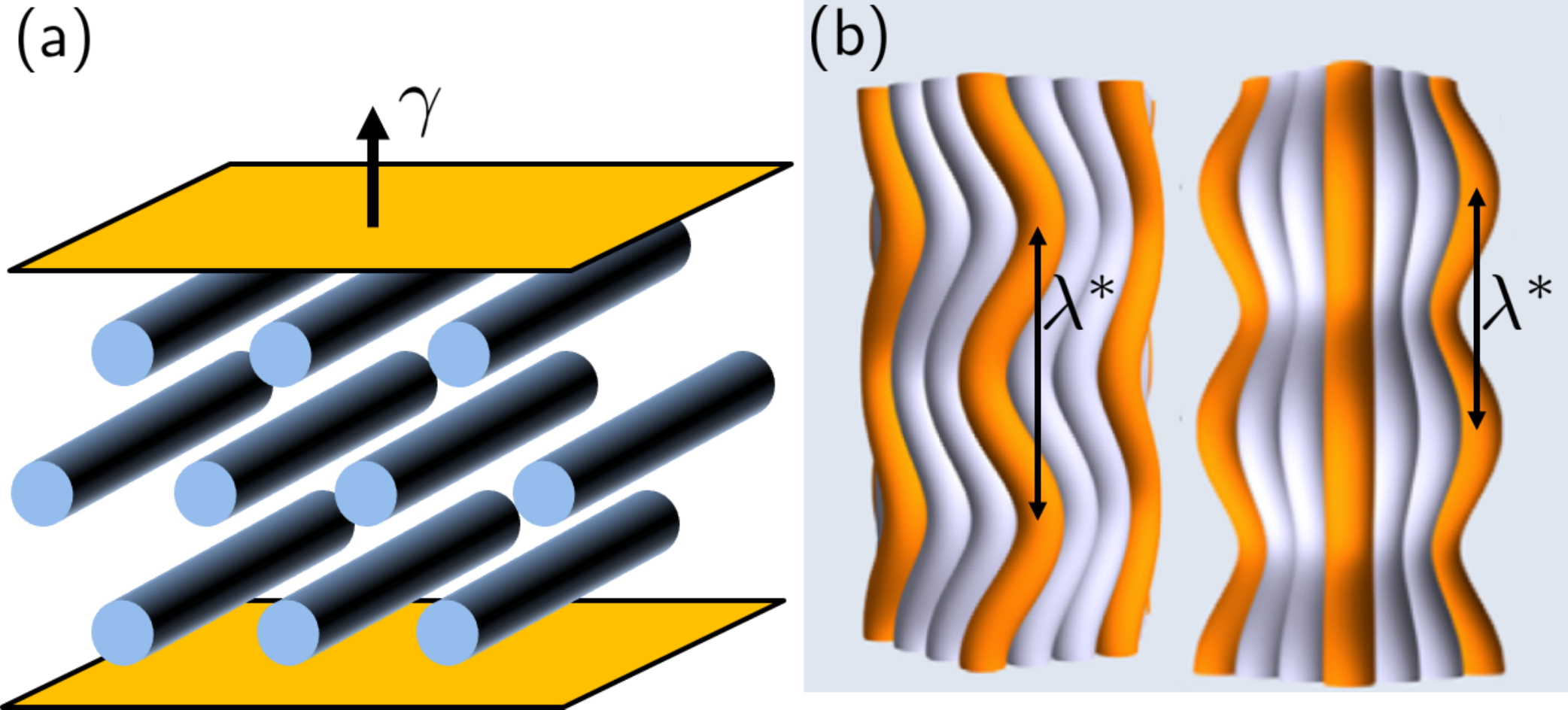}
\caption{(a) Uniaxial strain $\gamma$ applied perpendicular to the cylinders in a hexagonal phase of the diblock copolymer results in a HH undulatory instability of the cylinders, similar to the ones shown in (b) for  bundles of elastic fibers. The bundles exhibit undulation instabilities as in the columnar phases of diblock copolymer (and discotic liquid crystals). The buckling in the fiber bundles, with characteristic size $\lambda^*$, comes from geometrical frustration  resulting from the incompatibility of disclinations in the cylindrical packing and the equal cylinder spacing.  Figure is adapted from   \cite{bundle3}. \label{fig:columns}}
\end{figure}

The columnar, or hexagonal, phases of diblock copolymers also exhibit HH-like instabilities.  Applying a uniaxial strain perpendicular to the length of the cylinders may induce undulations as the cylinders try to maintain the same spacing under strain (Fig.~\ref{fig:columns}(a)). The resultant instability in the cylinders, illustrated schematically in Fig.~\ref{fig:columns}(b), may be analyzed in the same fashion as the lamellar system \cite{bundle1,bundle4}. 

A related instability is also found in bundles of elastic filaments \cite{bundle3}. There, the instability arises when one has a defect in the hexagonal packing of fibers. The packing defect, a disclination, is incompatible with the equal spacing of the cylinders in the packing. The cylinders then buckle to relieve this frustration \cite{bundle3}. Depending on the type of disclination, one can find various deformation modes, two of which are shown in Fig.~\ref{fig:columns}(b). This is yet another example where geometrical frustration leads to a spatial modulation -- the central theme of this review.

\subsection{Columnar liquid crystals}\label{SecColumnar}

The HH instability has also been postulated as the striation mechanism for columnar phases. Cagnon, Gharbia, \textit{et al.} were the first to observe an undulation instability in columns of a \textit{thermotropic}, discotic liquid crystal forming stripes under both compression and dilation of the system, reminiscent of the HH instability of smectics under dilation \cite{cagnon1984column,columnar-3}. They used the HH model to discover that the curvature elastic modulus of thermotropic, columnar liquid crystals can be six orders of magnitude larger than that of thermotropic smectics and nematics, possibly due to column entanglements. 

A decade after the work of Cagnon, Gharbia \textit{et al.}, Oswald, \textit{et al.} observed similar undulatory behavior in hexagonal, \textit{lyotropic} liquid crystals, with strain introduced by a directional growth apparatus, in which the sample, sandwiched between two glass plates, is pulled across a pair of hot and cold ovens \cite{oswald_nonlinear_1996}. Compared to the dilation experiments, the lyotropic system of Oswald \textit{et al.} underwent undulatory instabilities due to thermal effects, thereby experiencing mechanical stress in both vertical and in-plane directions. Their measurements and calculations further indicated that the columns in their system are not correlated at large distances. However, whether that conclusion can be drawn for \textit{thermotropic} systems remains unknown due to experimental difficulty in obtaining thermally induced striations in thermotropic, discotic liquid crystals. Furthermore, isolating the formation of stripes through macroscopic dilation of lyotropic systems is also challenging because of difficulties in mitigating water evaporation -- another major source of stripe instabilities.

\begin{figure}[ht]
\includegraphics[width=0.48\textwidth]{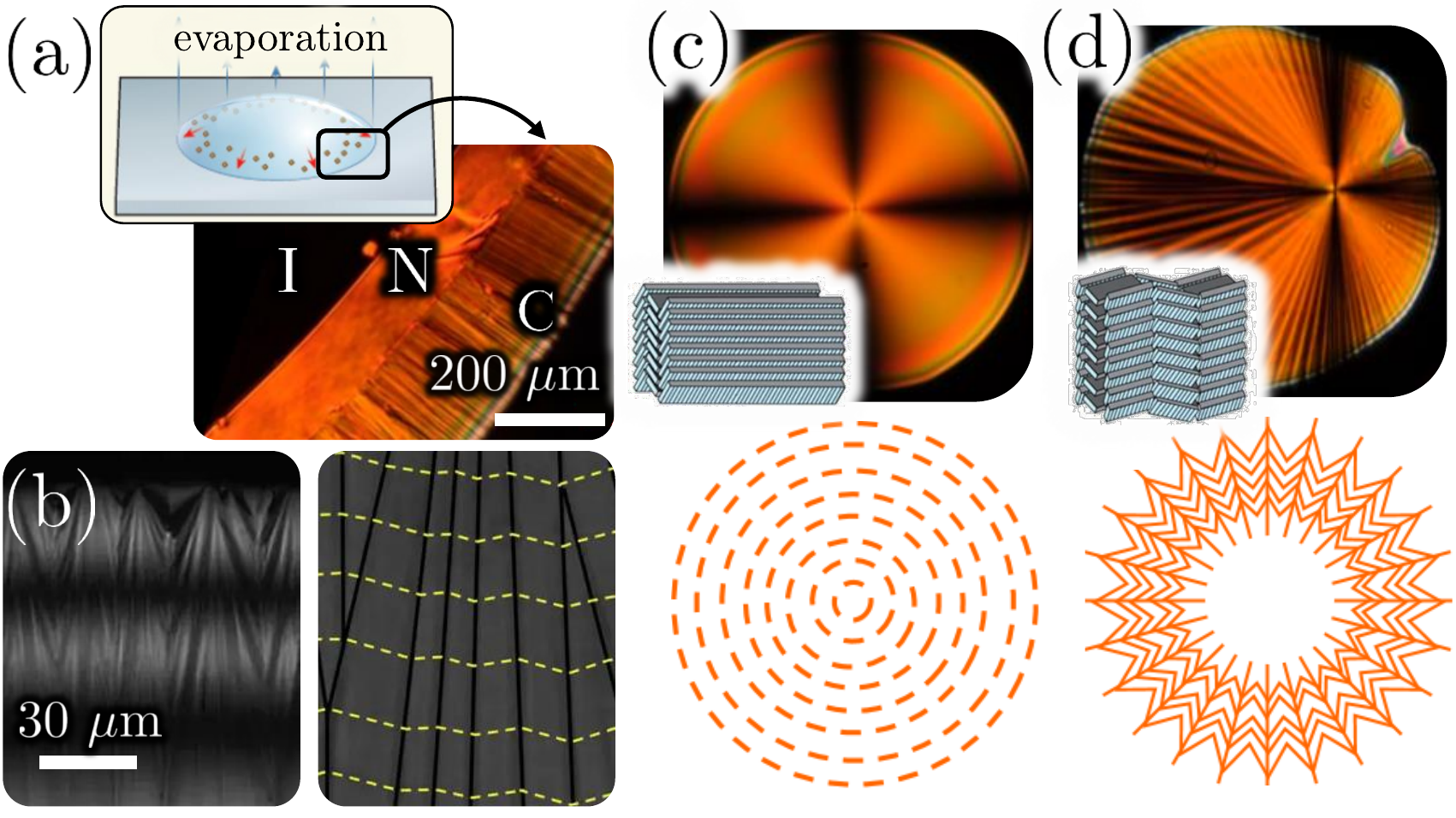}
\caption{\label{Chromonics-Coffee} (a) Evaporation of a particle suspension within a sessile droplet occurs more rapidly at the droplet edges, driving particles to the contact line, resulting in the ``coffee ring" effect. (Reproduced from \cite{larson2017twenty}.) The coffee ring effect for an aqueous solution of Sunset Yellow results in phase coexistence, with isotropic (I) near the center, then nematic (N), and finally columnar (C) when moving radially outward (bottom). (b) Domain walls are formed visible under polarizing microscopy (left), resulting from the buckling of columns (yellow line, right schematic). (Reproduced from \cite{davidson2017}.) (c) The coffee ring effect is slowed when the droplet is immersed in oil instead of air, resulting in the columns forming a neat nematic phase. (d) Further evaporation leads to a controlled herringbone texture from the buckling of columns (bottom schematics, reproduced from \cite{lydon2010chromonic}). Cross-polarized micrographs courtesy of Kunyun He.}
\end{figure} 

Water evaporation has been suggested as the source of undulatory instabilities for lyotropic systems in more recent experiments. Kaznatcheev \textit{et al.} studied a lyotropic liquid crystal that forms columns in the chromonic phase \cite{odl-chromonic}. Lyotropic chromonic mesophases are typically formed by plank-like molecules with aromatic cores surrounded by polar groups that can also form columns. In water, the molecules form charged columns by stacking face-to-face in order to hide their aromatic cores. Because the inter-disk association is through weak, noncovalent interactions, the assembled columns are polydisperse, with their average lengths dependent upon the molecular concentration of the disks, the disk ionic strength, the depletant concentration, and the temperature \cite{tortora2011chiral}. 

Kaznatcheev \textit{et al.} used a sulfonated benzo[de]benzo[4.5]imidazo[2,1-a]isoquinoline[7,1] dye as lyotropic, chromonic liquid crystal and observed stripes appearing after film deposition, exposed to air. The stripe direction was perpendicular to the column direction, indicating that the stripes resulted from buckling of the columns. They described the striations with a HH model, hypothesizing that the evaporation of water creates mechanical stress in the system by decreasing the separation between adjacent columns. The excess space caused by the evaporation must be filled by either new columns or by tilting the columns. Creating new columns would generate dislocations that then propagate throughout the system, which is energetically costly and slow. However, tilting of the columns could occur rapidly, so would then be more favorable, again reminiscent of the classic HH instability.

Investigating lyotropic, chromonic systems with gradients of concentration from water evaporation is desirable to better validate the HH model as the mechanism of stripe formation. The so-called ``coffee ring" effect achieves this, in which a sessile droplet of a particle suspension has an evaporation rate dependent upon the radial distance to the center of the droplet, with the highest evaporation rate at the droplet's contact line (Fig.~\ref{Chromonics-Coffee}(a), top). This evaporation gradient drives particles towards the droplet's outermost rim, subsequently generating a radial concentration gradient of particles. A sessile droplet of the lyotropic, chromonic dye, Sunset Yellow, undergoing the coffee ring effect exhibits a concentration gradient of the mesogen, resulting in the coexistence of phases within the droplet [Fig.~\ref{Chromonics-Coffee}(a), bottom]. The columnar phase near the contact line has radially-aligned stripes. Davidson \textit{et al.} measured the light adsorption due to linear dichroism, revealing that the average director orientation is parallel to the contact line, drawn in Fig.~\ref{Chromonics-Coffee}(b), right \cite{davidson2017}. Domain walls of the columnar phase are also visible in Fig.~\ref{Chromonics-Coffee}(b), left, indicating the presence of undulations that bend the columns during the evaporation process.

The evaporation of water can be slowed by replacing the surrounding air with oil. This is accomplished by introducing a non-ionic triblock polymer surfactant with hydrophilic polyethylene oxide in the ends and a hydrophobic polypropylene oxide in the center, such as Pluronic 31R1. The surfactant aids the wetting of the aqueous Sunset Yellow droplet on glass within hexadecane. By slowing the water evaporation rate, a smooth nematic phase of the columns can be obtained, shown in Fig.~\ref{Chromonics-Coffee}(c). Further evaporation then generates a uniform, herringbone texture in the droplet [Fig.~\ref{Chromonics-Coffee}(d)]. As the solution is progressively concentrated, the existing columns are extended, creating a differential strain in the mesophase that results in undulations and buckling of the columns, illustrated at the bottom of Fig.~\ref{Chromonics-Coffee}(d). The fracturing of columns is likely a consequence of bend deformations being more energetically costly than discontinuities, suggesting stronger inter-columnar association at high dye concentrations.

\begin{figure}[ht]
\includegraphics[width=0.46\textwidth]{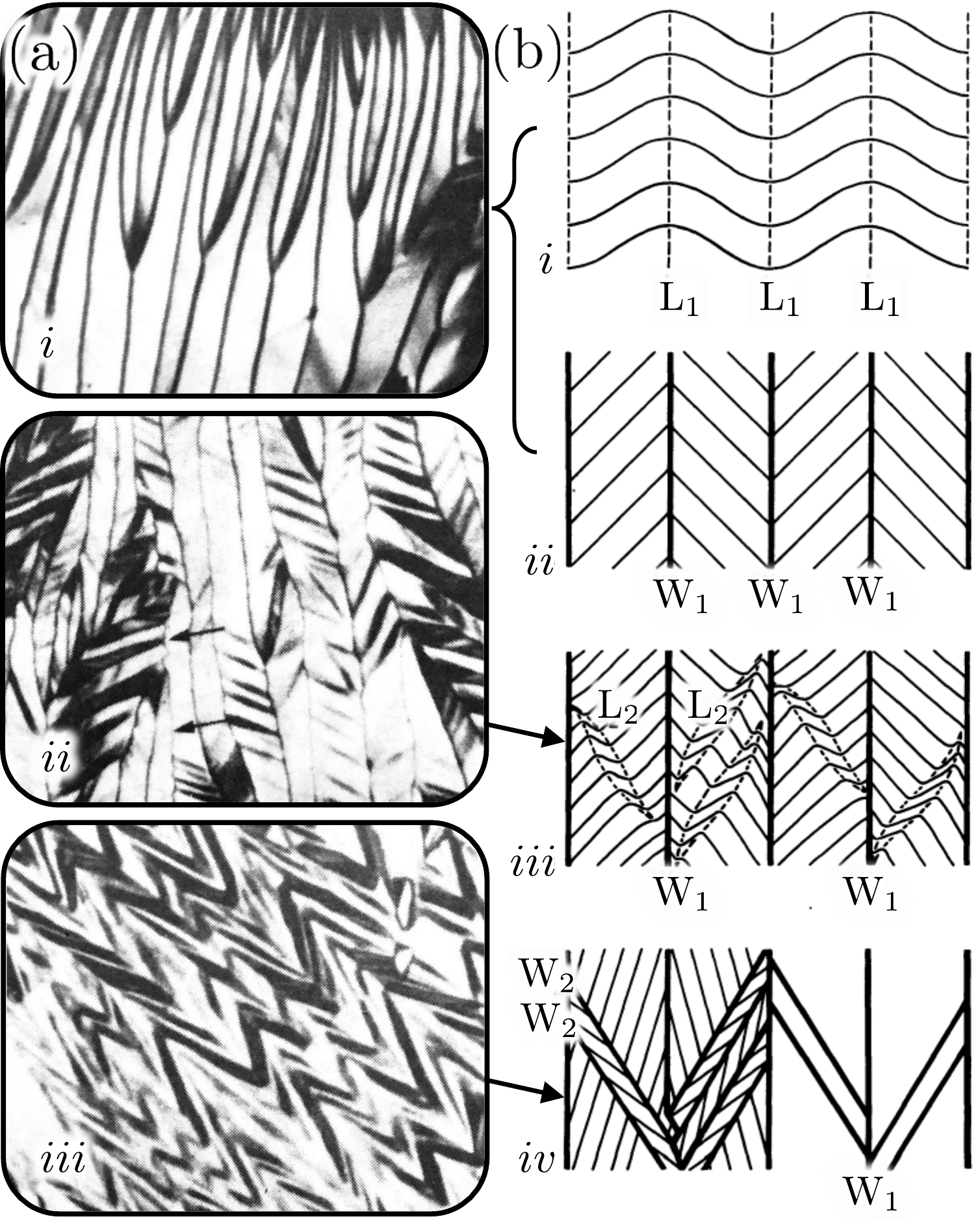}
\caption{\label{Chromonics-PBLG} (a) Textures of the hexagonal columnar phase of PBLG are shown in polarizing optical micrographs at 120$\times$ magnification. From (i) to (iii), the texture evolves from undulations to a herring-bone pattern. In (ii), regions of maximum curvature of the PGLA become walls of discontinuity, and new undulations appear within elongated domains. (b) Illustration of the textural transformation from an undulating pattern (i) to a herring-bone pattern (iv). Molecular orientations are represented by thin, continuous lines, walls of bend deformations (L) are indicated by dashed lines, and walls of discontinuity (W) are drawn as thick lines. Bend walls, L$_1$, transform into domain walls, W$_1$, as the molecular concentration increases. The process is repeated to form secondary domains, where bend walls L$_2$ transform into secondary discontinuities, W$_2$. Reproduced from \cite{livolant_liquid_1986}.}
\end{figure} 

The herringbone texture can also be found in phases of more complex molecules, like biological polymers, including DNA. The polarized optical micrographs of condensed xanthan, poly($\gamma$-benzyl-l-glutamate) (PBLG), and DNA have been investigated by Livolant and Bouligand, where the transition from undulations to a herring-bone pattern could be observed, shown for PBLG in Fig.~\ref{Chromonics-PBLG} \cite{livolant_liquid_1986}. The formation of secondary domains of periodicity within the herring-bone pattern as described by Livolant and Bouligand is evocative to the formation of secondary domains within smectic shells, detailed in Sec.~\ref{SmecticShells}. Condensed DNA also exhibits the herring bone texture \cite{livolant-dna-1}. Livolant \textit{et al}. confirmed with electron microscopy, and x-ray diffraction that highly concentrated, 50-nm DNA molecules have columnar longitudinal order and hexagonal lateral order, and can also form undulating patterns \cite{livolant-leforestier}. Recent studies of the evaporation of DNA suspensions, exemplified by \cite{smalyukh_structure_2006} and \cite{yoon-dna}, further produced DNA textures that should also be describable with the HH model. The HH instability is prevalent even in water-based liquid crystals.

\subsection{Biological materials}

\begin{figure}[ht]
\includegraphics[width=0.48\textwidth]{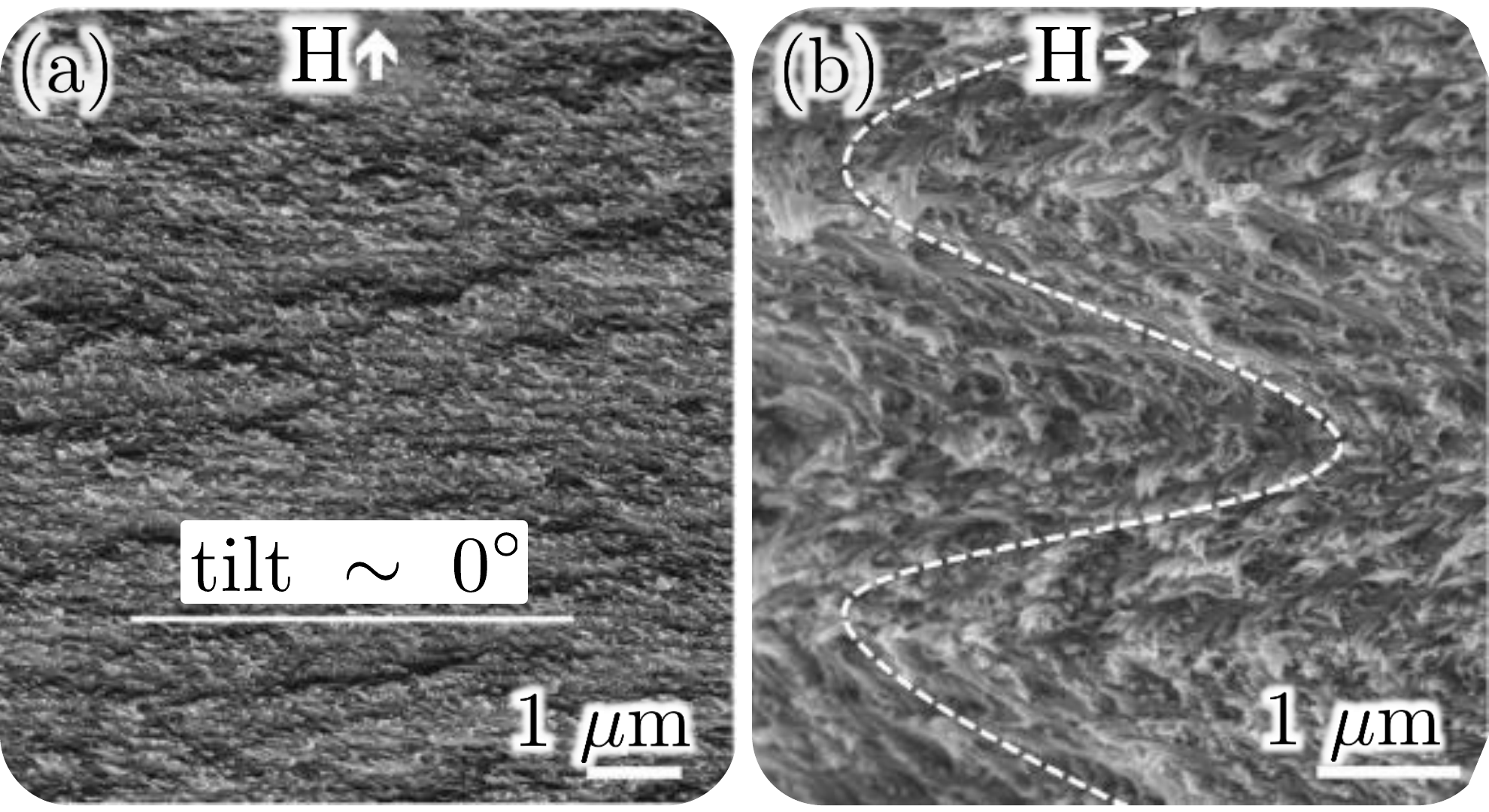}
\caption{\label{Bio-CNC} Scanning electron micrographs of cross sections from evaporated cellulose nanocrystal films. (a) Applying a vertical magnetic field (indicated by \textbf{H}) upon drying yields a single-domain, homogeneous cellulose nanocrystal film, with the pitch axis parallel to the magnetic field direction. (b) Applying a magnetic field to align the pitch axis horizontally generates a zig-zag pattern after evaporation, indicating buckling of the cholesteric pseudolayers. Reproduced from \cite{vignolini-cnc}.} 
\end{figure}

Undulation instabilities can also be seen in biological systems at intermediate length scales, such as within systems of particle-like fibrils, such as chitin, found in the exoskeletons of beetles and crustaceans, and cellulose, found within plants. Both chitin and cellulose, as with the majority of biological materials, have chiral building blocks. When concentrated beyond a threshold concentration, these biopolymers can form particles that self-assemble into colloidal, cholesteric liquid crystals \cite{bouligand, rey}. The cholesteric pseudolayer reorientation and formation of focal conic domains at a curved interface is seen on the surface of jeweled beetle shells due to the cholesteric ordering of the constituent chitin, providing a mechanism for their structural coloring and optical response \cite{beetle,rey}. Cellulose nanocrystals, derivable via acid hydrolysis from bacteria, cotton, wood, tunicate, and more, can also be concentrated to form a cholesteric phase \cite{lagerwall-cnc}. Colloidal suspensions of cellulose nanocrystals can be spread and evaporated to form a solid, dry film with photonic properties, forming a polydomain, cholesteric structure with a pitch in the visible wavelength range. Large magnetic fields can be used during evaporation to form a single domain, aligning the pitch along the direction of the magnetic field, as shown in Fig.~\ref{Bio-CNC} \cite{vignolini-cnc}. When Frka-Petesic \textit{et al.} applied a horizontal magnetic field during drying, aligning the cholesteric helix perpendicular to the plane of evaporation, they found a zig-zag pattern in the film [Fig.~\ref{Bio-CNC}(b)]. Although mechanical stress in the system is applied parallel to the layers in this case, the zig-zag pattern can also be thought of as a result of an HH-type mechanism. Evaporation during the processing of cellulose nanocrystal films could further introduce hydrodynamic stresses that can undulate and strain the cholesteric pseudolayers \cite{drying-cnc}.

\begin{figure}[ht]
\includegraphics[width=0.37\textwidth]{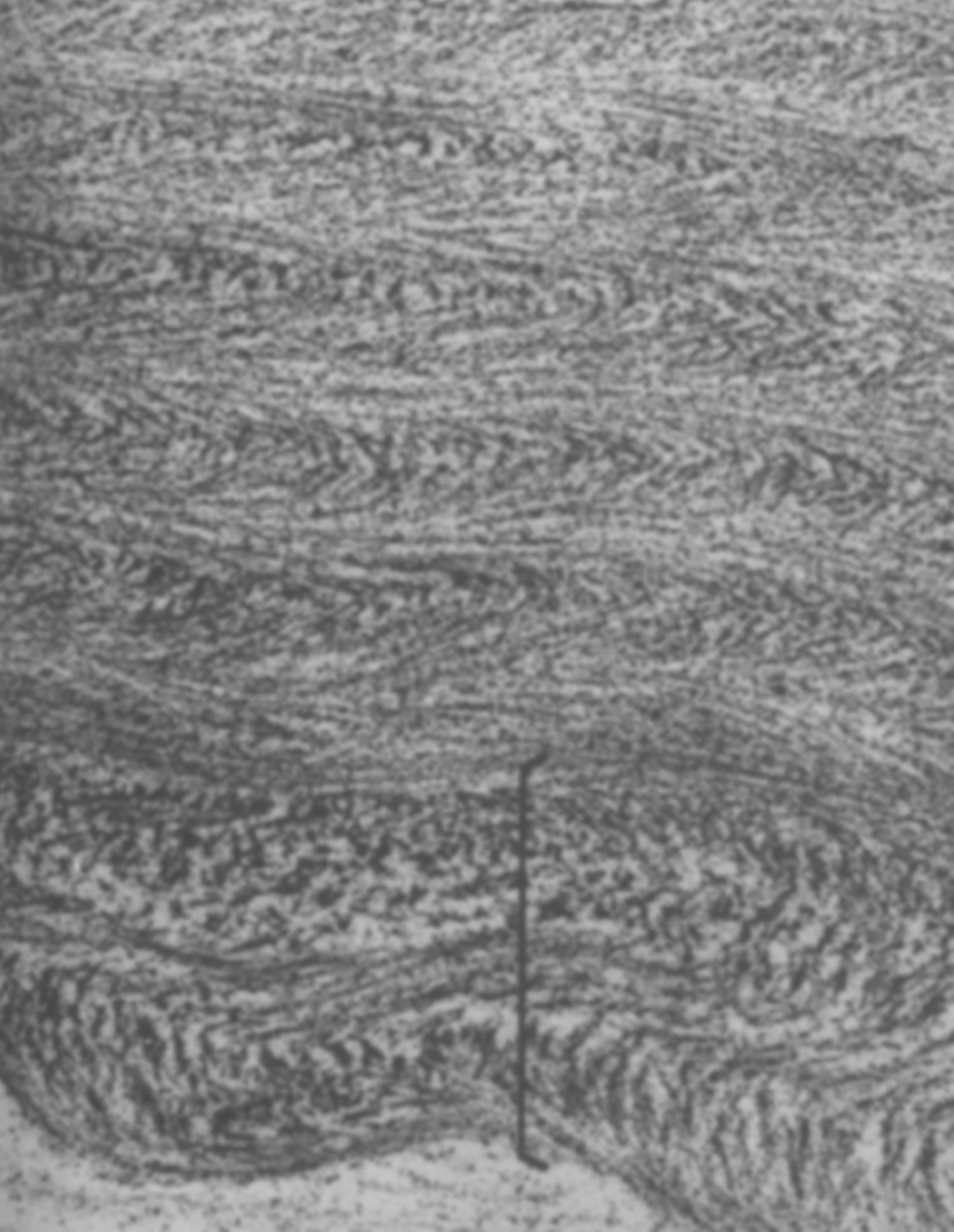}
\caption{\label{plantcellwall} Transmission electron micrograph of the elongating zone of mung bean seedlings (\textit{Vigna radiata}). The cholesteric pseudolayers of the cell wall, visible through the Bouligand arches of the cross section, undulate near the interface where growth of the cell wall takes place (bottom). Reproduced from \cite{plantcellwall}.}
\end{figure}

\begin{figure}[ht]
\includegraphics[width=0.37\textwidth]{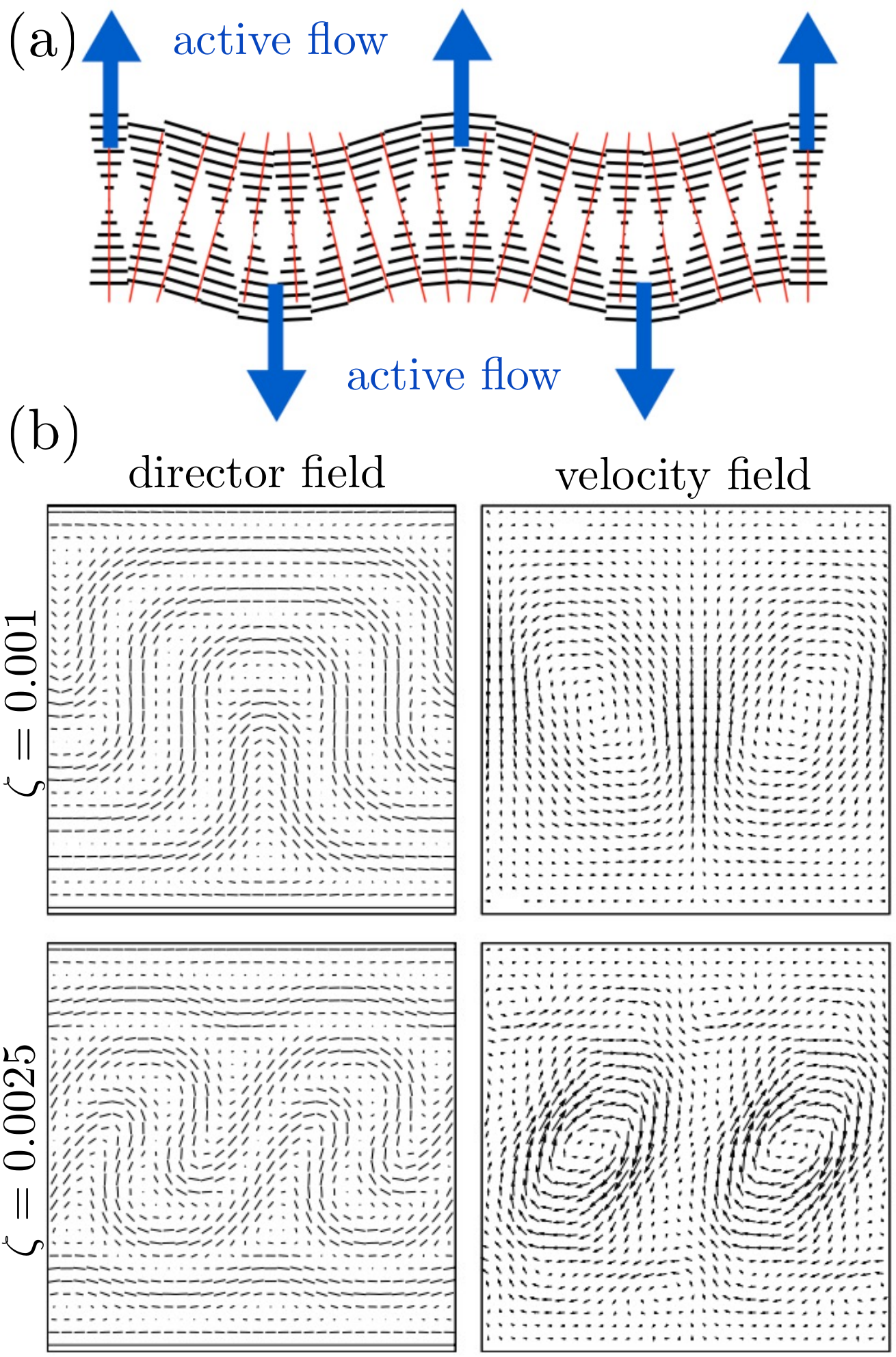}
\caption{\label{Bio-Active} (a) Sketch of the HH mechanism in extensile, active cholesterics. Black lines show the projection of the director field onto the plane, while red lines represent the splayed pitch axis. Blue arrows show the active flow direction, which increases the distortion and drives the undulation instability. (b) Simulation results for an extensile, active cholesteric confined in a quasi-two-dimensional geometry with flat walls. Homogeneous planar anchoring is set for both the top and bottom surfaces. The rightmost column plots projections of the director field onto the plane, and the leftmost column plots the corresponding velocity fields. $\zeta$ is proportional to the concentration of active particles, and is positive for extensile materials and negative in contractile ones. For both $\zeta = 0.001$ and $\zeta = 0.0025$, the profiles are steady states of the system. Reproduced from \cite{active-cholesteric}.}
\end{figure}

Generally, biological systems are not only often \textit{chiral}, but also \textit{active} and thereby out-of-equilibrium \cite{bouligand,plantcellwall,nacre,rey-lcmodelsbio,beetle,mitov-softmatter}. The development of the primary cell walls of plants is a striking example, shown in Fig.~\ref{plantcellwall}. Activity, including forces generated during growth processes, introduces hydrodynamic stresses that strain the chiral ordering of the system (Fig.~\ref{Bio-Active}(a)). Whitfield \textit{et al.} investigated cholesterics from the framework of active liquid crystals, integrating force-dipole stresses into a passive, chiral nematic formulation \cite{active-cholesteric}. In their work, Whitfield \textit{et al.} found that extensile stresses can trigger HH layer undulations in cholesterics. The steady state director fields and their corresponding velocity fields for varying extensile activity levels are plotted in Fig.~\ref{Bio-Active}(b). Both director fields exhibit pairs of $\lambda^{\pm}$ pitch defects, reminiscent of defects in cholesterics shells. Pairs of $\lambda^{\pm}$ defects often result from the HH instability in cholesterics, as detailed in Sec.~\ref{sec:Cholshell}. Kole, \textit{et al.} advanced this work by showing how active stresses in a cholesterics couple uniquely to the chirality of the material, generating elastic forces tangent to the layers \cite{kole-activeCLC}. This ``odder than odd'' elasticity from chiral activity leads to HH-like undulations that produce a two-dimensional array of hydrodynamic vortices. Whether passive or active, the HH mechanism is a valid mechanism of pattern formation in biological materials. 

\subsection{Magnetic systems \label{magsystems}}

Thin magnetic films present an interesting two-dimensional version of the HH instability. Such films can be fabricated from epitaxial garnet or a thin cobalt slice \cite{magsmectic3}. In certain cases, these films form magnetic domains in the form of stripes or hexagonal arrays of bubbles with a characteristic size $\lambda$, analogous to the smectic layer spacing or the spacing between cylinders in a hexagonal phase of a block copolymer. These domains form when  long-range dipolar magnetic interactions, which favor antiparallel alignment of magnetic spins, compete with the usual, short-range ferromagnetic interaction that tends to align neighboring spins. This is a typical scenario of short-range, attractive and long-range repulsive interactions necessary to form systems that exhibit modulated phases. The spatial modulations may then take the form of stripes, with properties analogous to smectic liquid crystals or block copolymers. As we have previously summarized, such modulated phases exist in a wide range of systems including phase-segregating lipids, block copolymers, and ferrofluids \cite{modulatedrev1,modulatedrev2}.

\begin{figure} [h]  
\includegraphics[height=2.3in]{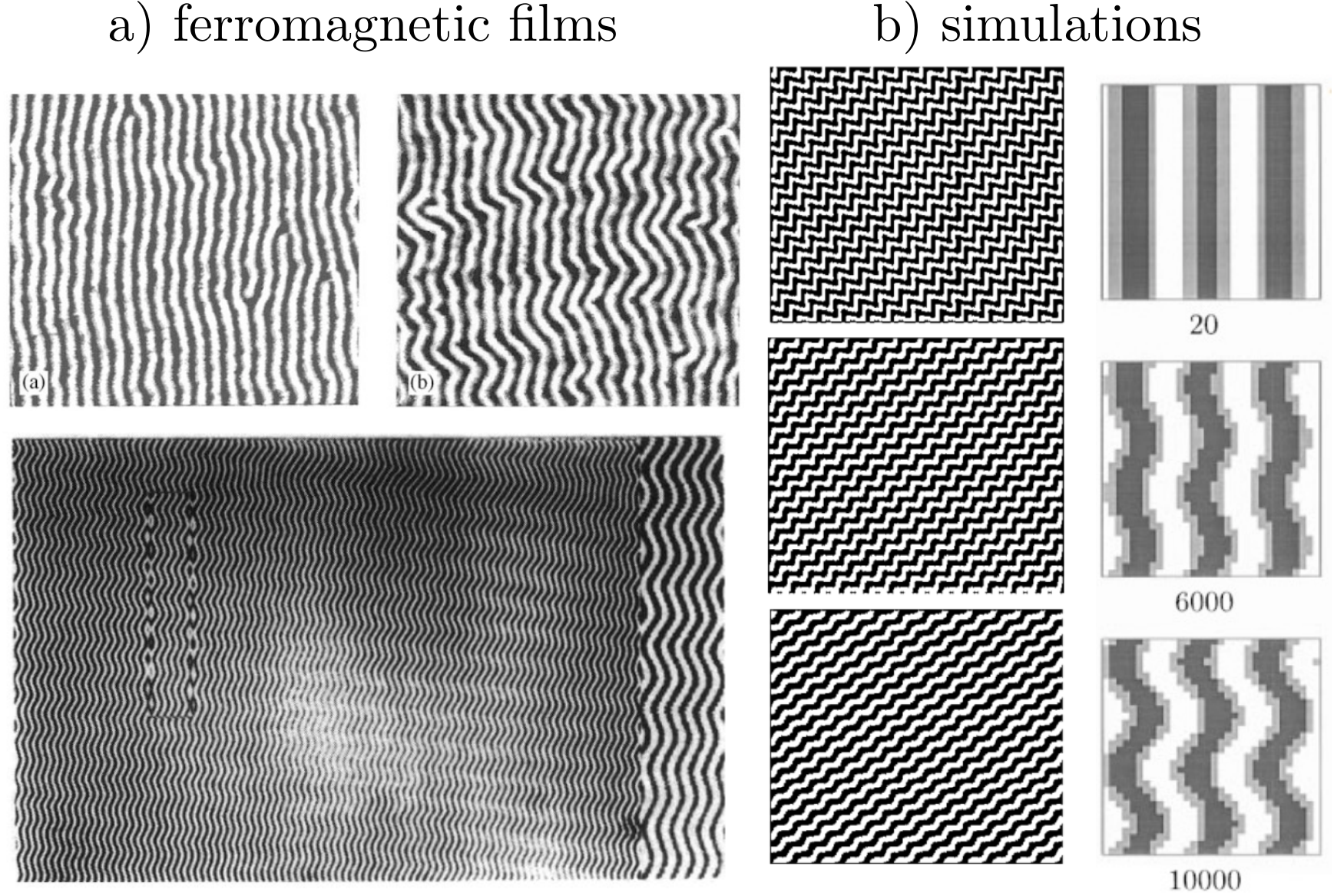}
\caption{(a) Undulating magnetic domains under the influenced of a cycled magnetic field [top two panels from \cite{magsmectic3}] and a temperature change [bottom panel from \cite{seul_evolution_1992}]. Changing the field or the temperature effectively dilates the magnetic stripe domains, inducing an HH-like instability. (b)  Simulations of thin ferromagnetic films [left three panels from \cite{magsmectic2}] and two-dimensional block copolymers with an analogous free energy [right three panels from \cite{doishear}] .\label{fig:magnets}}
\end{figure}

To understand the instability in a ferromagnetic film, consider a coarse-grained magnetization field $M(\mathbf{x})$ describing the magnetization in the thin film at some spatial coordinate $\mathbf{x}=(x,y)$. The free energy for $M(\mathbf{x})$ will  will have  the general form
\begin{equation}
\begin{aligned} f_M & =\int \mathrm{d}^2 \mathbf{x} \bigg[ \frac{D}{2}\,|\nabla M|^2+\frac{r}{2}\, M^2 + \frac{u}{4} \,M^4 \\& +\mu\int \mathrm{d}^2\mathbf{x}'M(\mathbf{x})g(\mathbf{x}-\mathbf{x}')M(\mathbf{x}')\bigg]\end{aligned} \label{eq:FEmag}
\end{equation}
where $g(\mathbf{x}-\mathbf{x}')$ is a Green's function for the dipolar interactions, and $D$, $r$, $u$, and $\mu$ are phenomenological constants related to the material properties.  We expect generally that its Fourier transform is $g(\mathbf{q}) \approx - g_1 |\mathbf{q}|$, which gives us the necessary instability for the formation of a modulated phase with characteristic wavelength $t = 2 \pi/q^*\approx 16 \pi^3 D/(g_1 \mu)$ \cite{modulatedrev2}. In general, there are two types of patterns: an array of circular domains and uniform stripes. In the case of the stripe ground state, the free energy in Eq.~\eqref{eq:FEmag} can be shown to be equivalent to the smectic free energy  in two dimensions \cite{magsmectic1,magsmectic2}. There is then an analogue of the HH instability where a magnetic field is applied, which has the tendency to change the characteristic size $t$ of the domains. Cycling this field has the same dilational effect as a mechanical strain in a smectic system, reviewed in Sec.~\ref{History-Sm}. Thus, the HH instability can be realized in thin magnetic films. An example of the domain shapes one finds under such magnetic field cyclings are shown in Fig.~\ref{fig:magnets}.

\section{Conclusion}

With this review, we shine a spotlight on the applicability of the HH instability to a broad range of materials with periodic ground states. By surveying phenomena in cholesteric and smectic liquid crystals, we illustrate geometrical frustration in lamellar systems as a result of sources ranging from applied fields to boundary conditions. The frustration is then relieved by the HH instability, where undulations produce periodic structures with wavelengths orthogonal to and larger than that of the ground state. 

By considering examples of cholesteric and smectic shells, where the liquid crystal is confined between two concentric and spherical, fluid interfaces, we highlight the role of topological constraints, anchoring conditions, boundary deformability, and curvature. These factors can both trigger the HH instability and shape the resulting patterns. While topological frustration necessitates the existence of discontinuities from the global curvature, the HH instability only cares about how the system looks locally. Topological constraints can dictate that a frustration exists, but the exact reaction to the frustration is a question of energetics and local geometric incompatibilities. The HH instability is then, in its nature, a response to local geometrical frustration.

The generality of the HH mechanism is evident from undulation instabilities appearing in periodic systems beyond the classic thermotropic, lamellar phases. These include twist-bend nematics, lyotropic liquid crystals, and polymers, as well as biological and magnetic materials. After accounting for fluid boundaries, the HH instability can also describe phenomena in living matter, where fluid interfaces are pervasive and activity can strain lamellar structures.

We anticipate the HH instability to become increasingly valuable for understanding the organization of layered materials. The phases formed by bent-core rods are an enduring area of investigation, newly invigorated by the experimental realizations of the splay-bend nematic \cite{2020-molec-splay-bend,2020-colloid-splay-bend,2021-splay-bend-dijkstra}. Future studies on the structures formed by these spatially modulated phases will almost certainly rely upon the HH model, as exemplified by the striations of twist-bend nematics. Moreover, as the field of active liquid crystals progresses, experimental realizations of active cholesterics and active smectics will emerge. The latest theoretical frameworks already invoke the HH mechanism to characterize lamellar distortions from active stresses \cite{active-cholesteric,kole-activeCLC}. Furthermore, cholesteric liquid crystals remain widely employed in optical and elastomeric materials. With undulations being common in the dynamics of cholesterics, the HH instability has the potential to be leveraged for tunable properties in advanced technologies. Indeed, recent work exploited the field-induced undulations of cholesterics to develop dynamic and switchable diffraction gratings and surface coatings \cite{ryabchun-applications-1,ryabchun_cholesteric_2018,ryabchun-applications-2,ryabchun-applications-3}. The HH instability is a generic but often overlooked method of pattern formation that has been and will continue to be integral to the structuring of periodic systems. 

\section{Acknowledgments}

This work would not have been possible without Maurice Kleman's monumental contributions to the field of liquid crystals.  We dedicate this review to his memory and as a tribute to his insight. We thank Kunyun He and Daeseok Kim for their micrographs and useful discussion. G.D and T.L.L. were supported by the French National Research Agency (JCJC Program, Grant 13-JS08-0006-01). T.L.L. acknowledges funding from the French National Research Agency (AAPG Program, Grant 18-CE09-0028-02).  L.T. and R.D.K. were supported in part by NSF Grants DMR-1262047 and DMR-1720530.  This work was supported by a Simons Investigator grant from the Simons Foundation to R.D.K. L.T. acknowledges funding from the Simons Society of Fellows of the Simons Foundation (Grant No. 579910) and from the Marie Curie Individual Fellowship project \textit{EXCHANGE\_inLCs} (Grant No. 892354). M.O.L. acknowledges partial funding from the Neutron Sciences Directorate (Oak Ridge National Laboratory), sponsored by the U. S. Department of Energy, Office of Basic Energy Sciences.

\end{document}